 \documentclass[iop]{emulateapj}
 
\slugcomment{Accepted for publication in the Astrophysical Journal}

\shorttitle{The Sparsest Clusters With O Stars}
\shortauthors{Lamb, J. B.,  et al.}

\usepackage{epsfig}
\begin{document}

\newcommand{\Mcl}{M_{\rm cl}}
\newcommand{\Mcllo}{M_{\rm cl,lo}}
\newcommand{\Nlo}{N_{*,{\rm lo}}}
\newcommand{\mmax}{m_{\rm max}}
\newcommand{\mmaxtwo}{m_{\rm max,2}}
\newcommand{\mratio}{m_{\rm max,2} / m_{\rm max}}


\title{The Sparsest Clusters With O Stars}


\author{J. B. Lamb\altaffilmark{1}, M. S. Oey\altaffilmark{1}, J. K. Werk\altaffilmark{1,2}, and L. D. Ingleby\altaffilmark{1}}
\altaffiltext{1}{Department of Astronomy, University of Michigan, 500 Church St., Ann Arbor, MI 48109, USA}
\altaffiltext{2}{Department of Astronomy, Columbia University, 550 West 120th St. New York, NY 10027 USA}



\begin{abstract}
 
There is much debate on how high-mass star formation varies with
environment, and whether the sparsest star-forming environments are
capable of forming massive stars.  To address this issue,
we have observed eight apparently isolated OB stars in the SMC
using {\it HST}'s Advanced Camera for Surveys.  Five of these objects
appear as isolated stars, two of which are confirmed to be runaways.  The remaining
three objects are found to exist in sparse 
clusters, with $\lesssim 10$ companion stars revealed, having masses of
1 -- 4 $M_\odot$.  Stochastic effects dominate in these sparse clusters, so we
perform Monte Carlo simulations to explore how our observations fit
within the framework of empirical, galactic cluster properties.  We
generate clusters using a simplistic --2 power-law distribution
for either the number of stars per cluster ($N_*$) or cluster mass 
($\Mcl$).  These clusters are then populated with stars randomly
chosen from a Kroupa IMF.  We find that simulations with cluster
lower-mass limits of $\Mcllo \geq 20M_\odot$ and $\Nlo \geq 40$
match best with observations of SMC and Galactic OB star populations.
We examine the mass ratio of the second-most massive and most massive
stars $\mmaxtwo/\mmax$, finding that our observations all exist below the 20th
percentile of our simulated clusters.  
However, all of our observed clusters lie within the 
parameter space spanned by the simulated clusters, although some are
in the lowest 5th percentile frequency.  These results
suggest that clusters are built stochastically by randomly sampling
stars from a universal IMF with a fixed stellar upper-mass limit.  In
particular, we see no evidence to suggest a $\mmax - \Mcl$ relation.
Our results may be more consistent with core accretion models of star
formation than with competitive accretion models, and they are
inconsistent with the proposed steepening of the integrated
galaxy IMF (IGIMF).

\end{abstract}





\section{Introduction}
\label{intro}

Most observational properties of galaxies and stellar
populations are directly influenced by 
star formation.  Massive stars, although small in number,
disproportionally affect observables such as the integrated light from
galaxies, feedback effects, star formation rates, and many others.
However, there is significant debate regarding the conditions under
which massive stars form.  The competitive accretion model of star
formation requires that a population of low-mass stars must form in
the presence of a high-mass star (e.g. Bonnell et al. 2004), while the core accretion model of
star formation allows for massive stars to form in relative isolation (e.g. Krumholz et al. 2009).  Empirical studies are
similarly divided on the connection between cluster mass and massive star formation.  A physical relationship between the two (e.g. Weidner \& Kroupa 2006, hereafter WK06) would indicate that massive stars {\it always} form in clusters, while random statistical sampling (e.g. Elmegreen 2000) would indicate that massive stars {\it preferentially} form in clusters.  One observational method to
differentiate between these theories is to examine the sparsest
environments where massive stars are found.  A targeted study of field
massive stars can quantify limitations on the minimum stellar
groupings needed for massive star formation.  Such a study, which
we present in this work, provides direct 
observational constraints for the two competing theories of star
formation. 

The core accretion model suggests that stars of all masses form by a
fragmentation process in molecular clouds, where cores collapsing due
to self-gravity represent the mass available to form an individual
star or stellar multiple system (e.g. Shu et al. 1987).  In
this model, massive stars must necessarily form from massive cores;
however, it is unclear how such massive cores (up to hundreds of Jeans
masses) can collapse without further fragmentation.  Analytic models
by Krumholz \& McKee (2008) suggest that sufficiently dense clouds having
surface densities $\ge 1$ g/cm$^{-2}$ will trap stellar and accretion
radiation that heats clouds and prevents further
fragmentation.  Additionally, 3-D hydrodynamic simulations by Krumholz
et al. (2009) reveal that self-shielding occurs along filaments
resulting from gravitational and Rayleigh-Taylor instabilities, thereby
channeling gas onto massive stars despite radiation pressures that 
dominate gravitational forces.  These simulations result in the
formation of a high-mass star or multiple system with a small
companion population of low-mass stars.  Similarly, Spaans \& Silk
(2000) show that the star formation properties of a gravitationally
collapsing molecular cloud are highly dependent upon the equation of
state of that cloud.  For a cloud polytropic equation of state given by
$P \propto \rho^\gamma$ where $P$ is the thermal pressure and $\rho$
is the gas density, they concluded that $\gamma >1$ yields a peaked
stellar initial mass function (IMF) rather than a power-law distribution. 
Li et al. (2003) conduct further simulations and find that molecular
clouds with $\gamma >1$ will most likely result in the formation of
massive, isolated stars. 

In contrast, the competitive accretion model suggests that fragmentation only produces low-mass stars, with high-mass stars formed by winning a competition for the remaining gas (e.g. Zinnecker 1982).  In this scenario, the mass of a star is highly dependent upon the star-forming environment, with high-mass stars preferentially located at the bottom of the gravitational potential where the majority of a cluster gas reservoir gets funneled (Bonnell et al. 2001).  This model of massive star formation requires that massive stars form in a clustered environment, with an explicit relation between the mass of a cluster ($\Mcl$) and the mass of the most massive star in the cluster ($\mmax$) given by $\Mcl \propto \mmax^{1.5}$ (Bonnell et al. 2004).  Thus, competitive accretion forms massive stars along with a fully populated cluster of lower mass companion stars (Bonnell et al. 2007).

One of the primary differences between the observational predictions
of these models is in the 
formation of high-mass stars in low $\Mcl$ environments.  Competitive
accretion argues that formation of a high-mass star in a low-mass
cluster is extremely difficult, while core accretion places no formal
constraint on cluster mass.  The competitive accretion model implies
that the IMF is not a universal
property of star formation, but instead tends to limit $\mmax$
for a given $\Mcl$.  However, the IMF has been robustly
verified for a wide range of star-forming environments, leading many
to argue that $m_{\rm up}$, the upper stellar mass limit, is a universal property of star
formation, regardless of environment (see Elmegreen 2000, 2006,
2008).  In the case of a universal IMF, the relationship between $\mmax$
and $\Mcl$ is determined by the statistical mean (Oey \&
Clarke 2005).  Not everyone agrees with the universality of the IMF; for example, WK06 argue for the existence of
a deterministic $\mmax$-$\Mcl$  relation using analytic models.  They
further their argument by aggregating a data sample of Galactic clusters
from which they find a strong correlation between $\mmax$ and $\Mcl$,
closely following the derived $\mmax$-$\Mcl$  relation
in Weidner \& Kroupa (2004).  Weidner et al. (2010a), who update and
greatly expand the observational data set, conclude using a variety of
statistical tests, that it is highly unlikely that the sample of
Galactic clusters is generated from random sampling 
of a universal IMF. 

However, Selman \& Melnick (2008), using the same data from
WK06, argue that the correlation of $\mmax$ with $\Mcl$ may be caused by the quick dispersal of clusters dominated by a single massive star due to gravitational instabilities.  Since these objects would no longer be identifiable as clusters, such a dispersal effect would bias the WK06 cluster sample against clusters that formed with a flatter-than-Salpeter mass function, leaving behind only those clusters that follow a more standard Salpeter mass function.  Maschberger \& Clarke (2008) complement the WK06 data set with a sample of very small
clusters from Testi et al. (1997) and find that the resultant ensemble of clusters does not significantly deviate from
the expectations of a universal stellar IMF, when examining the correlation between the number of stars in a cluster ($N_*$) and $\mmax$.  They argue that analyses using $N_*$ instead of $\Mcl$ are more reliable since $N_*$ is a directly observable quantity, while $\Mcl$ must be inferred.  They caution that observational and sample selection effects can greatly influence the correlation of $\mmax$ with $\Mcl$ or $N_*$ and that much more observational data is needed to reach a conclusion.  

A $\mmax$-$\Mcl$  relation, if it exists, has broad
implications for cumulative stellar populations of galaxies.
The power-law form of the cluster mass function is robust, similar to
the IMF, with empirical derivations from
a wide range of Galactic objects and environments, generally consistent
with a power-law slope of --1.7 to --2.3 (\S 3.2).  Coupling the cluster
mass function, which is highly weighted towards low-mass clusters, with
a deterministic $\mmax$-$\Mcl$ relation can have a  large effect on
the integrated galactic initial mass function (IGIMF) for stars.  The primary
consequences include a decrease in the expected number of OB stars
within galaxies and an overall steepening of the IGIMF for the
composite stellar population of a galaxy (Kroupa \& Weidner 2003;
Weidner \& Kroupa 2005).  A steepened IGIMF appears to successfully
reproduce a variety of poorly-understood observationally-derived
relations, including the dwarf galaxy mass-metallicity relation
(K{\"o}ppen et al. 2007),  global correlations between H$\alpha$ to UV
flux ratios and galaxy mass (Hoversten \& Glazebrook 2008; Lee et al. 2009; Meurer et al. 2009), and
sharp radial surface brightness truncations in H$\alpha$ compared to
more extended-UV emission in the outer disks of nearby galaxies
(Thilker et al. 2007; Pflamm-Altenburg et al. 2009). Such
observational relations appear to arise naturally from clustered star
formation and the  $\mmax$-$\Mcl$ relation implicit to a steepened
IGIMF.  Maschberger et al. (2010) found evidence of a
steepened IGIMF in the competitive accretion simulations of Bonnell et al.
(2003, 2008), linking these two theories under their common assumption
of a  $\mmax$-$\Mcl$ relation.  Considering the
far-reaching implications of a steepened IGIMF, it is of utmost
importance to examine its validity using observational constraints of
isolated O stars.

One observational method to test the assertion of a $\mmax$-$\Mcl$ relation is to look for
isolated, massive star formation. Field O stars are abundant in the literature (e.g., Massey et al. 1995) and may account for 25-30\% of the O star population in a galaxy (Oey, et al. 2004).
While many of these stars are likely to be runaway stars from clusters, the remainder of
field stars with no evidence of companions would be difficult to incorporate into the $\mmax$-$\Mcl$  relation proposed by WK06 and inherent to the theory of competitive accretion. In a study of Galactic field O stars, de Wit et al. (2004, 2005) find that $4 \pm 2$\% of all Galactic O stars appear to have formed in isolation, without the presence of a nearby cluster or evidence of a large space velocity indicative of a runaway star.  This value is in agreement with the 5\% of isolated O stars (defined as O stars without any companion O or B stars) found from Monte Carlo simulations of clusters (Parker \& Goodwin 2007).

In this paper, we examine the stellar environment around field O stars to probe the
limiting cases where O stars form in the sparsest stellar groups, or
in near isolation.

\section{Observations}

\subsection{{\it HST} Imaging Observations}
\label{hstsec}

We target field O stars in the SMC for this study because this nearby galaxy
offers a view unobscured by gas and dust, allowing clear
identification of the field massive stars and any low-mass companions.
Our targets are taken from the work of Oey et al. (2004), who applied a
friends-of-friends algorithm to photometrically identify OB star
candidates, thereby identifying clusters and field stars in this sample.
For this study, all of the targets were spectroscopically verified as O or early
B stars, and all appeared isolated in ground-based imaging.
In a pilot SNAP program, we exploit the 0.05$\arcsec$/px spatial
resolution of the Advanced Camera for Surveys (ACS) Wide-Field Camera
aboard the {\it Hubble Space Telescope (HST)} 
to search for low-mass stars associated with the target OB stars.
Unfortunately, Cycle 14 had an unusually low SNAP return, and we
obtained observations of only eight targets.
Table \ref{targets} lists our sample.  Column 1 gives the star's ID
from the catalog of Azzopardi \& Vigneau (1975); the star smc-16 was
catalogued by Massey et al. (1995).  Columns 2, 3, and 4 list the
right ascension, declination, and {\it V} magnitude, respectively,
taken from Massey (2002).   Column 5 gives the spectral 
types, in some cases derived from our observations described below in \S
\ref{isosec}.  Column 6 gives our mass estimate derived from the
spectral type as described below, in \S \ref{imfsec}.  Column 7 gives
our measured heliocentric radial velocities (see \S 2.3). 

\begin{deluxetable*}{ccccccc}
  \tabletypesize{\small}
  \tablewidth{0pc}
  \tablecaption{SMC Field OB Stars}
  \tablehead{\colhead{Field OB Star} & \colhead{RA\tablenotemark{a}} & \colhead{Dec\tablenotemark{a}} & \colhead{{\it V}\tablenotemark{a}} & \colhead{Spectral Type\tablenotemark{b}} & \colhead{Mass ($M_\odot$)}  & \colhead{RV\tablenotemark{b} (km s$^{-1}$)}}
  \startdata

	smc16	& 01:00:43.94 & -72:26:04.9 & 14.38 & O9 V & $23 \pm  2$ & $121 \pm  21$	   \\
    	AzV 58	& 00:49:57.84 & -72:51:54.4 & 14.29 & B0.5 III	& $22 \pm  2$ & $146 \pm  11$  \\
	AzV 67 	& 00:50:11.13 & -72:32:34.8 & 13.64 & O8 V	& $37 \pm  3$ & $159 \pm  13$ \\
	AzV 106	& 00:51:43.36 & -72:37:24.9 & 14.18 & B1 II	& $18 \pm  1$ & $150 \pm  12$ \\
	AzV 186	& 00:57:26.99 & -72:33:13.3 & 13.98 & O8 III((f))\tablenotemark{c}	& $33 \pm  3$ & $159 \pm  10$   \\
    	AzV 223	& 00:59:13.41 & -72:39:02.2 & 13.66 & O9 II\tablenotemark{d} & $32 \pm  2$ & $189 \pm  7$\tablenotemark{e}	  \\
	AzV 226	& 00:59:20.69 & -72:17:10.3 & 14.24 & O7 III((f))\tablenotemark{c} & $35 \pm  3$ & $146 \pm  21$	 \\
	AzV 302	& 01:02:19.01 & -72:22:04.4 & 14.20 & O8.5 V\tablenotemark{e} & $27 \pm  2$ & $161 \pm  11$\tablenotemark{e}	 \\
    \enddata
    \tablenotetext{a} {From Massey (2002).}
    \tablenotetext{b} {Observed with the IMACS multislit spectrograph on the 6.5m Magellan/Baade telescope, unless otherwise stated.}
    \tablenotetext{c} {From Massey (2009).}
    \tablenotetext{d} {From Evans (2006).}
    \tablenotetext{e} {Observed with the MIKE echelle spectrograph on the 6.5m Magellan/Clay telescope.}
    \label{targets}
\end{deluxetable*}

\begin{figure*}
	\begin{center}
	\includegraphics[scale=.31,angle=0]{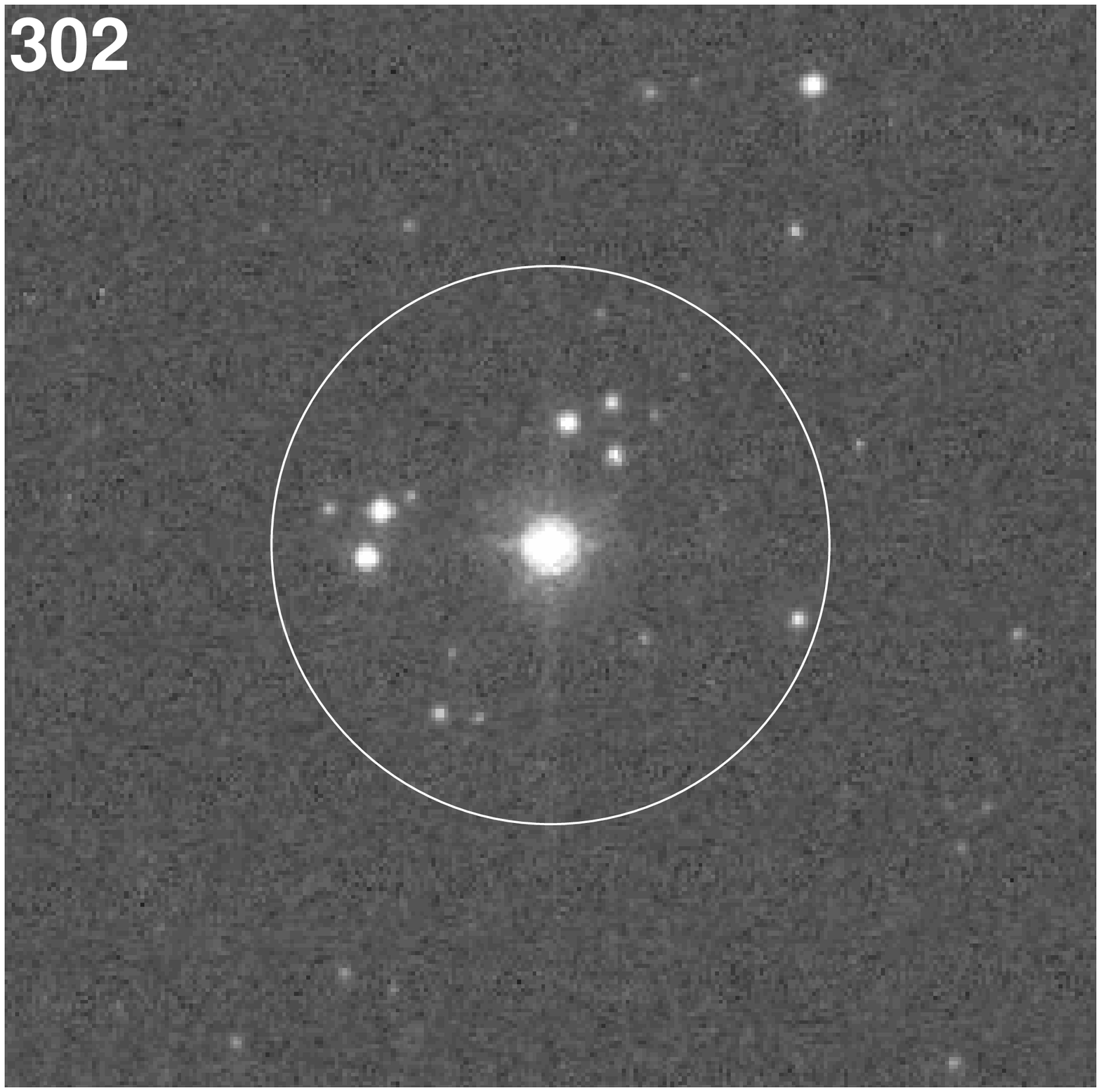}
	\includegraphics[scale=.31,angle=0]{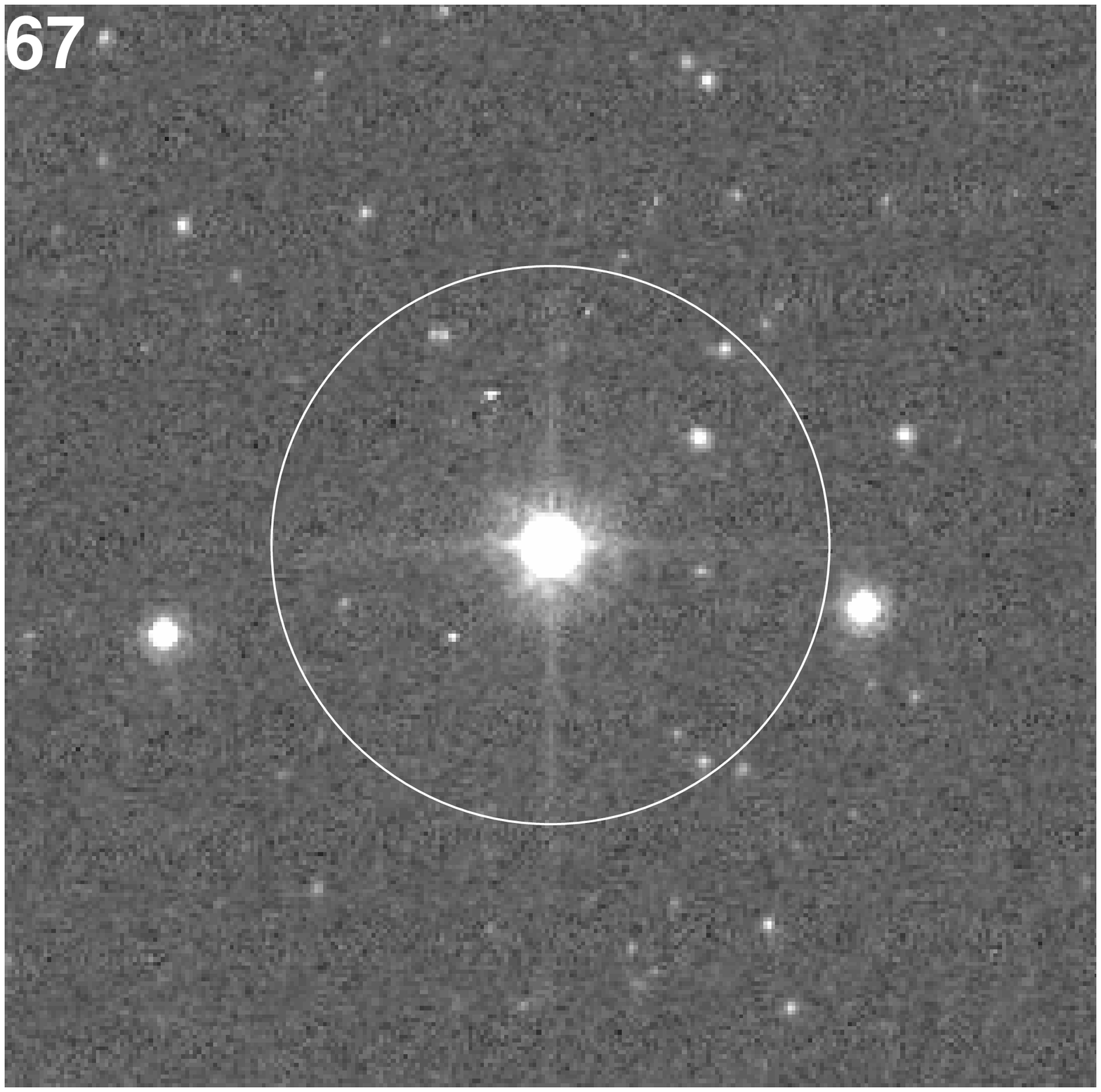}
	\includegraphics[scale=.31,angle=0]{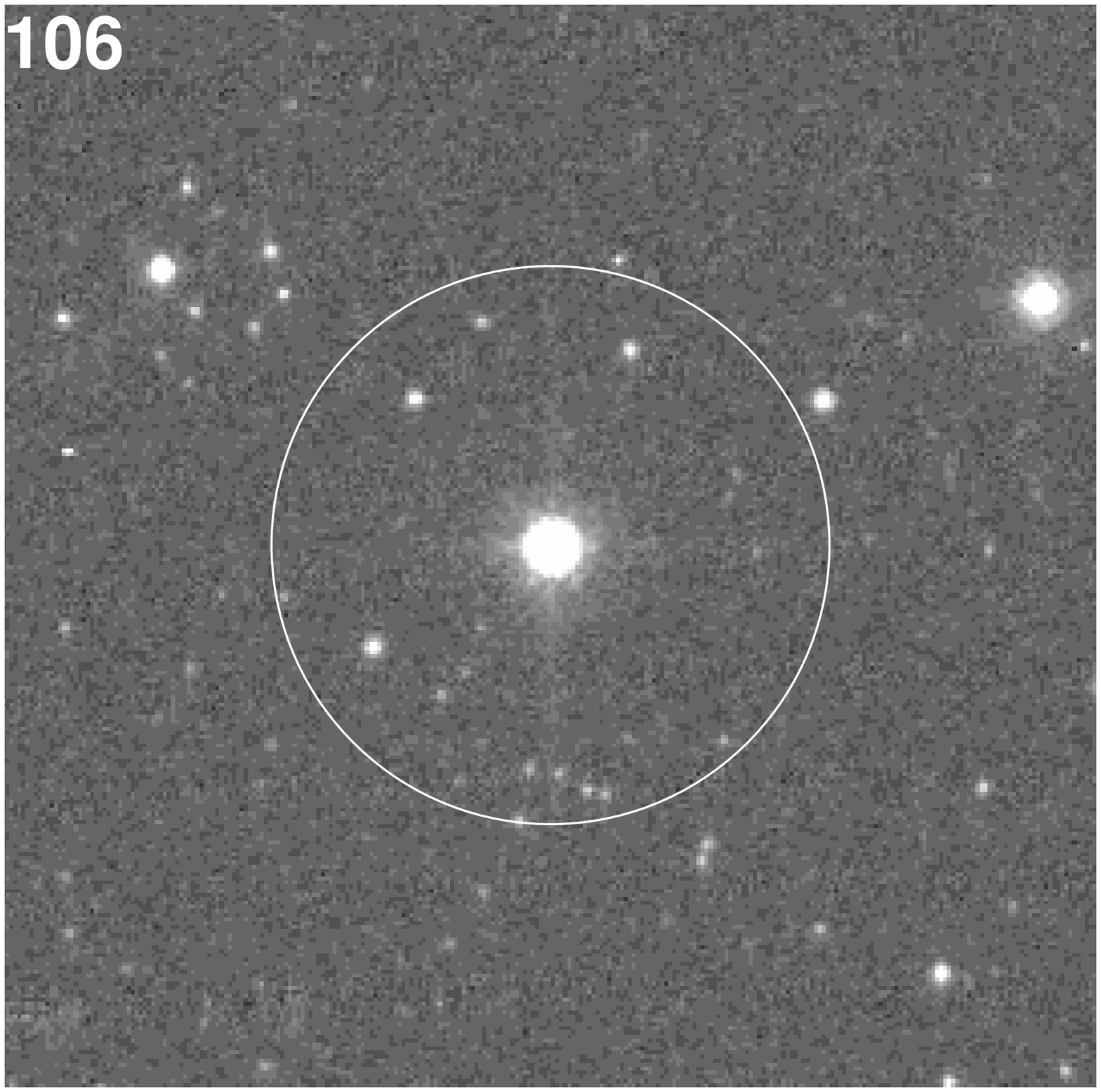}
	\includegraphics[scale=.31,angle=0]{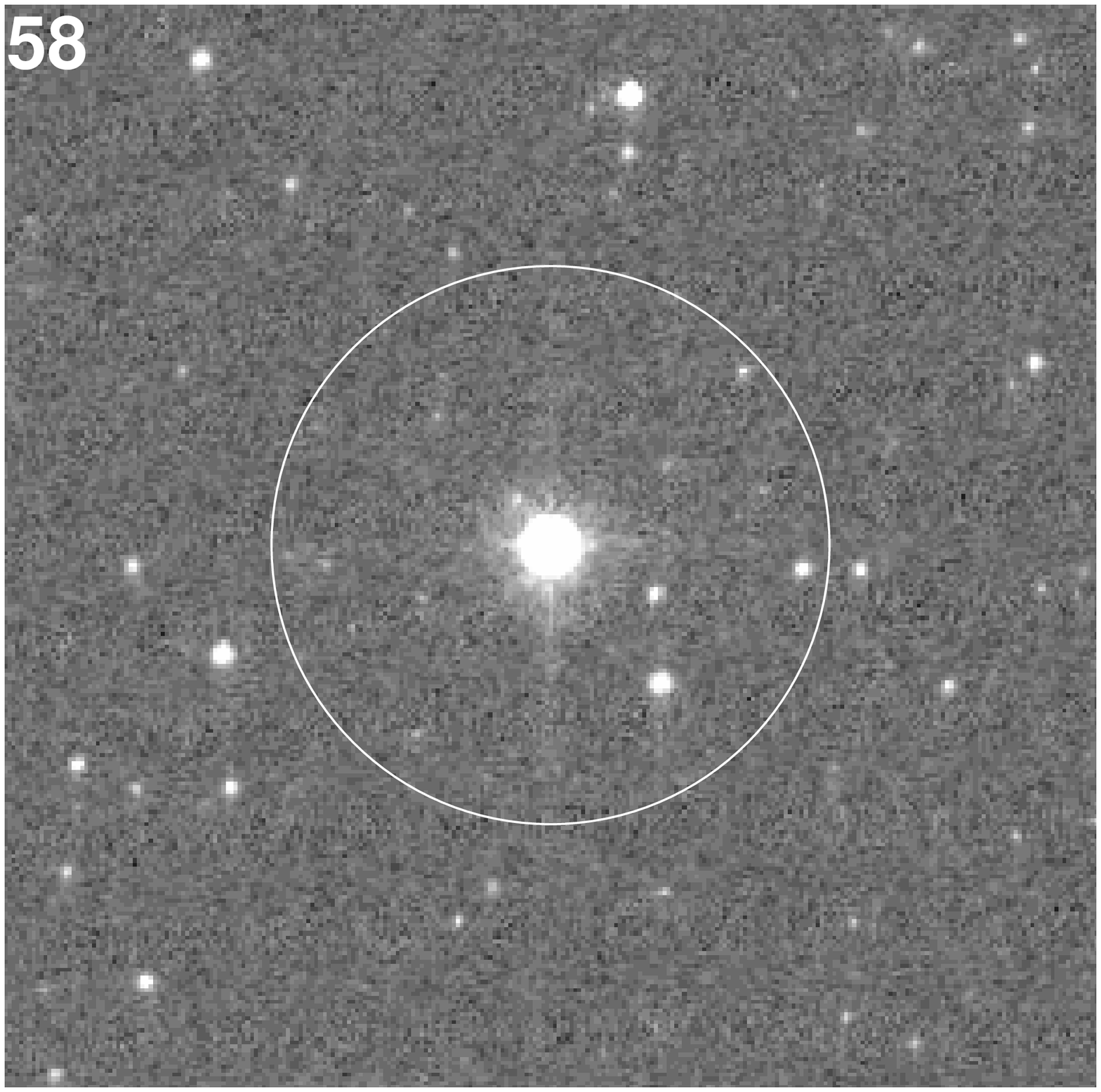}
	\includegraphics[scale=.31,angle=0]{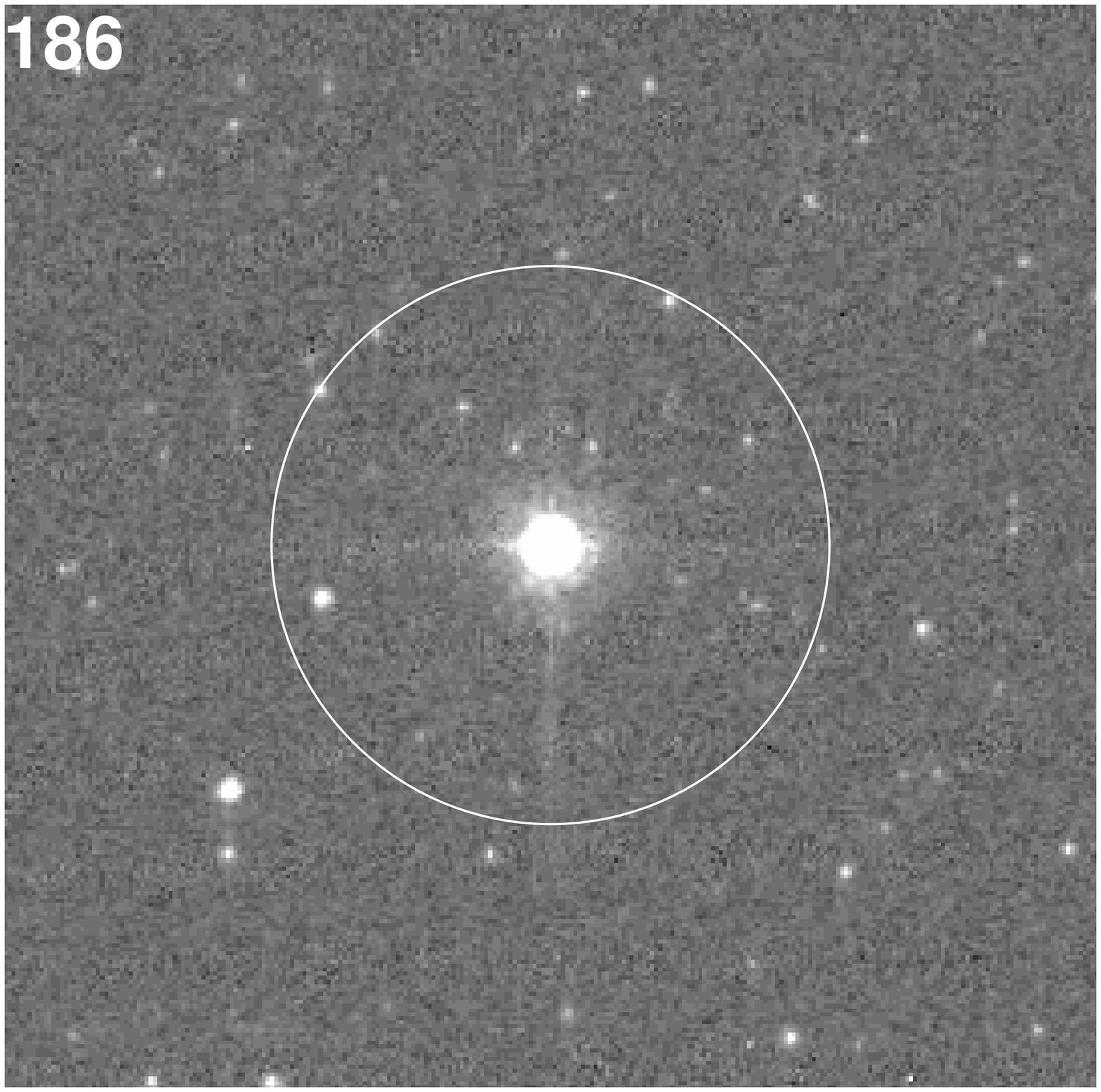}
	\includegraphics[scale=.31,angle=0]{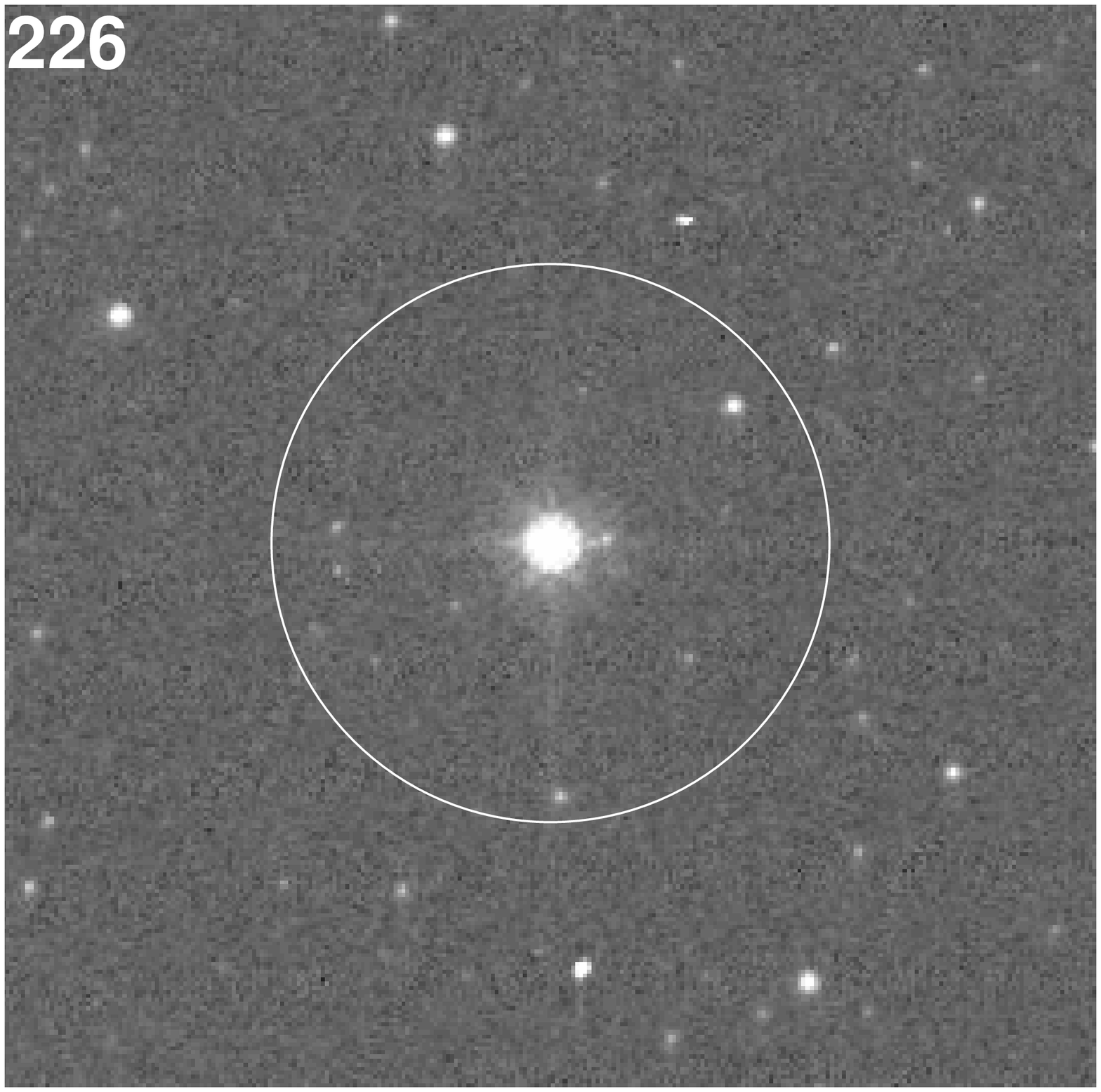}
	\includegraphics[scale=.31,angle=0]{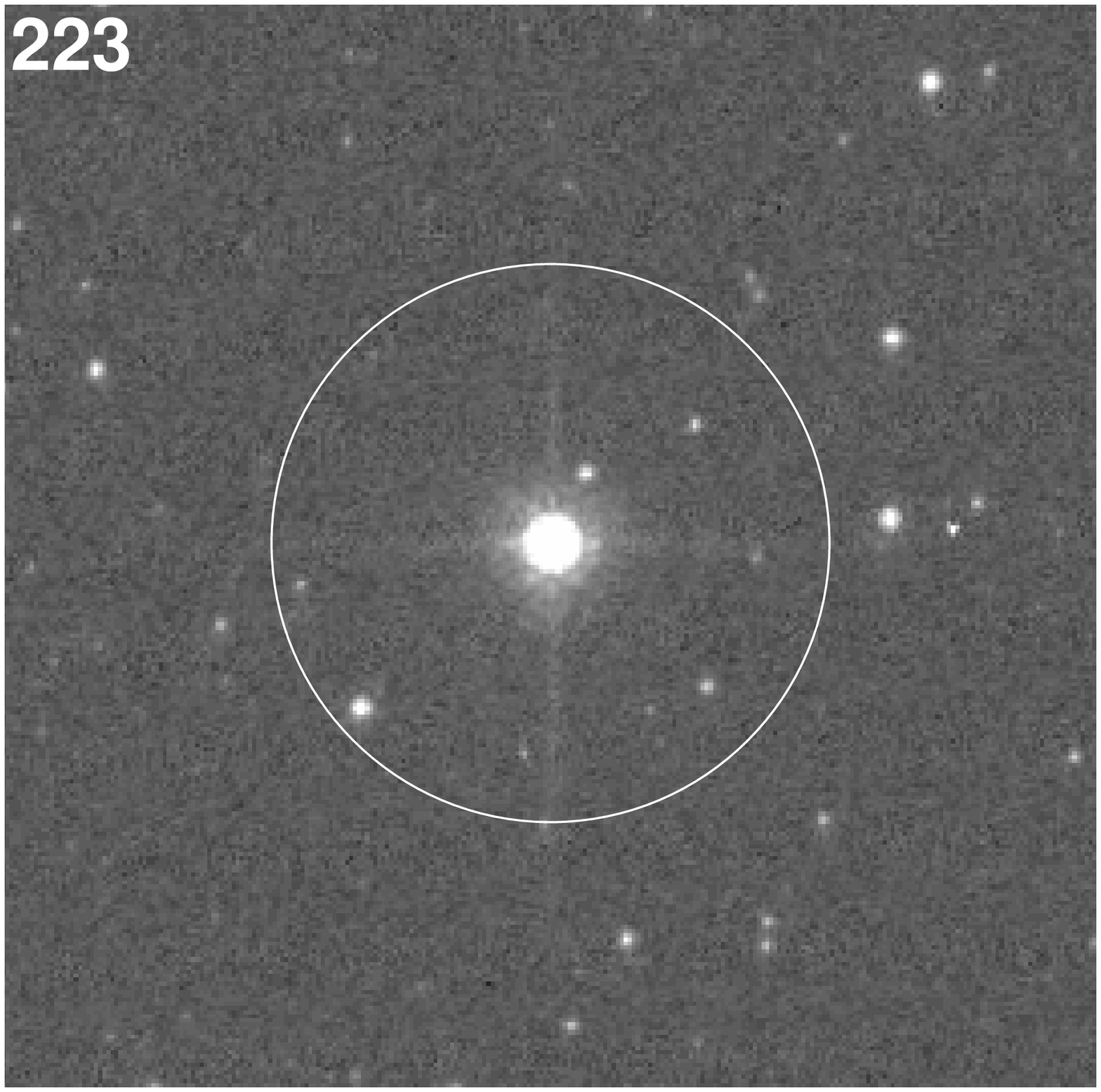}
	\includegraphics[scale=.31,angle=0]{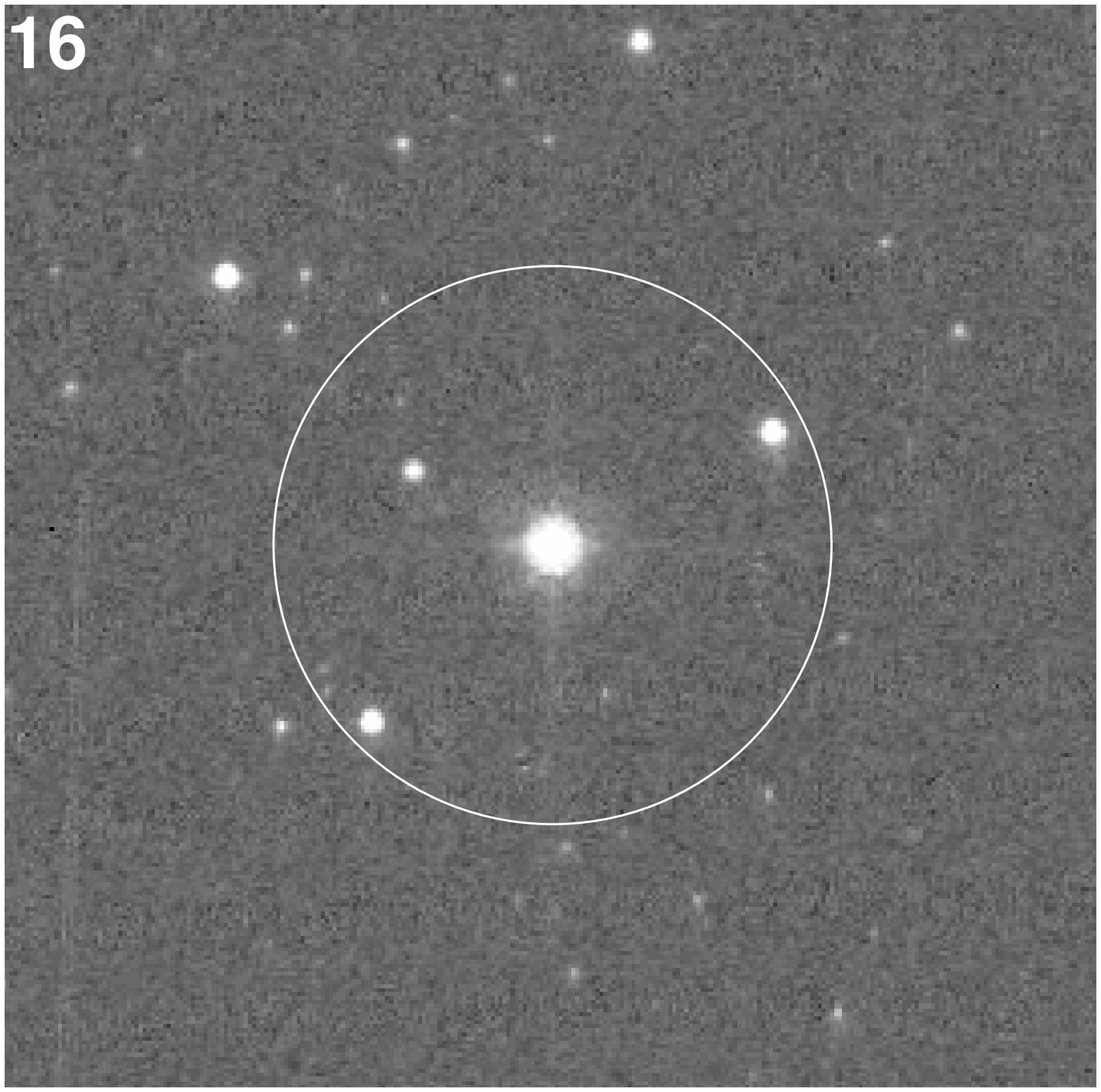}
	\caption{{\it HST} F814W images of each target star, with the star's ID number from the Azzopardi \& Vigneau (1975) catalog listed in the upper left of each panel.  The
	circle corresponds to a radius of 1 parsec.  The top row
	contains fields showing a stellar density enhancement.  The
	middle and bottom rows contains fields with no density
	enhancement.  The bottom row contains stars found to be
	runaways.} 
	\label{hst}
	\end{center}
\end{figure*}

We obtained exposures of 6 seconds in the F555W band and 18 seconds in the F814W band.  Figure \ref{hst}
shows the F814W images of each object, with a circle of radius one parsec (3.4 arcsec)
for reference, adopting an SMC distance of 60 kpc (Harries et al. 2003).  The F555W
exposures are complete down to 21st magnitude while the F814W are
complete down to 22nd magnitude.  

For each {\it HST} field, we use the IRAF DAOPHOT package to
identify stars and obtain their photometry, using a combination of
aperture photometry and PSF fitting.
The aperture photometry is more reliable due to an
undersampled PSF; however, PSF fitting was unavoidable in the case of
close companions.  The aperture photometry was done with an aperture
radius of six pixels, while PSF fitting was done with a two-pixel FWHM and
corrected to match the six-pixel aperture.  For the F555W images,
3--5\% of stars require PSF fitting, while in the F814W band, 10--15\%
require PSF fitting.  PSF fitting did not reveal close companions for any of the target stars; however,
the PSF subtraction of the target stars was very non-uniform, so companions cannot be entirely ruled out.

We used two separate methods to identify possible companion stars
associated with the targets:
(1) an analysis of the stellar density surrounding the
OB star and (2) a friends-of-friends algorithm.  For the first
method, we computed the stellar surface density of the field as a
function of radius from the target OB star.  We performed this
analysis using the F814W exposures, since they probe to a
fainter magnitude than the F555W images.  To measure the average
stellar density of the field,  we used an annulus centered on the OB
star, with an inner radius ranging from 5$\arcsec$ to 10$\arcsec$, 
and an outer radius 10$\arcsec$ beyond the inner radius; the 
annuli were positioned to avoid any obvious stellar clustering. 
We define a density enhancement to
occur when the observed stellar density is higher than the average
stellar density of the field, including the statistical uncertainty.  
The probability that the
observed stellar distribution matches the expected Poisson
distribution, for a given background stellar density, is:
\begin{equation}
\label{e:poisson}
f(k,x)=x^k *e^{-x}/k!
\end{equation}
where $x$ is the expected star count within a given radius and $k$ is
the observed star count within that radius.  A more useful
value, which we designate the ``field
probability'' $P(f)$, is the likelihood that the population follows the
stellar density of the field:
\begin{equation}
\label{fprob}
P(f)=1 - \sum_{0}^{k} f(k,x)
\end{equation}
The right term in equation~\ref{fprob} yields the probability of
observing more than $k$ stars within a certain radius, and so $P(f)$
represents the probability of obtaining the background field.
A smaller $P(f)$ therefore indicates an increased likelihood
of clustering.

\begin{figure*}
	\begin{center}
	\includegraphics[scale=.46,angle=0]{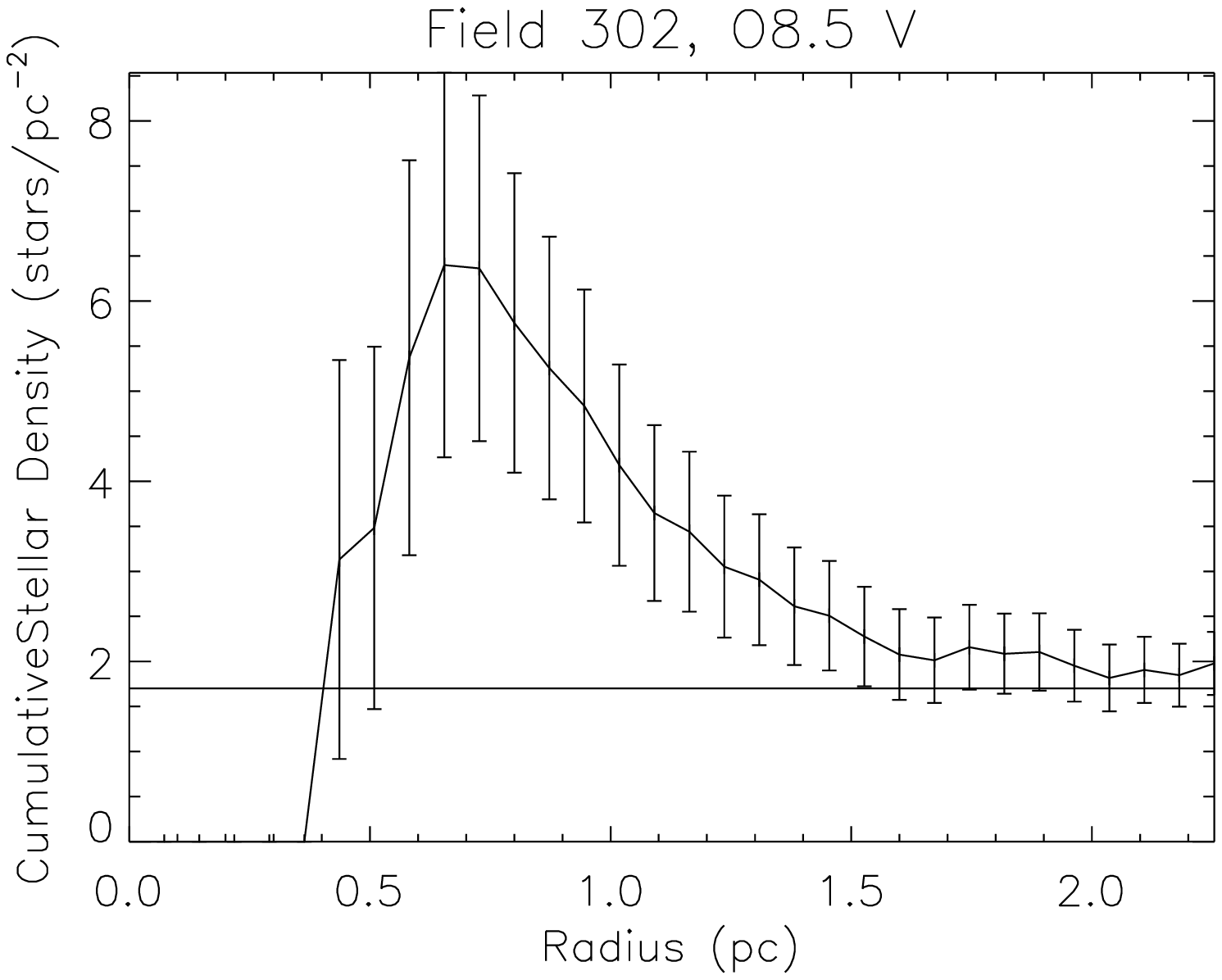}
	\includegraphics[scale=.46,angle=0]{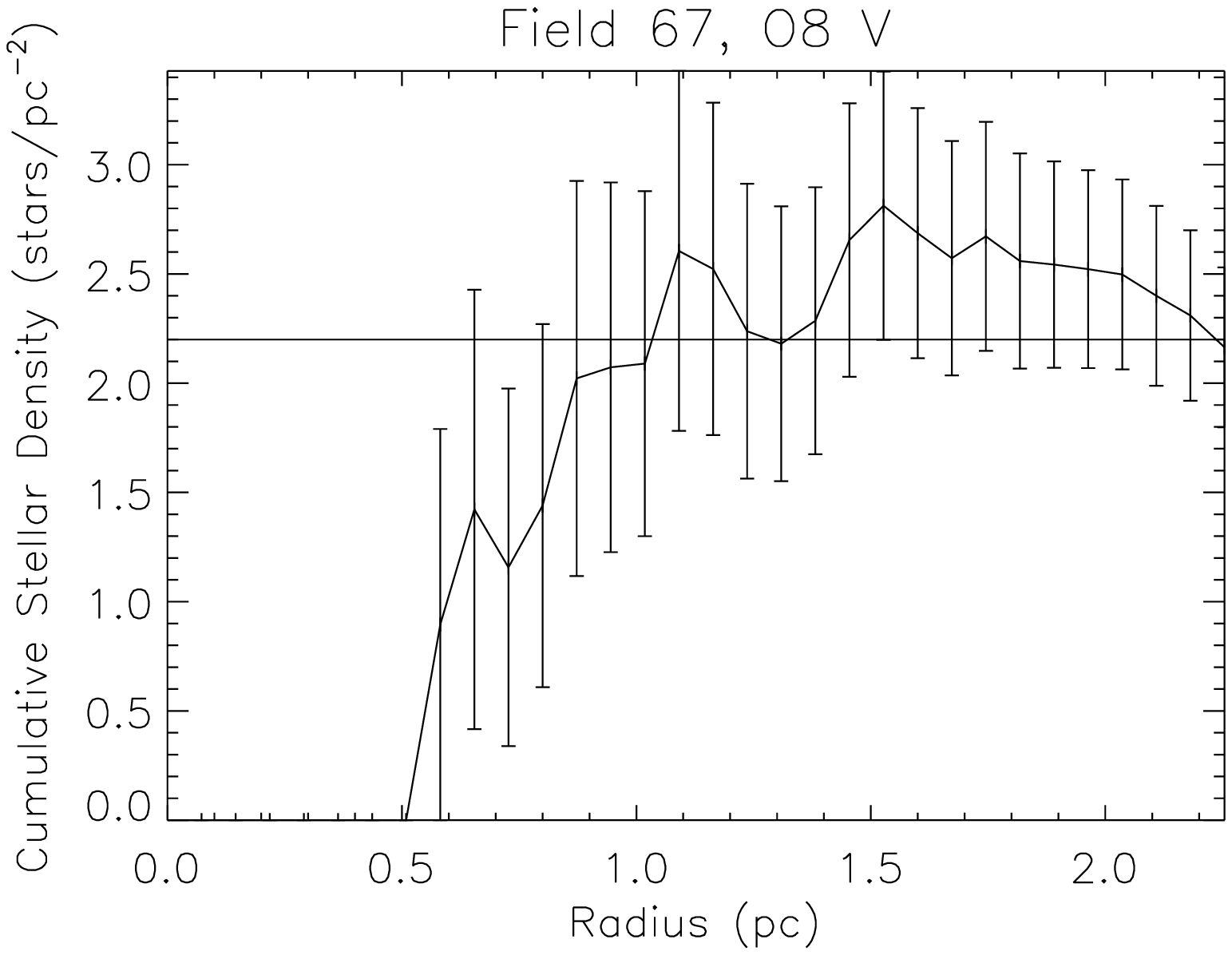}
	\includegraphics[scale=.46,angle=0]{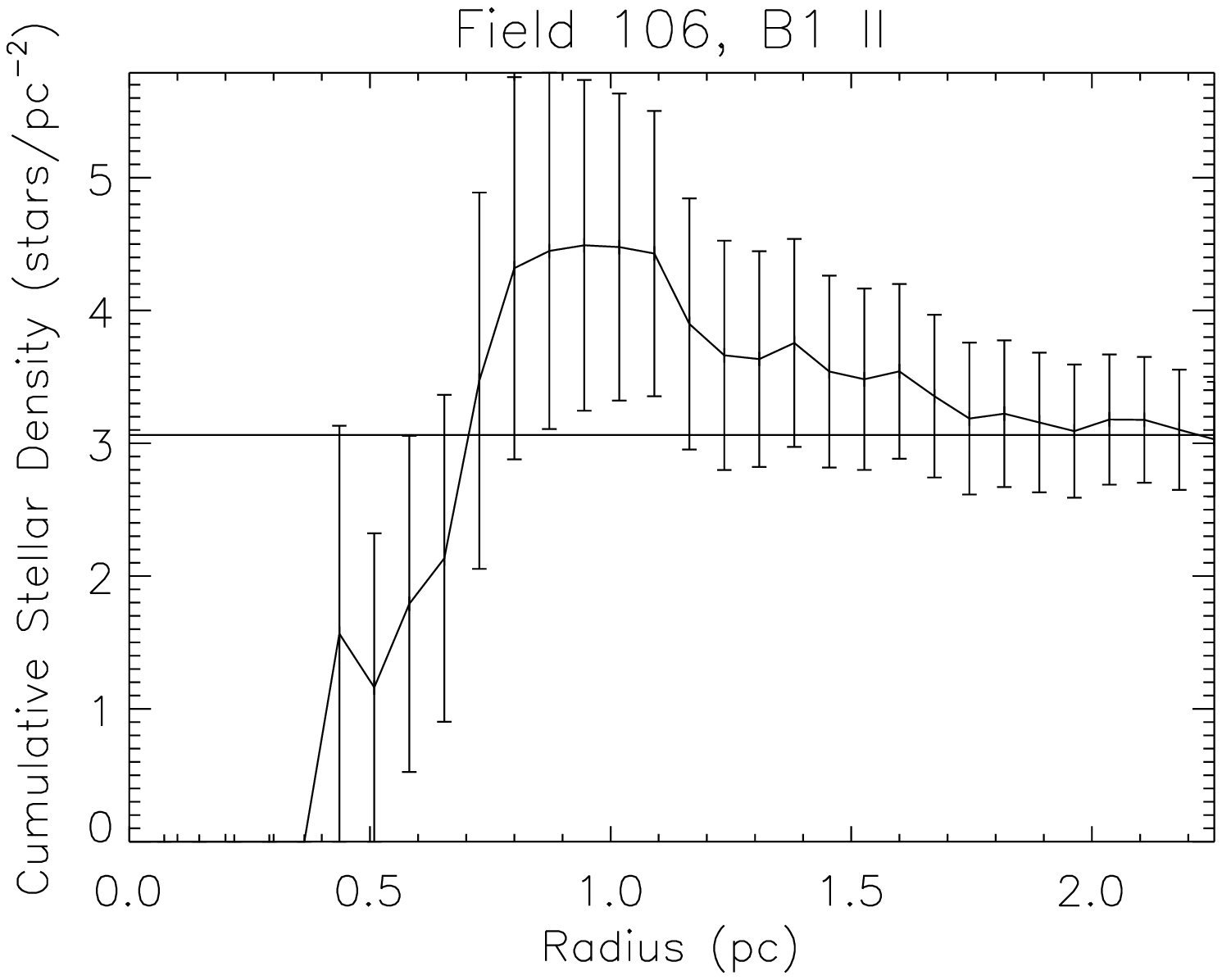}
	\includegraphics[scale=.46,angle=0]{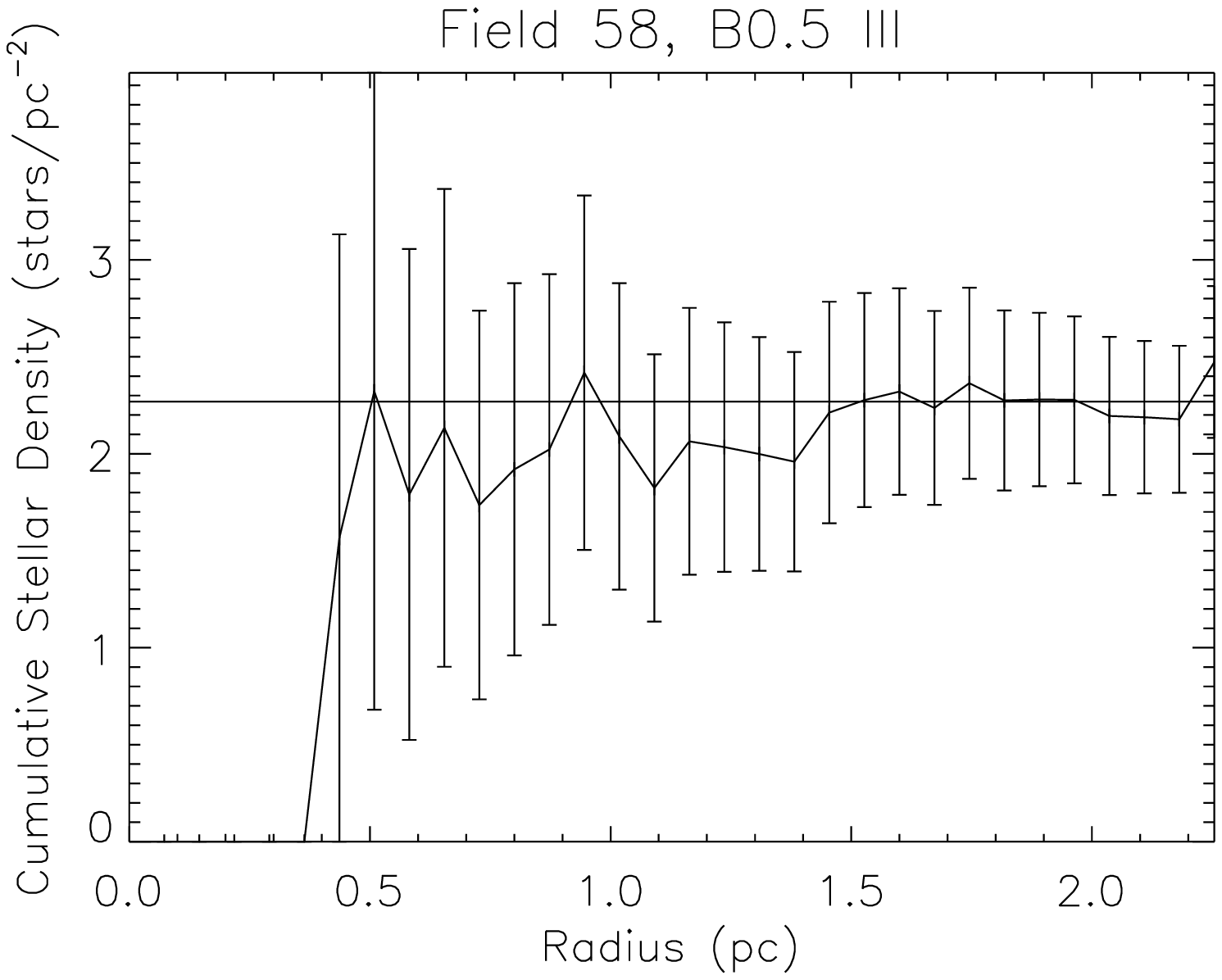}
	\includegraphics[scale=.46,angle=0]{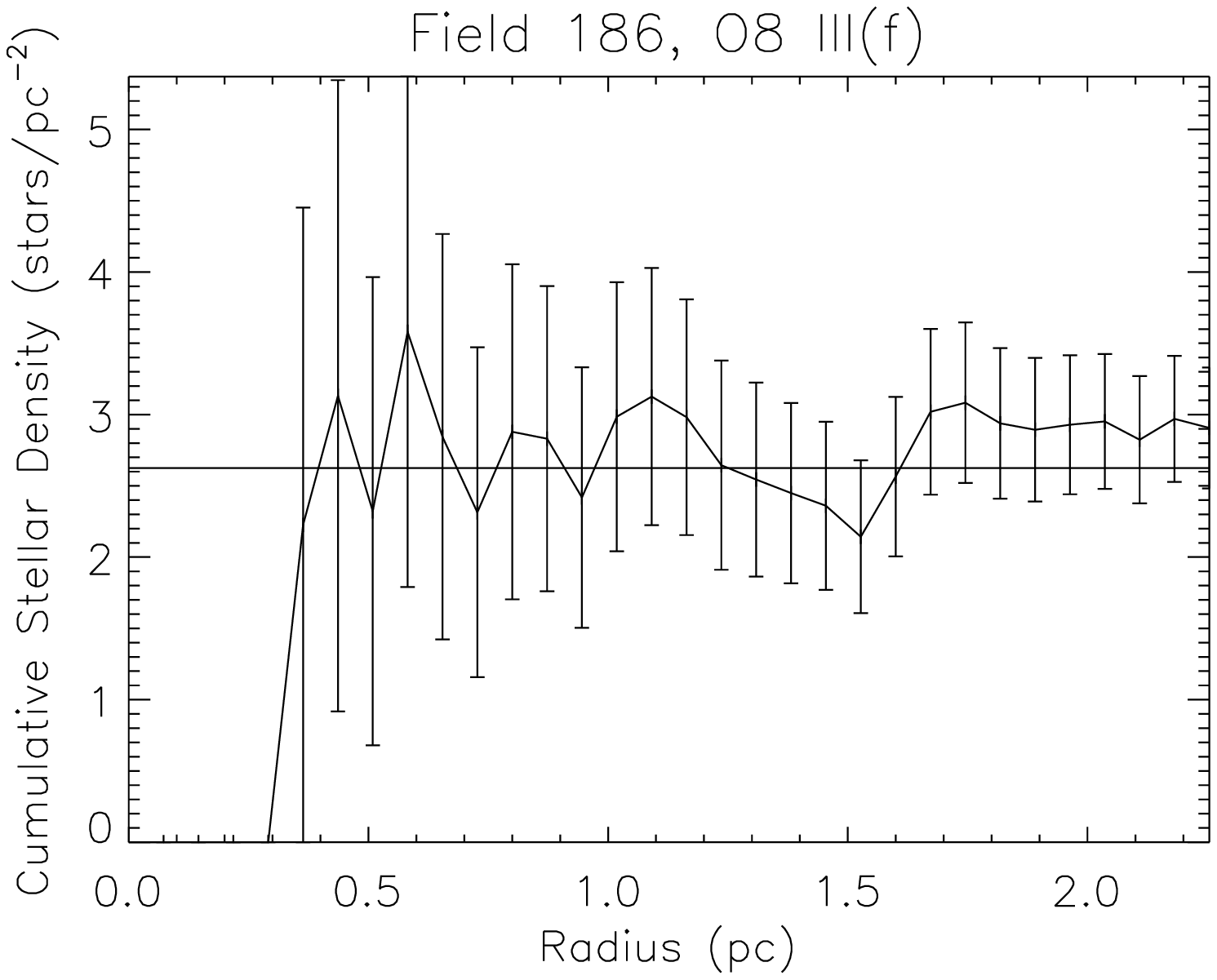}
	\includegraphics[scale=.46,angle=0]{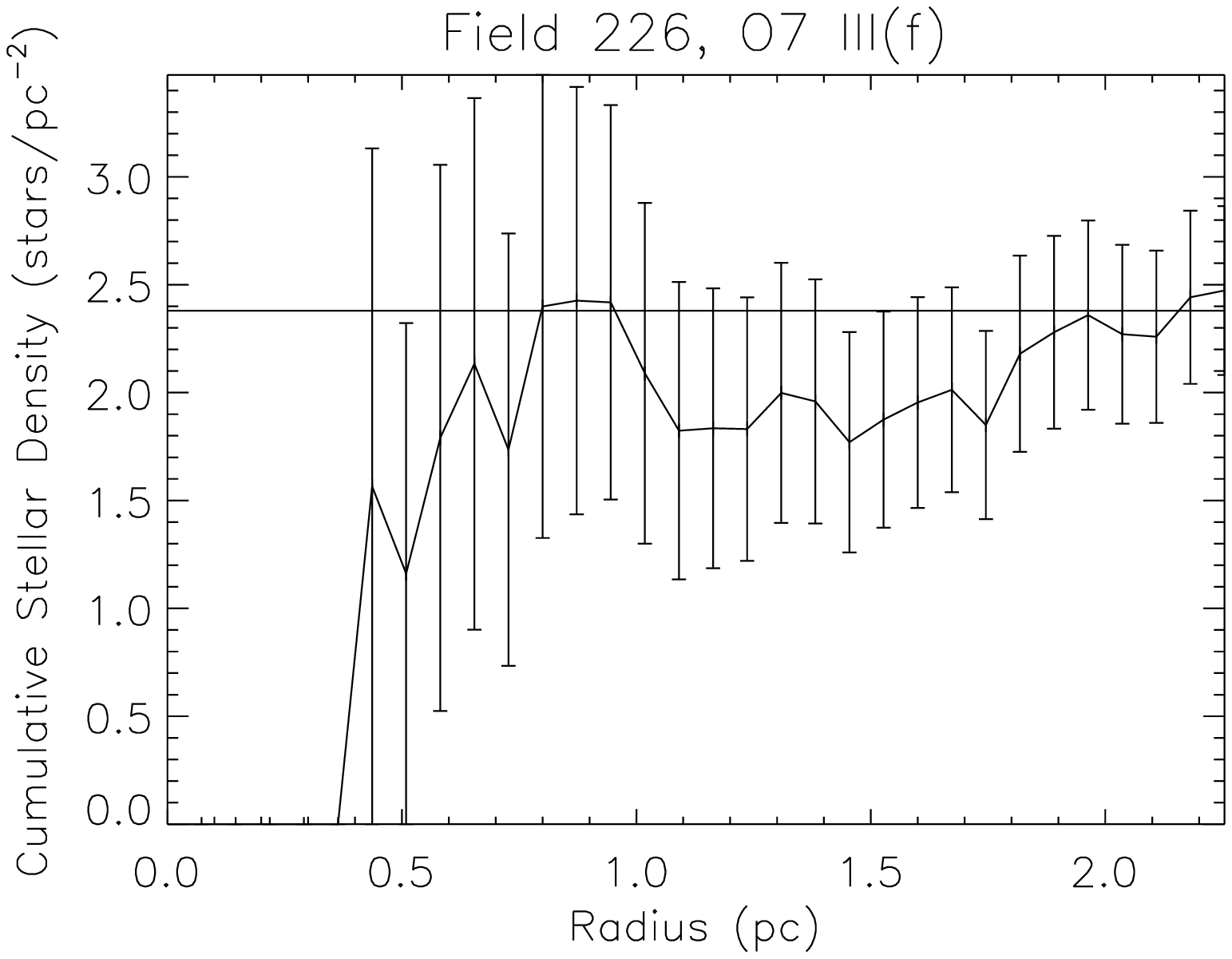}
	\includegraphics[scale=.46,angle=0]{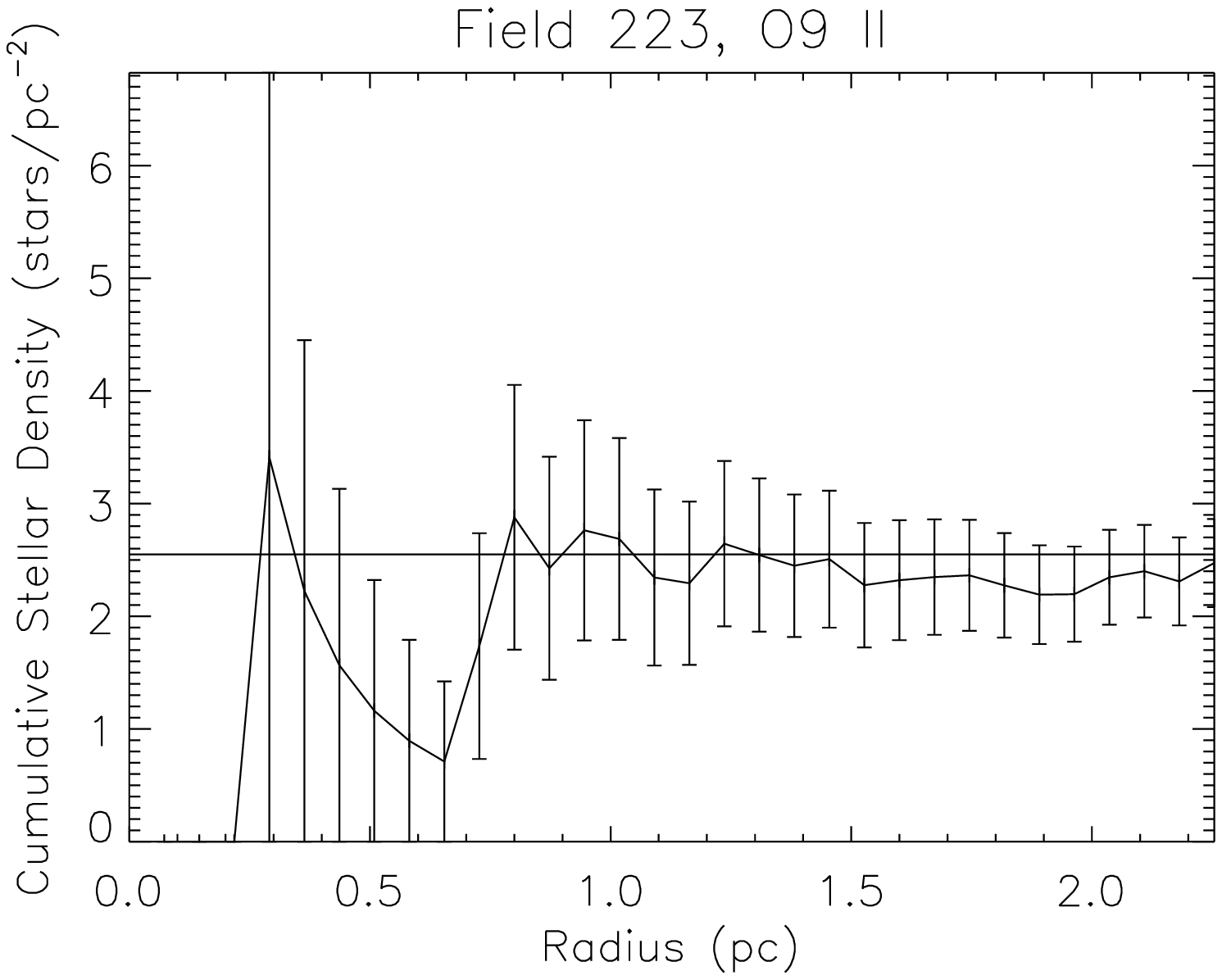}
	\includegraphics[scale=.46,angle=0]{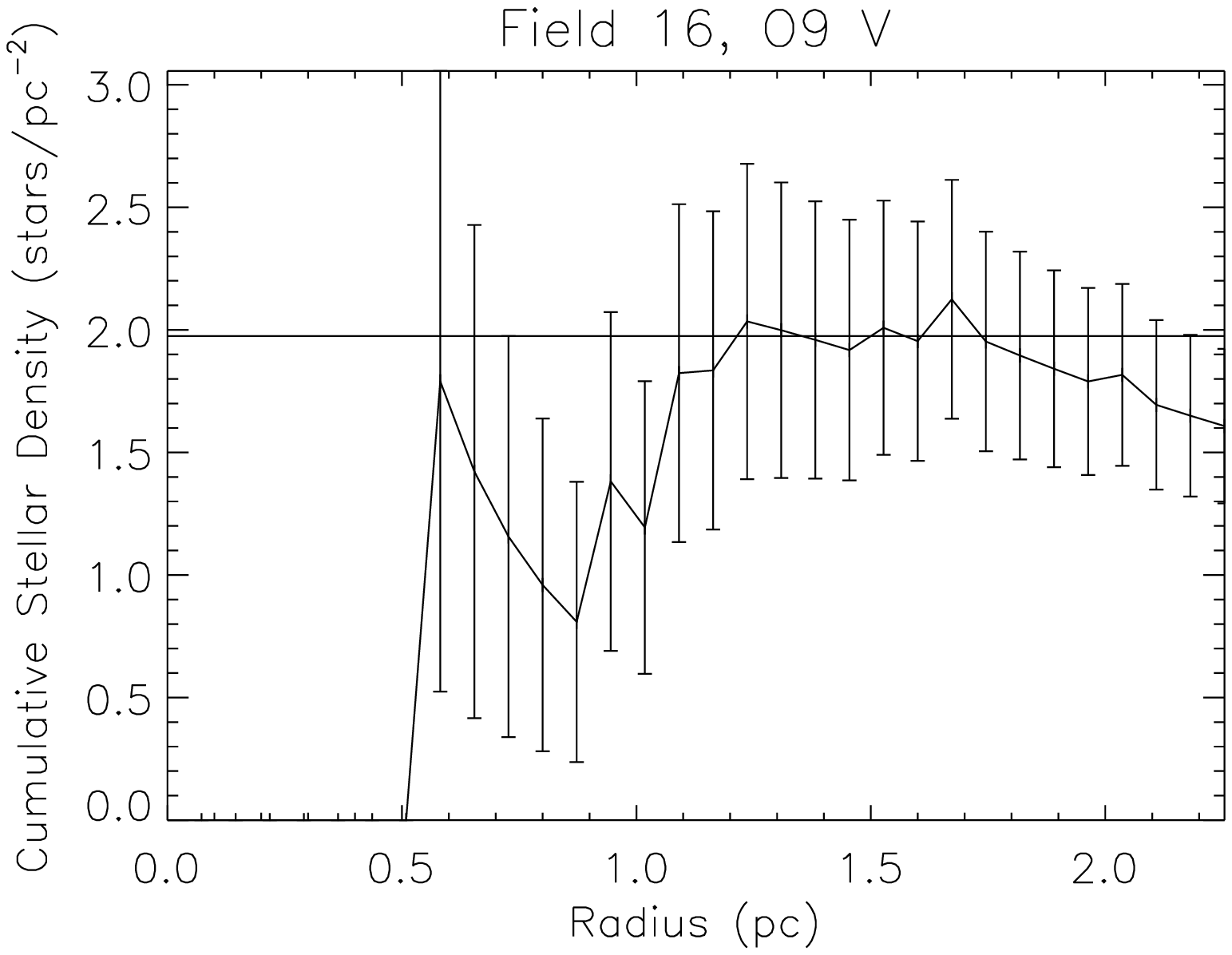}
	\caption{Cumulative stellar density as a function of radius
	from each target star.  Each field is titled with the target
	star's ID and spectral type.  The panels are
	shown in the same sequence as in Figure~\ref{hst}, with the first three
	stars showing a stellar density enhancement.} 
	\label{poisson}
	\end{center}
\end{figure*}

To examine the stellar environment near each target star, we plot the
cumulative stellar density as a function of radius from the target OB star
in Figure \ref{poisson}.  The horizontal lines show the measured background density of each
field.  We were unable to detect stars 
within the wings of the target star, which typically extended to a
radius of 0.2 pc.  Thus, the actual cumulative stellar densities may be higher than those observed.  
Table \ref{obsprop} lists the angular and physical radius $R_{\rm cl}$ at which $P(f)$ is minimized in columns 2 and 3, respectively. The value of $P(f)$ is listed in column 4 for targets showing a density enhancement.

\begin{deluxetable*}{ccccccc}
  \tabletypesize{\small}
  \tablewidth{0pc}
  \tablecaption{Cluster and Stellar Population}
  \tablehead{\colhead{Field} & \colhead{Angular Size (\arcsec)} & \colhead{$R_{\rm cl}$ (pc)} & \colhead{$P(f)$}  & \colhead{$l$ (pc)} & \colhead{Field IMF}}
  \startdata

	smc16	& ... & ... & ... & 0.52 &$\Gamma = -0.9 \pm  0.5$  \\
    	AzV 58	& ...  & ... & ... & 0.43 & $\Gamma = -1.2 \pm  1.1$   \\
	AzV 67 	& 10.3 & 1.5  & 0.114 & 0.50 & $\Gamma = -1.3 \pm  0.7$   \\
	AzV 106	& 6.9 & 1.0   & 0.047 & 0.43 & $\Gamma = -1.6 \pm  1.3$   \\
	AzV 186	& ... & ... & ... & 0.46 & $\Gamma = -0.9 \pm  1.2$   \\
    	AzV 223	& ... & ... & ... & 0.47 & $\Gamma = -1.1 \pm  0.5$   \\
	AzV 226	& ... & ... & ... & 0.48 & $\Gamma = -1.0 \pm  1.0$  \\
	AzV 302	& 4.8 & 0.7  & 0.0001 & 0.58 & $\Gamma = -1.0 \pm  0.7$ \\
    \enddata
    \label{obsprop}
\end{deluxetable*}

\begin{figure}
	\begin{center}
	\includegraphics[scale=.5,angle=0]{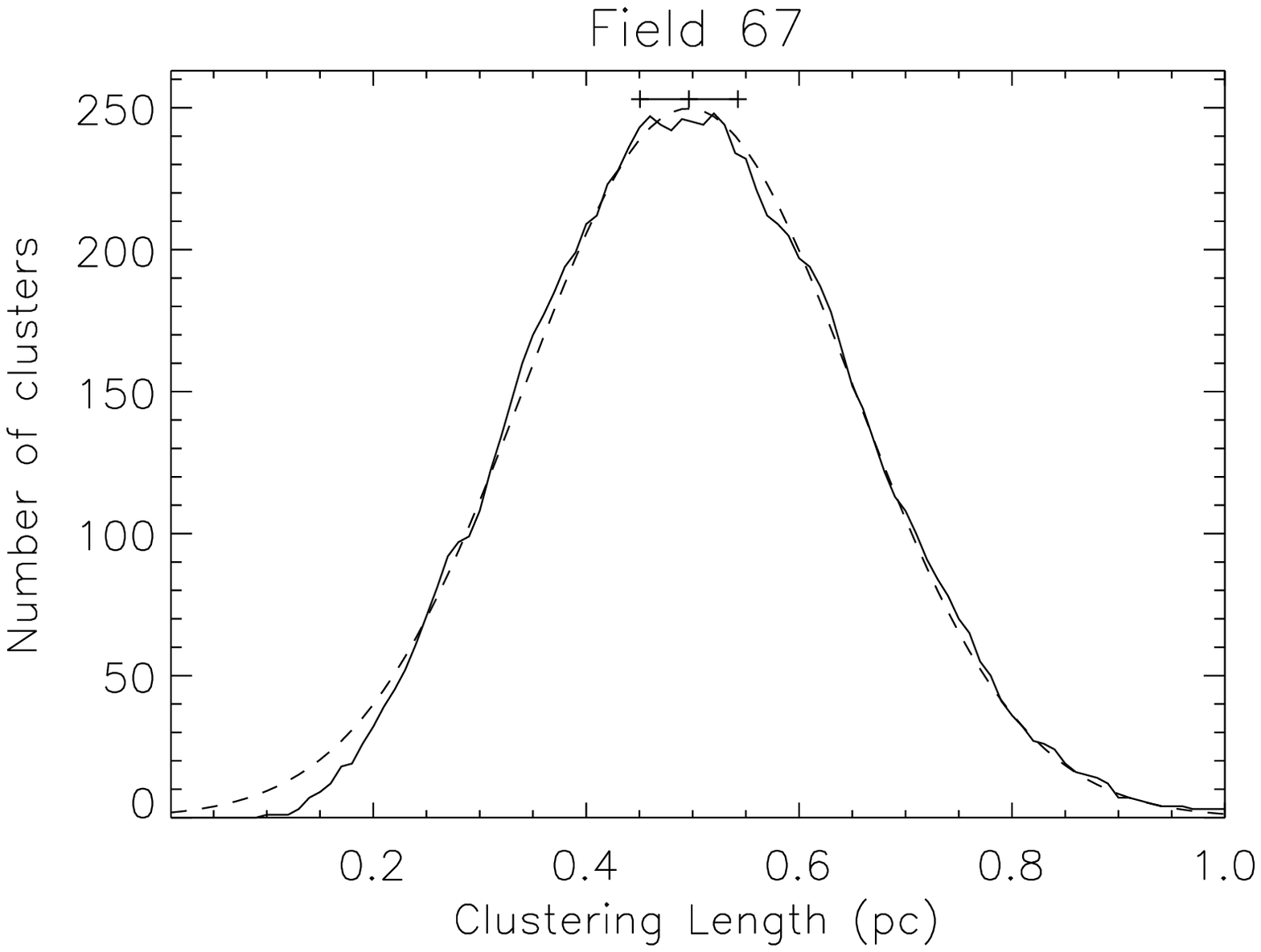}
	\caption{Number of clusters vs. clustering length, $l$, with a
	fitted Gaussian (dashed line) for a representative field, AzV
	67.  The bar shows the extent between one standard deviation
	below and above the peak value, and indicates the three
	values used for $l$ for each field.} 
	\label{gauss}
	\end{center}
\end{figure}

We also searched for density enhancements with a friends-of-friends
algorithm applied to our F814W images.  This method
defines group members to be all the stars within a fixed clustering length $l$
of another member of the same group.  Following Battinelli (1991), we
adopted a value for $l$ that maximizes the number of groups having
at least three stars.  The distributions of clusters vs. $l$ for each field are approximated well by normal distributions, and so we used gaussian functions to estimate $l$.  
Figure \ref{gauss} shows a representative example.  The
average $l$ for these observations is $0.48 \pm 0.05$
parsecs.  Table \ref{obsprop} lists the clustering length of each field in column 5.

\subsection{Minimal O Star Groups}

As seen from the data in Figure \ref{poisson} and Table \ref{obsprop}, there are three stars that show
robust evidence of small, associated stellar groups:
AzV 67, AzV 106, and AzV 302.  The field probabilities, $P(f)$, for these targets range
from 0.01\% to 11.4\%.  The friends-of-friends algorithm confirms the
existence of stars within the clustering length $l$ in each of these fields.
To examine the sensitivity to $l$, we also ran the
friends-of-friends code with values of $l \pm 0.05$ pc, where this
variation is equal to the standard deviation of $l$ in the sample (see above).
For the smaller values, this yielded companions only for AzV 302;
while the larger value resulted in associated stars for all targets in our
sample.  Thus, we are confident that the fitted peak values for $l$
are appropriate, and they support the identification of groups found by
identifying density enhancements.  We will refer to these
sparse groups as  ``minimal O star groups''.  AzV 186 is the only
field to appear clustered using the friends-of-friends algorithm, but
not the density enhancement algorithm.  We identify only
those fields that appear clustered using both algorithms, as minimal
O star groups.

\subsection{Isolated Field Stars}
\label{isosec}

The remaining four targets (AzV 58, AzV 223, AzV 226, smc16) show no evidence of
associated stars using 
either of the methods above.  These are candidates for massive stars
that formed in complete isolation.  However,
a substantial fraction of field O stars likely did not
originate in the field at all, but rather are runaway stars from
clusters (e.g., Blaauw 1961; Gies 1987; Hoogerwerf et al. 2001).  
Thus, we examine the likelihood that our isolated OB stars are runaways.

We are performing a complete spectroscopic survey of SMC field OB
stars that is now underway, using primarily the IMACS multi-slit spectrograph on
the 6.5-m Magellan/Baade telescope at Las Campanas Observatory (Lamb
et al, in preparation).
These data yield the stellar radial velocities and spectral
classifications, and we will ultimately estimate the runaway fraction for these
massive field stars.  In the course of this survey, we obtained
spectra of the target field OB stars in our {\it HST} imaging sample.
Table \ref{targets} gives our measured heliocentric radial
velocities and spectral classifications.  Several stars were
previously classified, and we either confirmed or revised the spectral
types, as shown.

We identify two of these stars, smc16 and AzV 223, as
runaways, defined as stars having $\ge$ 30 km s$^{-1}$ difference (de Wit et al. 2005) from the SMC systemic
velocity of $155$ km s$^{-1}$ (Staveley-Smith et al. 1997).  We note that Stanimirovi{\'c} et al. (2004) found the SMC to have a velocity gradient; however, the positions of our targets do lie between the 150 km s$^{-1}$ and 160 km s$^{-1}$ contours as plotted in Figure 3 of Stanimirovi{\'c} et al. (2004).  The two runaway stars will be removed from analysis in all subsequent sections, since
they are no longer located in the region of their formation.  This
leaves us with two, isolated, apparently non-runaway OB stars, AzV 58
and AzV 226.  Assuming an isotropic distribution of runaway
velocities, we still expect to miss a number of transverse 
runaways.  The missed fraction depends on the velocity distribution of
runaway stars; however, we estimate that typical ejection velocities
of 60 km s$^{-1}$ or 120 km s$^{-1}$ would cause us to miss two or one transverse
runaway(s), respectively.  Therefore, transverse runaways may account for both our remaining isolated stars.

\begin{figure*}
	\begin{center}
	\includegraphics[scale=.31,angle=0]{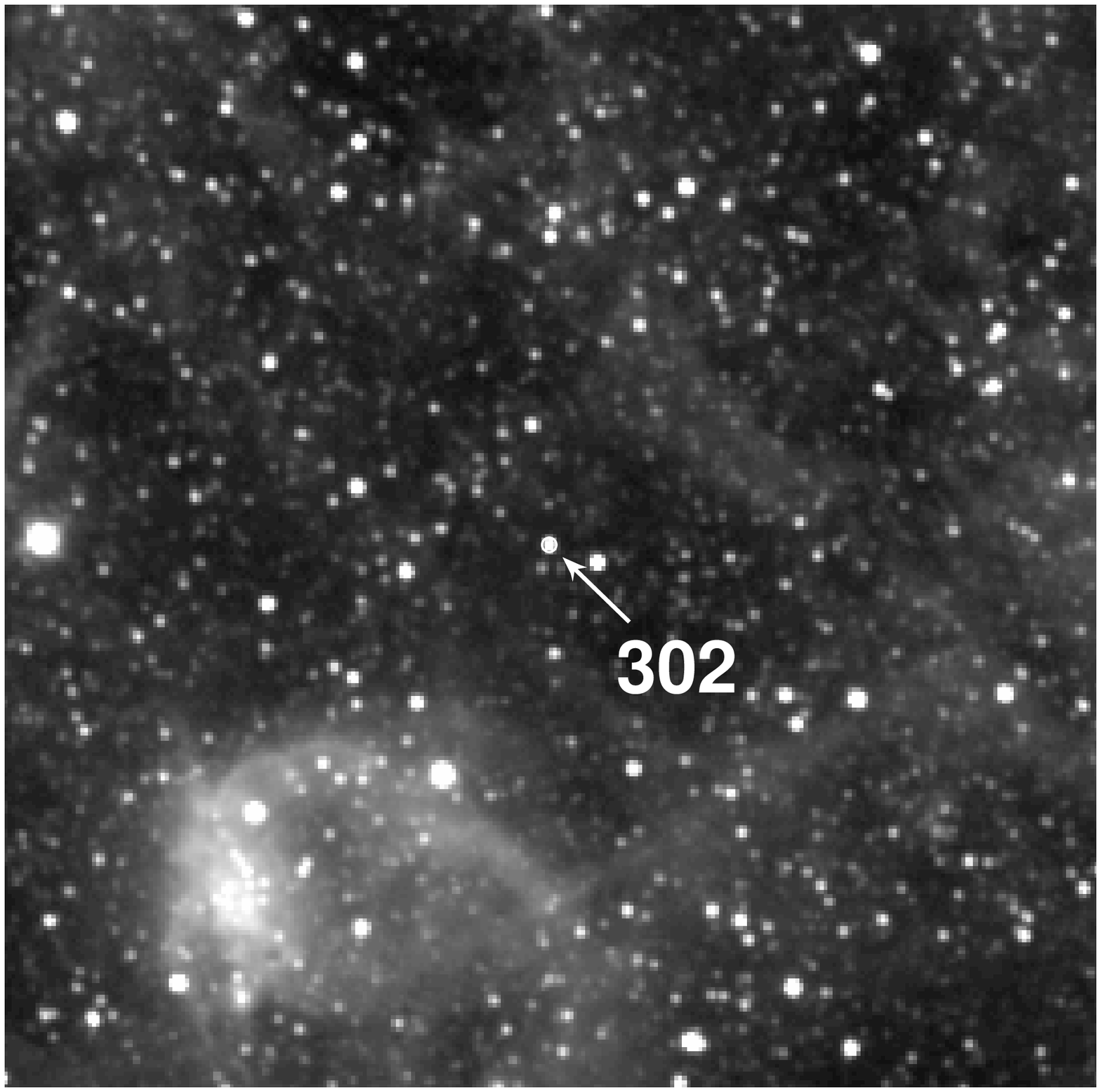}
	\includegraphics[scale=.31,angle=0]{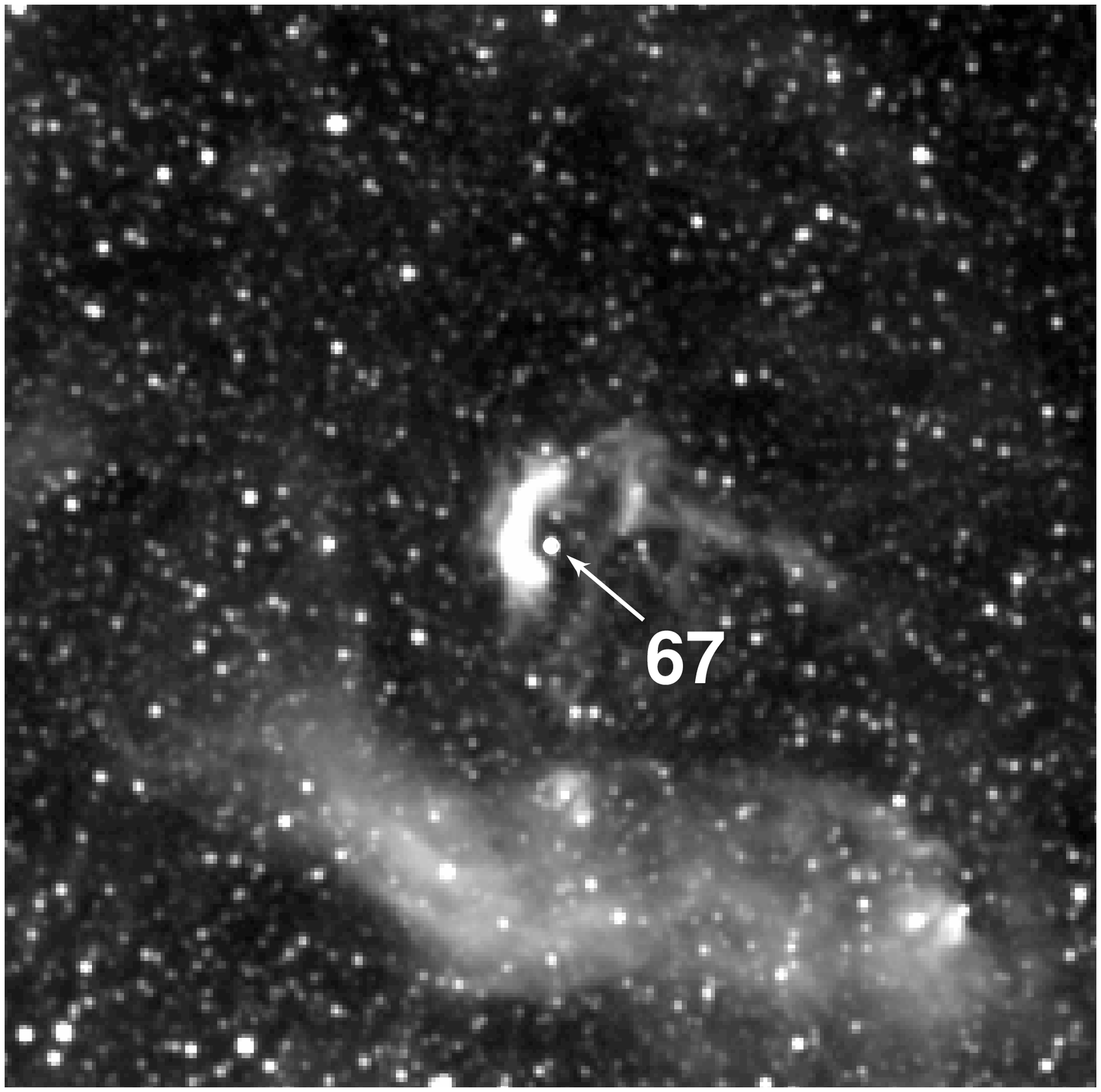}
	\includegraphics[scale=.31,angle=0]{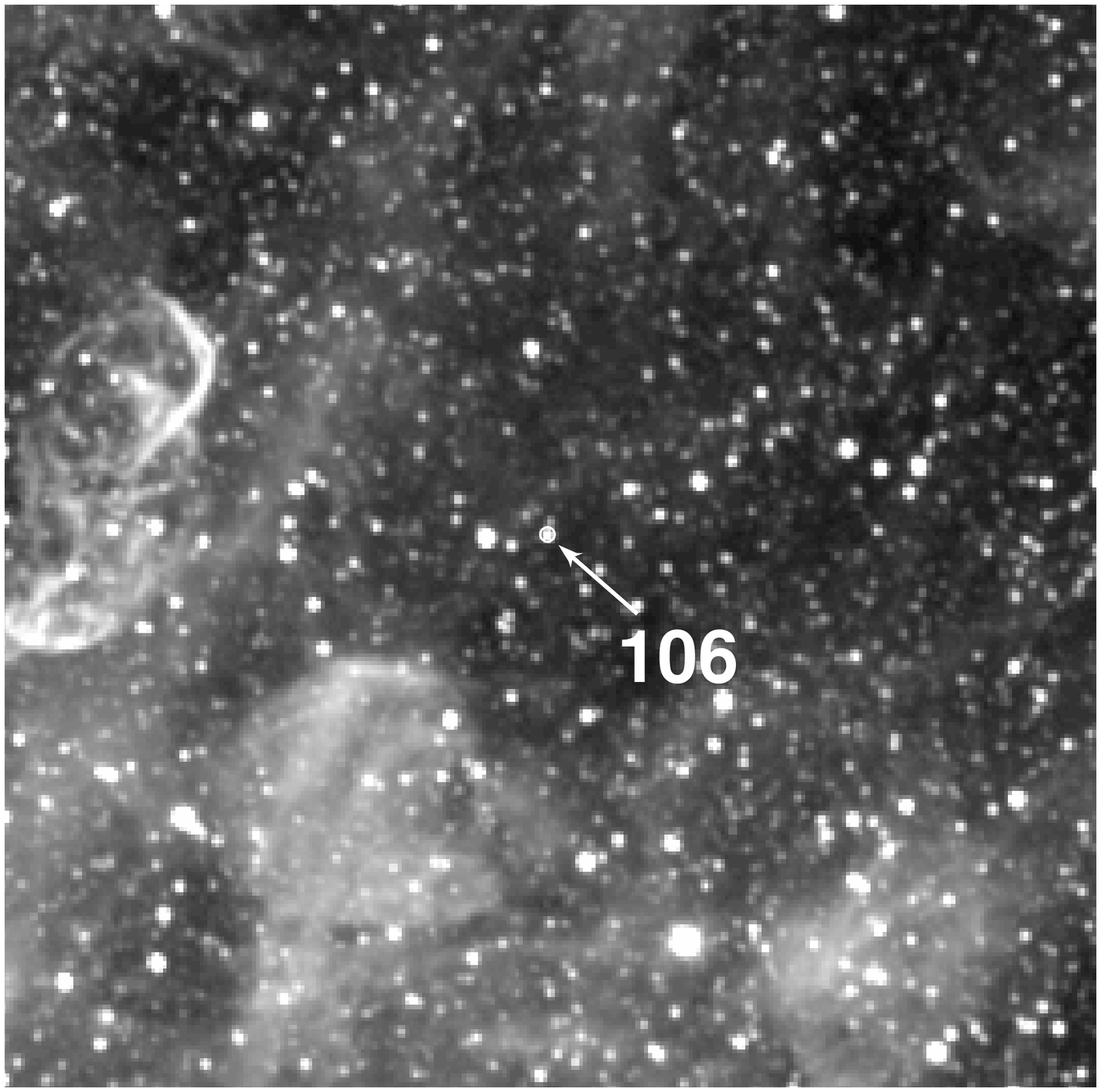}
	\includegraphics[scale=.31,angle=0]{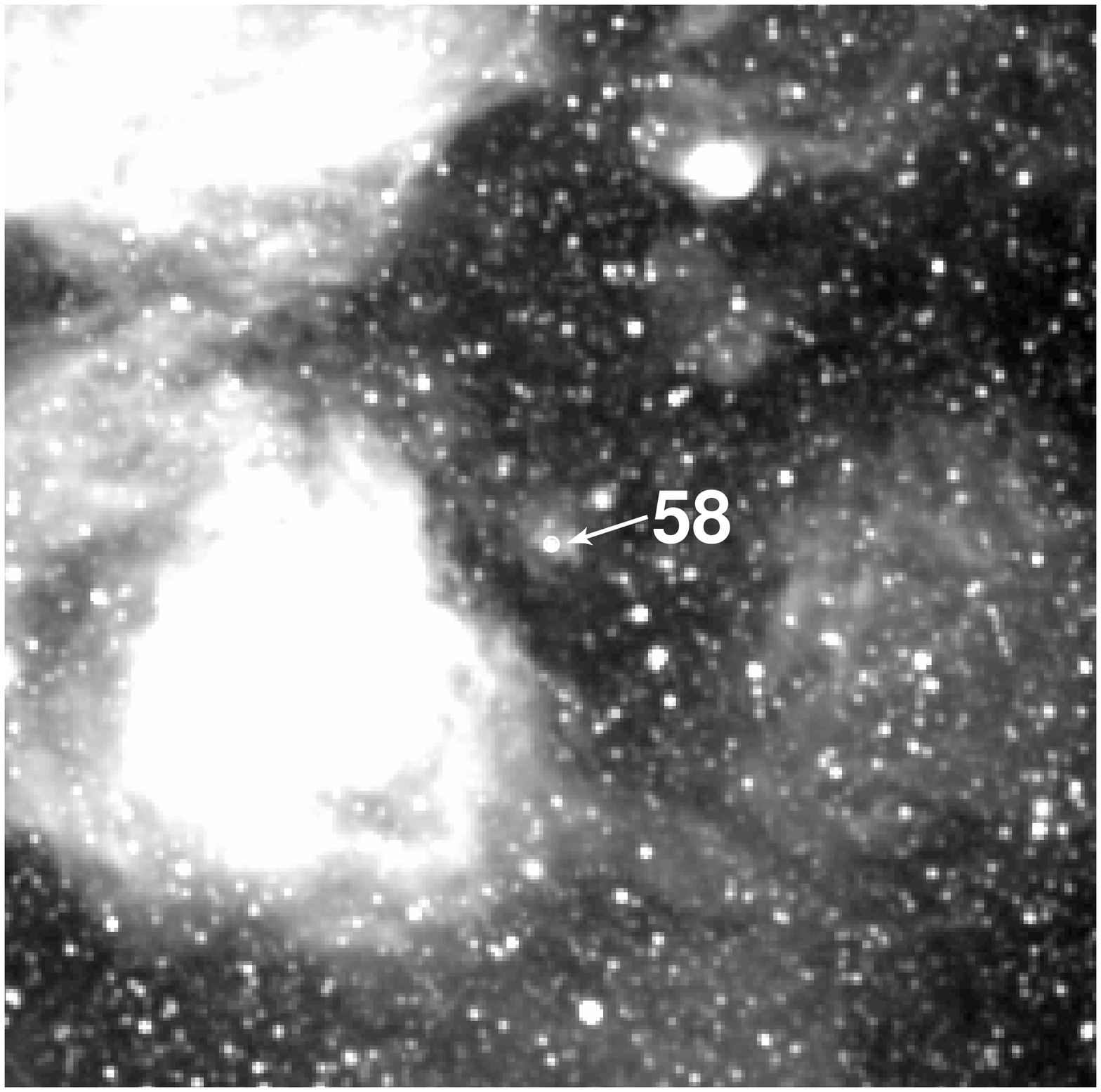}
	\includegraphics[scale=.31,angle=0]{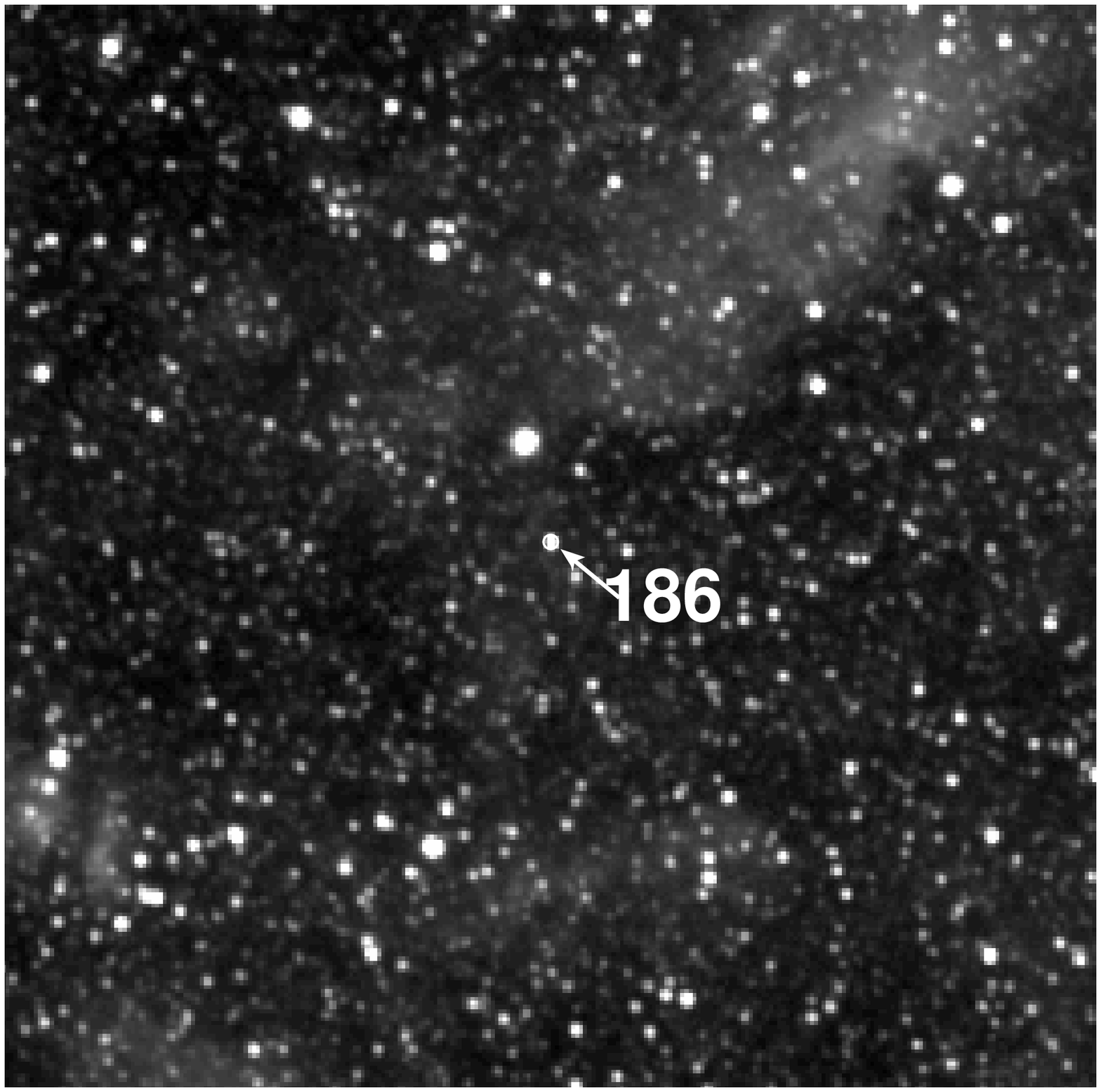}
	\includegraphics[scale=.31,angle=0]{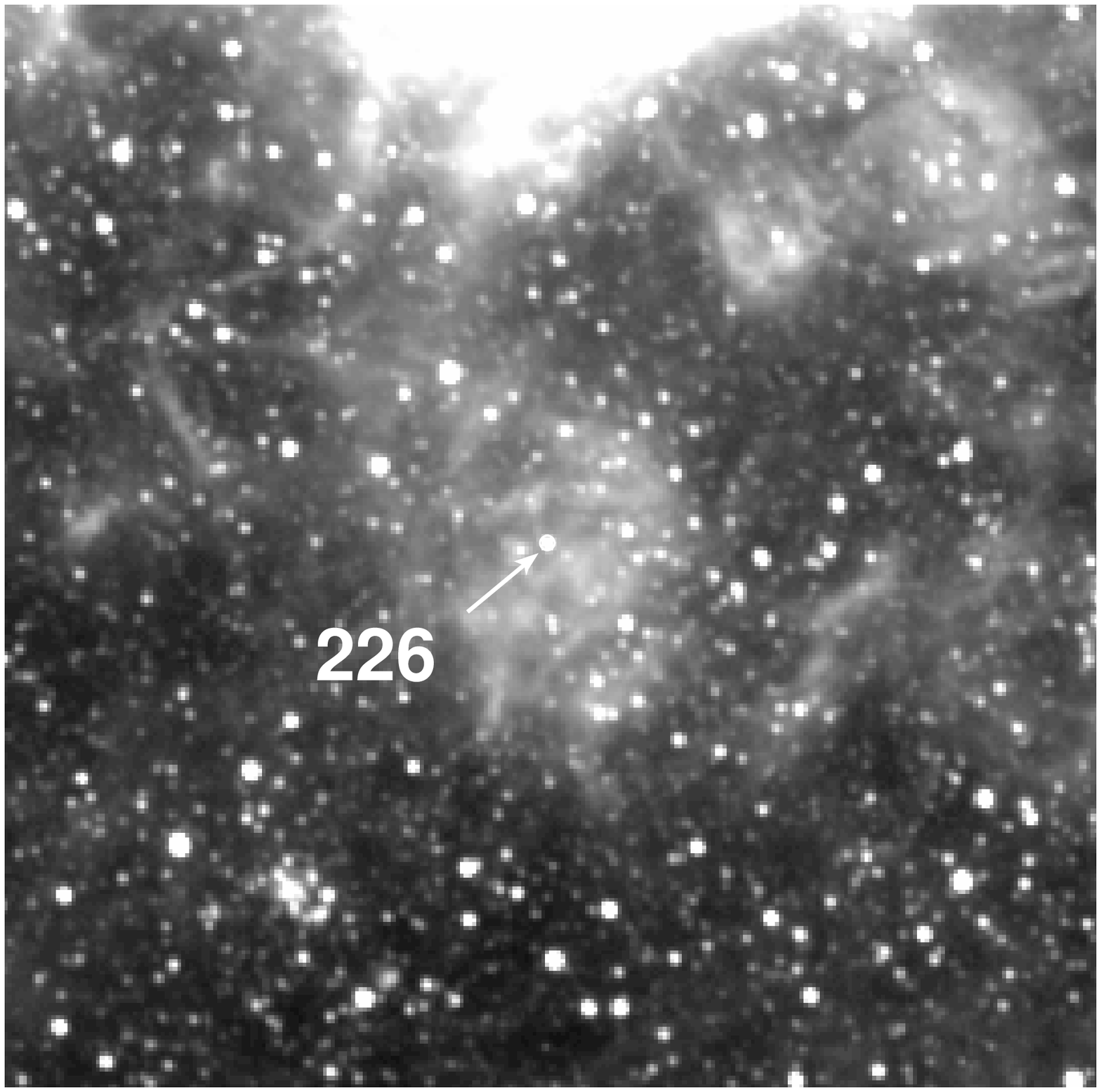}
	\includegraphics[scale=.31,angle=0]{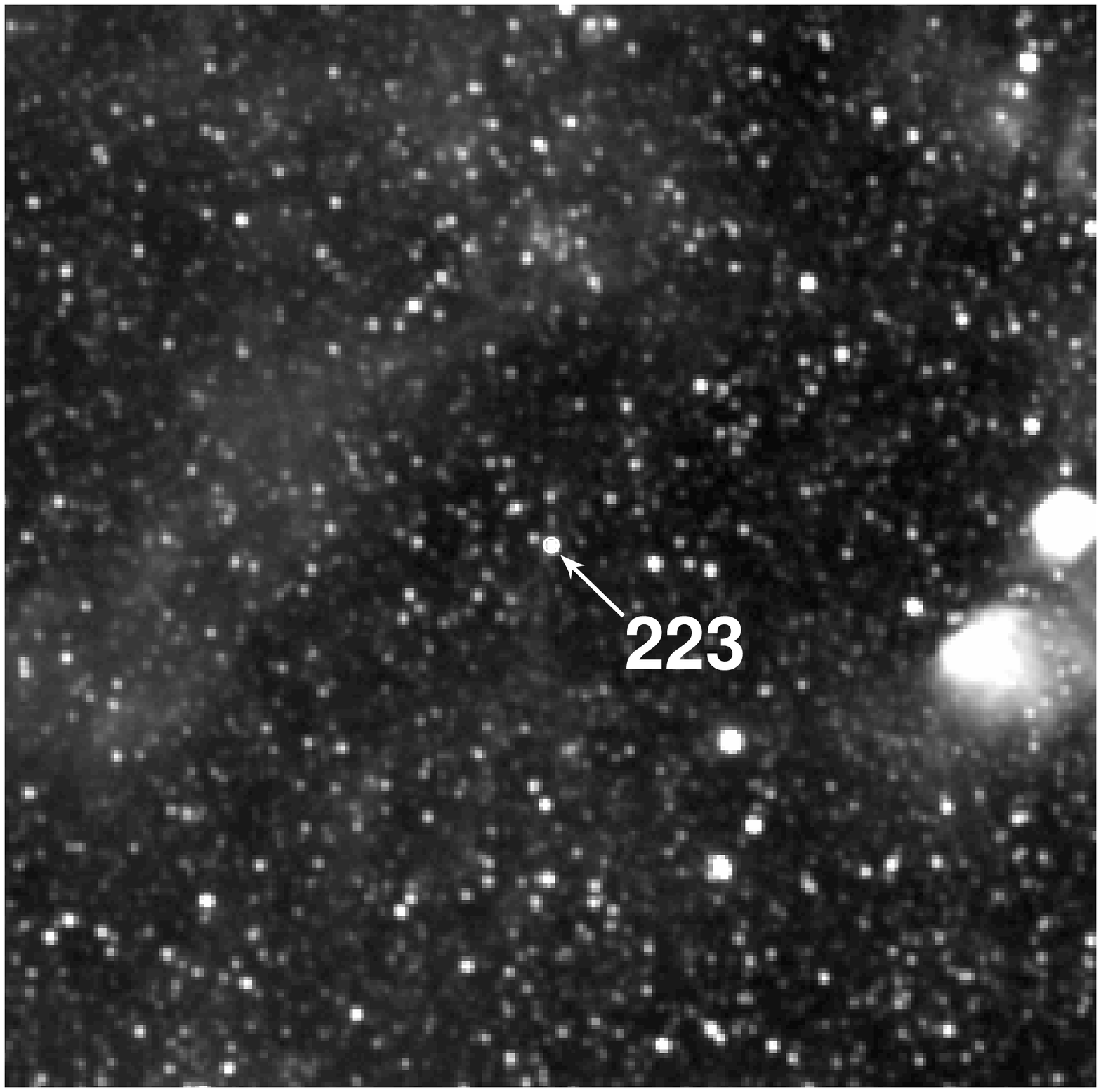}
	\includegraphics[scale=.31,angle=0]{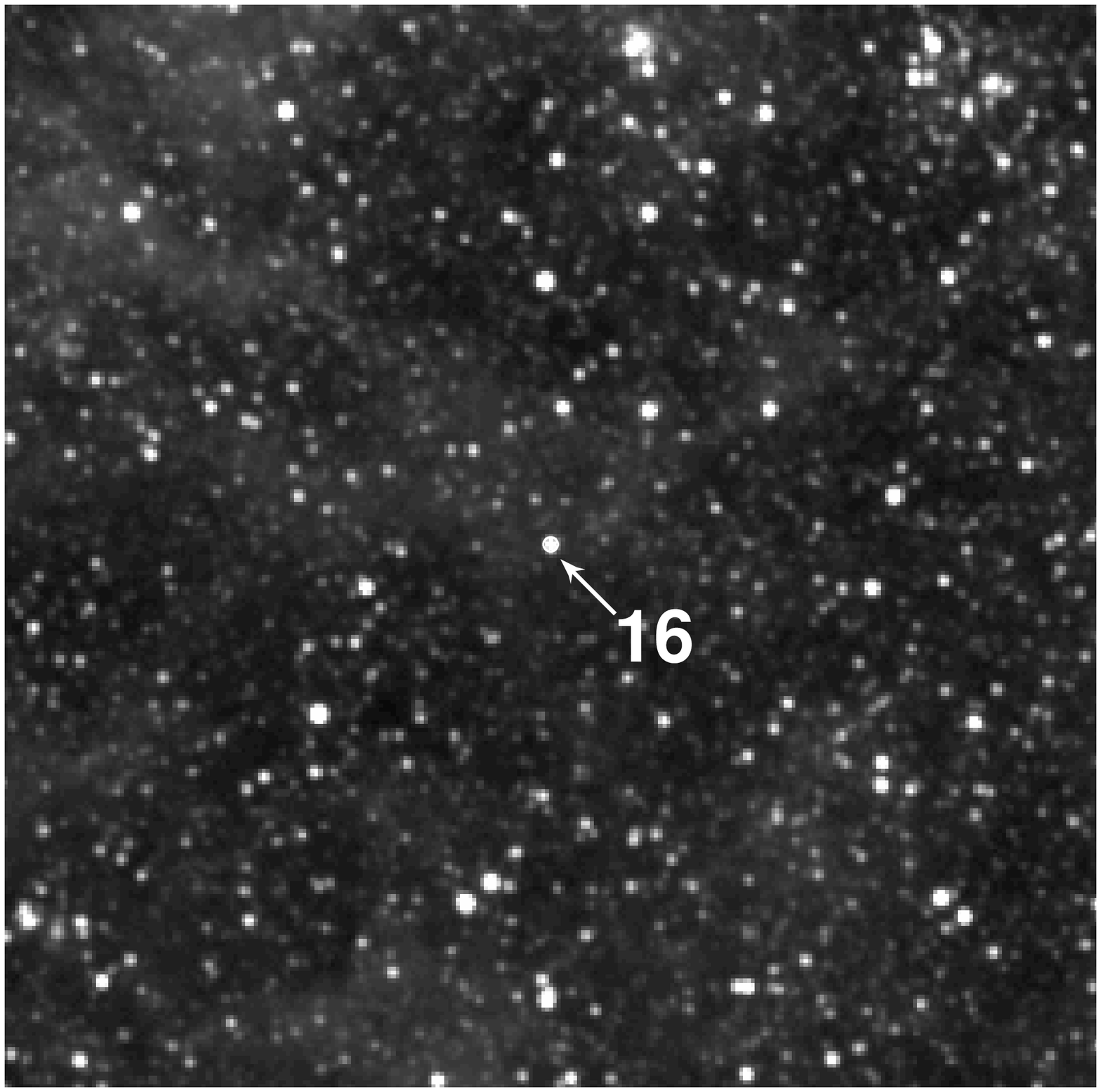}
	\caption{H$\alpha$ images from the Magellanic Cloud Emission
	Line Survey (MCELS), tracing the ionized gas in the region
	around each target star. The PSF of these observations is
	similar to the 1 parsec circles in Figure~\ref{hst}, and
	the field of view is 9$\arcmin$ $\times$ 9$\arcmin$.  The
	panels are sequenced as in Figure~\ref{hst}, with the top
	row showing fields with a stellar density enhancement.  The target stars are identified.} 
	\label{mcels}
	\end{center}
\end{figure*}

It is also informative to investigate the interstellar gas
around these field OB stars (Figure \ref{mcels}).  We examined
the ionized gas around our target stars, using
H$\alpha$ data from the Magellanic Cloud Emission Line Survey 
(MCELS; Smith et al. 2000).  Since gas is a necessary component of star
formation, the presence of gas can help to constrain which stars
may still be in the region of their formation.
As a control group, all three of our minimal O star groups show
H$\alpha$ emission within 2.0$\arcmin$ (35 pc), consistent with a physical
association within these groups as sparse, young clusters.  We also
find that both confirmed runaway stars are far removed from any
H$\alpha$ emission.  For the remaining isolated targets, 
the MCELS data show that AzV 58 and AzV 226 are located within {\sc Hii}
 regions in the line of sight (Figure \ref{mcels}), while AzV 186 is far from any {\sc Hii}
regions.  These results suggest that AzV 58 and AzV 226 may still be
in the region of their formation and thus they remain candidates for isolated
massive star formation. 
 
\subsection{The IMF}
\label{imfsec}

We construct color-magnitude diagrams (CMD's) from our ACS photometry, and use
these to evaluate the IMF for the minimal O star groups as well as the
field stars.  We converted the photometric results from the  F555W
and F814W bands to Johnson {\it V} and Cousins {\it I} 
bands, respectively, following the synthetic transfer equations
provided in Sirianni et al. (2005).  These {\it V} and {\it I} band
magnitudes were then extinction corrected using the SMC extinction
maps provided by the Magellanic Cloud Photometric Survey (MCPS;
Zaritsky, et al. 2002).  MCPS provides two sets of stars for estimating
extinction, the ``hot" stars ($12,000$ K $\le T_{eff} \le 45,000$ K)
and ``cool" stars ($5,500$ K $\le T_{eff} \le 6,500$ K).  We chose to
average the extinction values of ``cool" stars within 1$\arcmin$ of
the target star for the extinction calculation, as these objects better trace the sparse field population of our observations.  The ``hot" stars tend to trace active, clustered star formation with high gas content, yielding extinction values up to half a magnitude higher than those typically found in the ``cool" stars.

We then calculated absolute magnitudes using a
distance modulus of 18.9 (Harries et al. 2003).  Comparisons of our photometry with MCPS photometry
typically show differences less than 0.1 mag, with {\it
  I}-band matching more closely than $V$.  On average, our photometry
appears fainter than that of the MCPS in both {\it V} and {\it
  I} bands.  For stars fainter than 16th magnitude, Zaritsky et
al. (2002) find that stellar crowding may play a factor in
spuriously brightening the MCPS photometry when compared with the
OGLE catalog (Udalski et al. 1998), which they deem to be superior in
this regime.  Since nearly all stars in our fields are fainter than
16th magnitude, the difference between our photometry and MCPS
photometry is likely due to these crowding effects. 

We used Geneva stellar evolutionary tracks to extract mass estimates
from our photometry.  These tracks are based on the ATLAS9
no-overshoot models at SMC metallicity, calculated by Charbonnel et al. (1993) and have
been converted by Girardi et al. (2002) to Johnson-Cousins {\it UBVRI}
photometry.  The stellar masses are inferred by identifying the two evolutionary
tracks between which a star falls, on the $V$ vs $V-I$ CMD.
For the extremely blue, O and B target stars, the $V-I$
colors are degenerate, so effective temperature, $T_{\rm eff}$, is used instead of $V-I$ color.   
$T_{\rm eff}$ of our OB target stars are
based on our spectral classifications given in Table \ref{targets}, using
the conversions of spectral type to effective temperature for the SMC by Massey et
al. (2005) for the O stars, and Crowther (1997) for the B stars.

Following the formalism
of Scalo (1986) and analysis done by Massey (1995), we write the slope
of the IMF as:
\begin{equation}
\Gamma = d \log \xi(\log m)/d \log m
\end{equation}
where $\xi(\log m)$ is the mass function in units of stars born per
logarithmic stellar mass $m$ ($M_\odot$) per unit area (kpc$^2$) per unit time (Myr).  This
corresponds to a power-law mass spectrum given by:
\begin{equation}
n(m)\ dm \propto m^{\gamma} dm
\end{equation}
where $n(m)\ dm$ is the number of stars per unit mass bin and
the power law index $\gamma =  \Gamma -1$.  In this formalism, a Salpeter mass
function has a slope $\Gamma = -1.35$.  In order to derive $\xi(\log m)$, 
we counted the stars in each mass bin, corrected for
the size of the mass bin by normalizing to one dex in mass, and
divided by the area covered by the observations.  
For computing the IMF of the background field, we divided by the average age of
stars within each mass bin (Charbonnel et al. 1993) to account for differences in stellar
lifetimes as a function of mass.  Ages were calculated as an average between the lower and upper mass in each bin, weighted by the IMF.  This allowed us to measure an IMF under
the assumption of continuous star formation, rather than obtaining the
present-day mass function of the field.     

\begin{figure*}
	\begin{center}
	\includegraphics[scale=.5,angle=0]{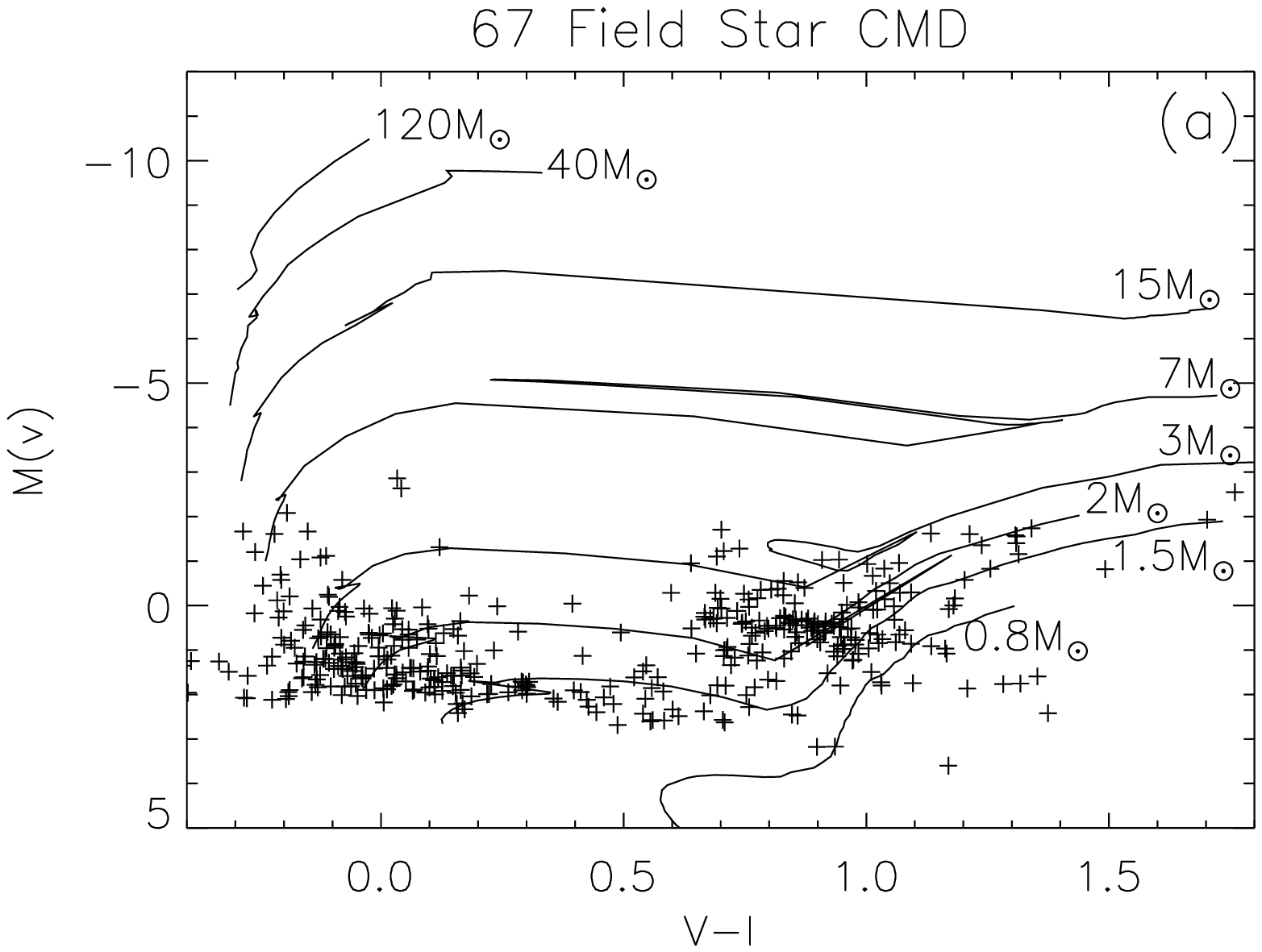}
	\includegraphics[scale=.5,angle=0]{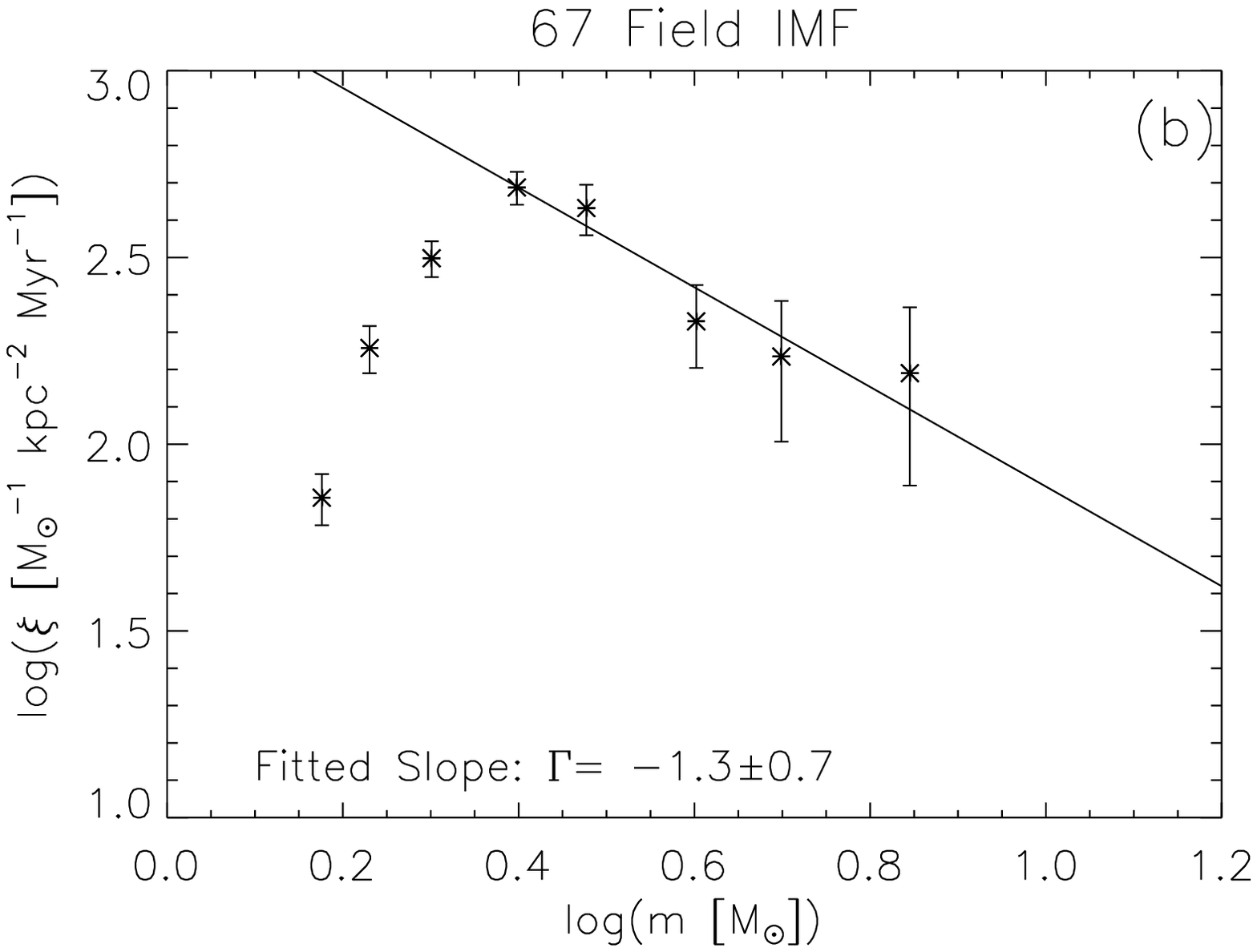}
	\caption{({\it a}) Color-magnitude diagram and ({\it b}) field IMF for a
          representative field, AzV 67.  Each of our observed fields
          is consistent with the Salpeter IMF within
          the uncertainty.  The error bars represent the Poisson
          uncertainties.  Mass bins below our
          completeness limit of $2M_\odot$ are excluded from the
          fitted line.} 
	\label{fieldcmd}
	\end{center}
\end{figure*}

The CMD and IMF for a representative, full field, that of AzV 67,
are shown in Figure \ref{fieldcmd}.  On the CMD plot, we draw SMC-metallicity evolutionary tracks (Charbonnel et al. 1993; Girardi et al. 2002)
to show the stellar mass ranges of the field population.  On the IMF plot, 
the error bars represent the Poisson uncertainty for each mass bin.  The
detection limit is $V=22$ in the F555W image, which corresponds to a mass of
$1.5M_\odot$.  The observations are incomplete up to 
$2M_\odot$.  
On the IMF plot, points corresponding to stellar mass bins below
$2M_\odot$ are excluded from the linear fitted line, which has been made weighting the data inversely by the errors. 
The IMF slopes for the field
population in each full, 202$\arcsec$ $\times$ 202$\arcsec$ ACS frame are given in
Table \ref{obsprop}, column 6.  Accounting for the stellar age
correction, each field exhibits an IMF
consistent with a Salpeter IMF, within the uncertainty.

Due to the small number of cluster members identified in both the
F814W and F555W exposures, it is not useful to derive 
IMFs for the three minimal O star groups individually.  Instead, we
created a composite CMD of all members from 
the three groups.  Because of field star
contamination, only a subset of the stars comprising these density 
enhancements are physically associated; the background field star density
implies a contamination of 1 -- 2 stars per target frame.  In deriving 
the IMF of these minimal O star groups, we exclude red giants, since
they are certainly field stars unassociated
with recent star formation.  Some main sequence stars may also be
field star contaminants, but they are indistinguishable from
true cluster members.  Table \ref{compphot} provides
photometry of the main sequence companion stars present in both $V$ and
$I$ images,  above the completeness limit of {\it V} = 21 and {\it I} = 22 magnitudes.  
Columns 1 and 2 list the right ascension and declination,
while Columns 3 and 4 list the {\it V} and {\it I} magnitudes.

\begin{deluxetable*}{ccccccc}
  \tabletypesize{\small}
  \tablewidth{0pc}
  \tablecaption{Photometry of Companion Stars}
  \tablehead{\colhead{O Star Group} & \colhead{RA} & \colhead{Dec} & \colhead{{\it V}} & \colhead{{\it I}}}
  \startdata

	AzV 67 & 00:50:10.39 & -72:32:30.9 &  $20.78 \pm  0.15$ & $20.77 \pm  0.06$	   \\
    	AzV 67 & 00:50:11.77 & -72:32:30.0 &  $19.85 \pm  0.07$ & $19.79 \pm  0.03$  \\
	AzV 67 & 00:50:11.95 & -72:32:32.3 &  $20.51 \pm  0.13$ & $20.67 \pm  0.06$ \\
	AzV 106 & 00:51:43.45 & -72:37:27.3 &  $19.51 \pm  0.05$ & $19.58 \pm  0.02$ \\
	AzV 106 & 00:51:44.02 & -72:37:23.9 &  $21.20 \pm  0.20$ & $21.15 \pm  0.08$   \\
    	AzV 302 & 01:02:18.74 & -72:22:04.7 &  $19.76 \pm  0.08$ & $19.52 \pm  0.03$	  \\
	AzV 302 & 01:02:18.81 & -72:22:01.1 &  $20.84 \pm  0.16$ & $20.68 \pm  0.06$	 \\
	AzV 302 & 01:02:18.83 & -72:22:01.7 &  $17.97 \pm  0.02$ & $17.98 \pm  0.01$	 \\
	AzV 302 & 01:02:19.40 & -72:22:02.2 &  $20.81 \pm  0.16$ & $20.46 \pm  0.04$	  \\
    \enddata
    \label{compphot}
\end{deluxetable*}

\begin{figure*}
	\begin{center}
	\includegraphics[scale=.5,angle=0]{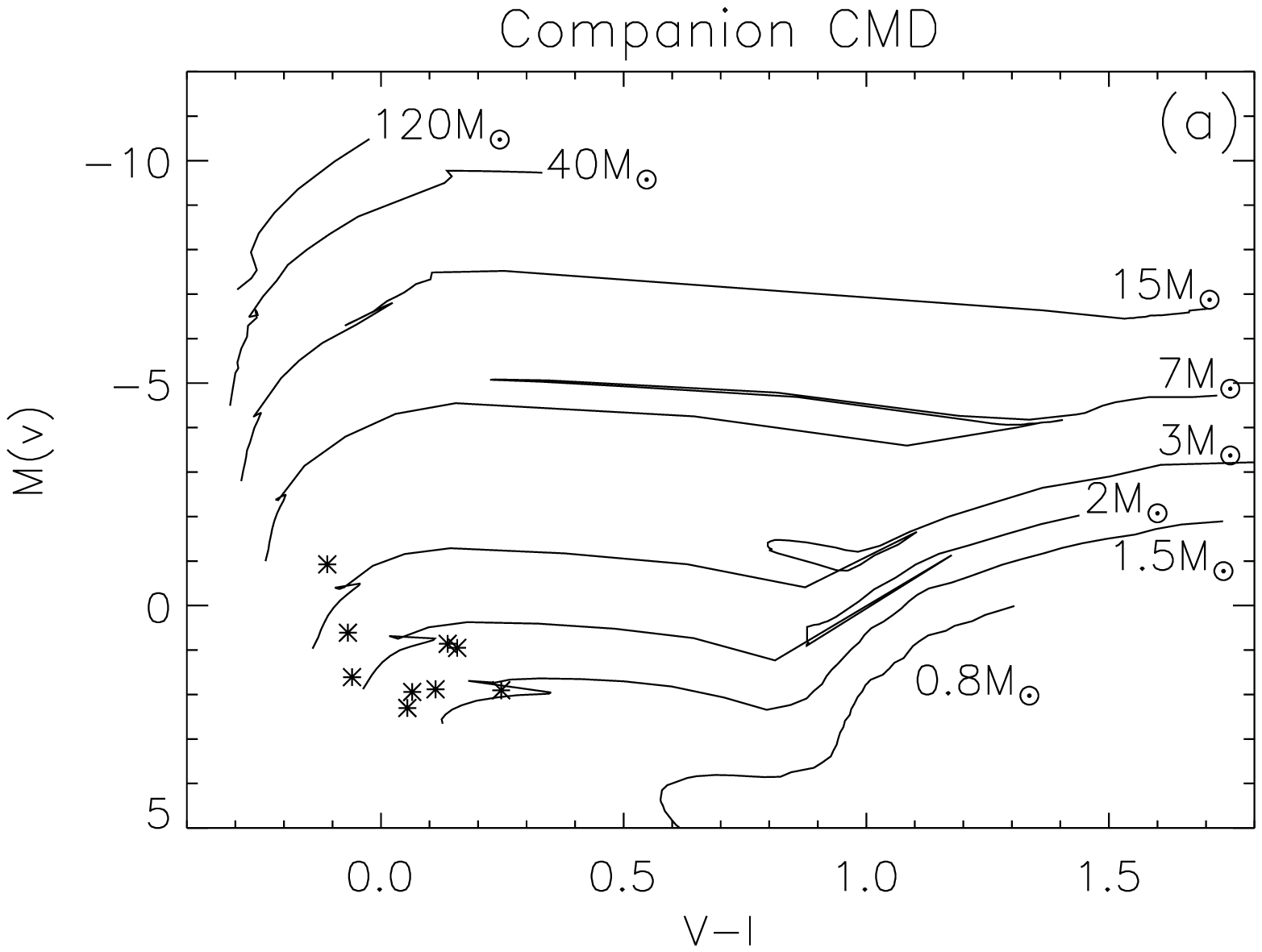}
	\includegraphics[scale=.5,angle=0]{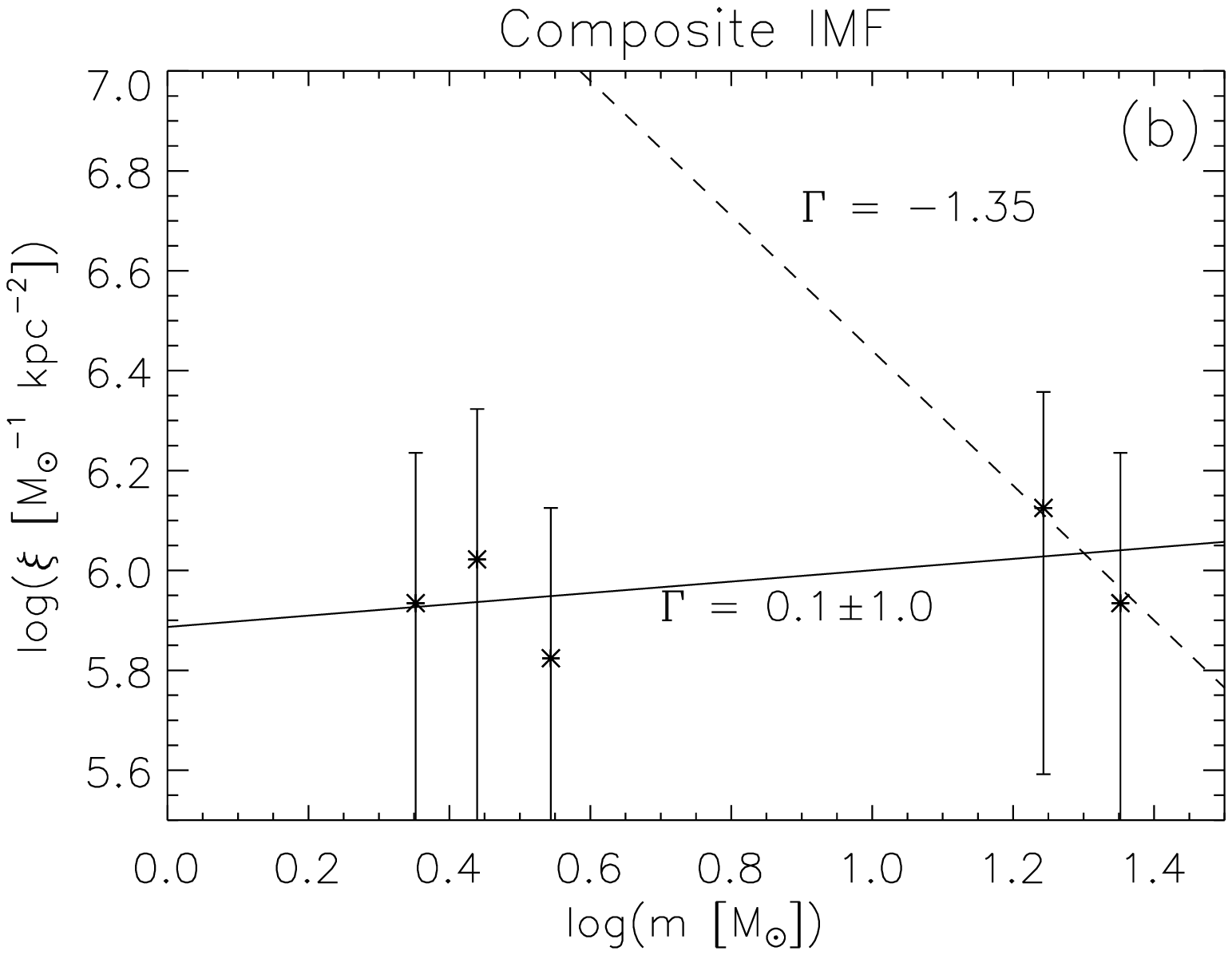}
	\caption{Shown in panel (a) is the CMD of the minimal O star
	groups with membership determined by the stellar density
	analysis.  Panel (b) shows the 
	IMF of the minimal O star groups including companions down to our
	completeness limit of $2M_\odot$ for the F555W band.  We plot a solid line, fit
	to the data, with $\Gamma = 0.1 \pm 1.0$.  For reference, a dashed line with $\Gamma = -1.35$, representing a Salpeter slope, is shown.} 
	\label{compimf}
	\end{center}
\end{figure*}

Figure \ref{compimf}{\it a} shows the
CMD for all companions found using the density
enhancement analysis.  The CMD for
companions found using the friends-of-friends algorithm is identical to that in Figure \ref{compimf}{\it a}
above our completeness limit of $2M_\odot$.  We plot the composite IMF of
these minimal O star groups in Figure \ref{compimf}{\it b}, excluding all
mass bins below our completeness limit.  We find one
companion star in each of the mass bins, $2M_\odot \leq m < 2.5M_\odot$,
$2.5M_\odot \leq m < 3M_\odot$, and $3M_\odot \leq m < 4M_\odot$.   
Upon generating IMFs of the companions found in the friends-of-friends analysis with clustering lengths $l \pm 1 \sigma$, we find
that changing the exact prescription for determining companionship has
little effect on the measured IMF for this population of minimal O
star groups.

The IMF slope for the composite population in the minimal O star
groups varies between $\Gamma = 0.1 \pm 1.0$ to $-0.2 \pm 0.9$ in the preceding
analysis, which is not far from a Salpeter slope of $\Gamma
= -1.35$ within the uncertainties.  However, the true slope may be even larger, since we do not correct for
contamination by main sequence field stars.  Due to the
presence of very massive stars and a small 
number of low-mass companions, the IMF is essentially predetermined
to be unusually shallow, but we note that it does not necessarily
represent a significant variation from the Salpeter value,
statistically, given the small numbers of stars.

\section{Monte Carlo Simulations}
\label{simsection}
This regime of sparse star formation is strongly dominated by
stochastic effects and, as described in \S 1, it offers an important discriminant between star formation theories.  A fundamental
question is whether the stellar IMF in clusters is largely independent
of parent cloud mass, and determined by only the most local physics
(e.g. Maschberger \& Clarke 2008; Krumholz et al. 2010).  If so, the IMF in the
lowest-mass clusters remains identical to that seen in higher-mass
clusters, and should be described by the simple random drawing of
discrete stars from an IMF represented as an ordinary probability
density function.  On the other hand, a scenario that is at least as
plausible is that the IMF is driven by, and limited by, the
mass of the parent molecular cloud (e.g. Bonnell et al. 2004; Weidner \& Kroupa 2004).  If so, then the stochastic effects seen in the lowest-mass clusters will be linked to the parent
cloud masses.  The objects in our study offer a unique opportunity to explore this
extreme parameter space.

\subsection{Analytic Probability}
\label{aprobsec}

We can quantify the likelihood
that these minimal O star groups conform to a standard but
under-sampled stellar IMF.  For a cluster of
$N_*$ stars, the probability that all stars are below mass $m_{\rm
  max}$ is given by:
\begin{equation}
\label{massprob}
P(\mmax, N_*) =\left[ \int_{m_{\rm min}}^{\mmax}  \phi(m)\,dm\
  \right]^{N_*} \quad ,
\end{equation}
where $\phi(m)$ is the IMF and $m_{\rm min}$ is its lower mass limit,
which we assume to be constant.
We adopt a Kroupa IMF (Kroupa 2001), whose form is:  
\begin{equation}
\label{kroupaimf}
\phi(m) dm \propto 
\left\{
\begin{array}{lr}
m^{-1.3} dm\  ~, ~ 0.08M_\odot  \leq  m  <  0.5M_\odot \\
m^{-2.35} dm\   , ~ 0.5M_\odot  \leq  m  <  150M_\odot \\
\end{array}
\right.
\end{equation}

For the minimal O star groups, we estimated $N_*$ as follows.  We
count the candidate member stars, identified as described in \S\S 2.1 --
2.2, and correct for the
expected number of field stars contaminating the cluster line of sight.  The
contamination is determined by the stellar density of the background
field as calculated in \S \ref{hstsec}
and the angular size of the cluster (Table \ref{obsprop}).  
We further correct the
observed numbers to account for stars below our completeness limit of
$1.5 M_\odot$ in the F814W images to get a final $N_*$ estimate over
the full mass range of the IMF.  
Table \ref{analytic} lists star counts for each field as follows.
Column 1 shows the star ID; columns 2 and 3 show the total number of
stars observed within the cluster raduis and the subset of those estimated to be field stars,
respectively; column 4 lists the resulting number of cluster members
above the detection threshold; and column 5 lists the
inferred $N_*$, integrating over the full stellar mass range from $m_{\rm lo}
= 0.08 M_\odot.$  We follow a similar process to estimate cluster mass by summing the masses
of all stars in the cluster and again integrating below our completeness limit down to $m_{\rm lo}
= 0.08 M_\odot.$  The cluster mass estimates are listed in column 6 and will be utilized for analysis in \S 5.

\begin{deluxetable*}{ccccrrr}
  \tabletypesize{\small}
  \tablewidth{0pc}
  \tablecaption{Observed and Estimated Cluster Membership}
  \tablehead{\colhead{Star} & \colhead{Observed} &\colhead{Field} & \colhead{Detected} & \colhead{Estimated} & \colhead{Estimated} & \colhead{$P$($\mmax$, $N_*$)} \\
   \colhead{} & \colhead{Stars} &\colhead{Stars} & \colhead{Members} & \colhead{$N_*$} & \colhead{$\Mcl$($M_\odot$)} & \colhead{}}
  \startdata

    	AzV 58	  & 1 & 0 & 1 & $\leq$19 & $\leq$43 & $\leq$0.03 \\
	AzV 67 	 & 24 & 15 & 9 & 171 & 103 & 0.11 \\
	AzV 106	 & 17 & 9 & 8 & 152 & 72 &  0.20 \\
	AzV 186	 & 1 & 0 &  1 & $\leq$19 & $\leq$72 & $\leq$0.01 \\
    	AzV 226	 & 1 & 0 & 1 & $\leq$19 & $\leq$77 & $\leq$0.01 \\
	AzV 302	 & 13 & 3 &  10 & 190 & 81 & 0.19 \\
    \enddata
    \label{analytic}
\end{deluxetable*}

We then use these $N_*$ values and the
$\mmax$ values from Table \ref{targets} to calculate $P(\mmax, N_*)$
(equation \ref{massprob}).  These
values are given in column 7 of Table \ref{analytic}.  For the minimal O star 
groups, we find that the likelihoods of these clusters containing
stars as massive as those observed range from 11 to 20\%.    In the case of those
isolated O stars with no observed companions, and assuming
that all cluster members are below our detection threshold, the likelihoods
that the observed stars formed with no companions above $1.5 M_\odot$
ranges from upper limits
of $\leq$1\% to $\leq$3\% (Table ~\ref{analytic}).  These likelihoods, while low, 
are not exceedingly so, suggesting
that the occurrence of these minimal O star groups, even those with no
stars above the detection threshold, is not especially unlikely given the
assumed parameters.  Our target selection was based on apparent
isolation in ground-based imaging, and so we expect our objects to
fall in this low-probability regime.
As a reminder, we note that these probabilities are 
based on maximized numbers of stars below the detection limit, in
all cases. 

\subsection{Numerical Simulations}
\label{numsims}

We now devise Monte Carlo simulations that generate clusters and cluster
members to explore the frequency of the observed minimal O star groups
in the context of a stellar clustering law or cluster mass function.  

We simulate the scenario for which the IMF is completely
independent of cluster mass.  We generate clusters using a stellar
clustering law having a default $\beta = -2$ power law slope
(hereafter the ``$N_*$ simulations''): 
\begin{equation}
\label{smcostars}
n(N_*) \ dN_* \propto N_*^{\beta} \ dN_* \quad ,
\end{equation}
where $n(N_*)\ dN_*$ is the number of clusters in the range $N_*$ to
$N_* + dN_*$.  
We set single stars, $N_* = 1$, as our minimum ``cluster'' and
set an upper limit for clusters at $N_* = 10^6$.   
Our use of the --2 power law index
is motivated by observations of a wide variety of
stellar populations including young, massive clusters (e.g., Hunter et
al. 2003; Zhang \& Fall 1999; Fall et al. 2009), super star clusters (Meurer et
al. 1995), globular clusters (e.g., Harris \& Pudritz 1994), and HII
regions (Oey \& Clarke 1998).  
Oey et al. (2004) found that this power
law applies smoothly down to individual field OB stars,
and we assume that the same clustering law holds true for all stellar
masses, down to our lower mass limit of $0.08M_\odot$.

Once clusters are generated following the $N_*$ clustering law
(equation~\ref{smcostars}), each 
cluster is randomly populated with stars using the IMF given by
equation \ref{kroupaimf}.  The IMF, including its stellar mass limits,
is constant for all clusters, thereby allowing
true, isolated O stars to be generated in $N_* = 1$ ``clusters".  For each
$N_*$ simulation, up to $10^7$ clusters are generated to ensure
that we create enough clusters to fully populate the cluster
parameter space.  

We also carry out simulations that generate clusters by total cluster
mass $\Mcl$, according to essentially the same power-law distribution
(hereafter the ``$\Mcl$ simulations''): 
\begin{equation}
\label{massfunction}
 n(\Mcl)\ d\Mcl \propto \Mcl^{\beta} \ d\Mcl\quad ,
\end{equation}
where $n(\Mcl)$ is the number of clusters in the range $\Mcl$
to $\Mcl + d\Mcl$.  We adopt an upper limit to the cluster mass function 
of $10^5M_\odot$.  

Most simulations of clusters heavily favor this method of populating
clusters by mass (e.g. WK06; Parker \& Goodwin 2007; Haas \& Anders
2010).  However, the exact prescription for populating stars up to the
target $\Mcl$ varies.  For our simulations,
we follow Parker \& Goodwin (2007) by populating the cluster with stars,
randomly sampled from the IMF, until the
cluster contains at least 98\% of its target mass in stars.  At this
point, if the last star generated pushes the cluster mass past 105\%
of the target cluster mass, the entire cluster is discarded.
The process of populating the cluster with stars is repeated
until the total stellar mass falls within 98\% to 105\% of the target
cluster mass.  If these conditions are not imposed, the
high-mass stars are too often generated as the last star in the
cluster, and the cluster mass function is not well preserved in the
final sample of clusters.
An investigation of the effects on cluster population
using different algorithms can be found in WK06 and Haas \& Anders (2010). 

We also perform variations of our simulations
using $\beta = -1.8$ and --2.3 power law slopes.  These are values
typically observed as empirical variation for real systems.
For example, observations of giant molecular clouds 
(GMCs) and self-gravitating clumps within GMCs show power-law mass
distributions with slopes $\beta \sim -1.7$ (see Bertoldi \& McKee 1992;
Rosolowsky 2005).  Observations of star clusters in the Large
Magellanic Cloud show measured values of
$\beta \sim -2.0$ to --2.3 (Hunter et al. 1993) and  $\beta \sim -1.8$
(Chandar et al. 2010).  Allowing $\beta$ to vary in our simulations
allows us to explore its effect on our results.

\subsection{The Cluster Lower-Mass Limit}
\label{disc1}

Our cluster simulations are designed to probe the limiting case of
sparse O star formation.  A critical parameter in this regime is
the lower limit placed upon cluster mass or membership number.

With respect to quantized star counts, observational evidence for a
--2 power law relation extending to  $N_* = 1$ clusters are limited to
samples of O and OB stars, probing stellar masses $\geq 20M_\odot$ and
$\geq 10M_\odot$ (Oey et al. 2004).  The form of the stellar
clustering law for a complete sample of stars below these masses is
unknown.  Our simulations extrapolate this --2 power law relation to
stellar masses two orders of magnitude lower than those observed.
However, it is possible that the clustering law has a turnover or
cutoff in the low number regime, or the --2 power law relation may
break down at a stellar mass higher than our lower mass limit of
$0.08M_\odot$.  

This low $N_*$ regime exists in an intermediate stage between
clustered formation and isolated formation, which Adams \& Myers
(2001) term ``group formation".  They estimate that the majority of
star formation occurs in this group formation regime, with $N_*$ =
10--100.  They also estimate a lower bound of $N_*$= 36 for a group to
evolve as a cluster, defined as having a relaxation time longer than
the crossing time of the group.  However, N-body simulations show that clusters
in this ``group formation'' regime are likely to lose a significant portion of their stellar members
on timescales of a few Myr (Bonnell \& Clarke 1999).  Interactions that cause 
the stellar losses preferentially affect low-mass stars, thus making the initial $N_*$ a difficult parameter 
to estimate from observations for clusters in this regime.  This effect may be present in the embedded cluster catalog from Lada \& Lada
(2003), which restricts membership to clusters with $N_* \geq 35$ and shows a turnover in the cluster mass function below $50 M_\odot$.  
Since the typical stellar mass is $\sim 0.5M_\odot$, the turnover that
they observe in the cluster mass function at $50 M_\odot$ corresponds
to $N_* \sim$ 100, right at the regime where the N-body simulations predicted stellar membership loss.  Therefore, while
the exact nature of the lower limit of the clustering law is still
unknown, a truncation at a lower value of $\Nlo$ in the range from $\sim$35 to 100 stars
or $\Mcl$ from $\sim$17.5 to 50$M_\odot$ is reasonable.   This lower limit is similar to that employed in previous
cluster simulations, which typically range from $5M_\odot$
(Weidner \& Kroupa, 2004) to $50M_\odot$ (Parker \& Goodwin, 2007).

With respect to the cluster mass function, extragalactic studies of
complete samples of clusters have probed to $\Mcl \sim 10^3
M_\odot$ (Chandar et al. 2010), limiting our knowledge of the form of
the mass function for complete samples below this mass.  Lada \& Lada
(2003) probe clusters below $10^3 M_\odot$ by compiling a catalog of
embedded clusters in the solar neighborhood.  Their results are consistent
with a $\Mcl^{-2}$ cluster mass function for clusters from $50
M_\odot$ to $10^3 M_\odot$.  Below $50 M_\odot$, they find a
statistically significant turnover in the cluster mass function.  This may indicate
that the lower limit of our $\Mcl$ simulations should be truncated or
the power law should turn over around a value $50 M_\odot$.

In the next sections, we compare different simulations
with observed statistics for SMC massive star clustering and 
isolated Galactic O stars.  In this comparison, we include three
separate lower $N_*$ limits for the clustering law at $\Nlo$ = 1,
40, and 100 stars, and two separate lower
$\Mcl$ limits of the cluster mass function at $\Mcllo$ = $20M_\odot$ and $50M_\odot$.
Since the typical stellar mass is $\sim 0.5M_\odot$, $\Nlo$ = 40 is the appropriate equivalent lower limit for $\Mcllo$ = $20M_\odot$ and that for $\Mcllo$ = $50M_\odot$ is $\Nlo$ = 100.  While these values are empirically motivated, they do also allow
the formation of stars  $>20M_\odot$, as required for our purposes.  In what follows, it is important to bear in mind that
the cluster parameterizations are extrapolated beyond observed mass ranges.

\subsubsection{Comparison with SMC Clustering Statistics}
\label{smcstat}

We compare the distribution of massive stars in the simulated
clusters with the actual, observed distribution of stars among
SMC clusters.  Oey et al. (2004) broke down the clustering of observed,
photometrically-identified massive star candidates with a
friends-of-friends algorithm.  They identified two samples, the
``O-star sample'' and the ``OB-star sample,'' corresponding to masses
$m\gtrsim 20M_\odot$ and $m\gtrsim 10M_\odot,$ respectively.
In our simulations, we consider only clusters containing one or more O
or OB stars, defined by the same respective mass ranges.  Table~\ref{oclusterlaw} summarizes our findings on massive star clustering.  The rows are divided into three sections which denote results from the $N_*$ simulations, the $\Mcl$ simulations, and the Oey et
al. (2004) observations.  For the simulations, $\Nlo$ or $\Mcllo$ is listed in column 1 and the power law slope is listed in column 2.  
Columns 3, 4, and 5 list the frequencies of having one, two, or more than two O stars, respectively, in the clusters; while columns 6, 7, and 8 list these frequencies for OB stars in the clusters.

\begin{deluxetable*}{ccccccccc}
  \tabletypesize{\small}
  \tablewidth{0pc}
  \tablecaption{Fraction of O (OB) star clusters having a given number of O(OB) stars}
  \tablehead{\colhead{$\Mcllo$ or $\Nlo$} & \colhead{$\beta$} & \colhead{1 O star}  & \colhead{2 O stars}  & \colhead{$>2$ O stars} & \colhead{1 OB star} & \colhead{2 OB stars}  & \colhead{$>2$ OB stars}}
  \startdata
	& & & & $N_*$ Simulations\tablenotemark{a} \\ \hline
	 1  & -1.8 & 0.75	& 0.10	& 0.151 & 0.73 & 0.11 & 0.16\\
	 40  & -1.8 & 0.61	& 0.15	& 0.24 & 0.53 & 0.18 & 0.29\\
	 100  & -1.8 & 0.53	& 0.18	& 0.29 & 0.40 & 0.20 & 0.40\\
    	 1  & -2.0 & 0.87	&  0.07	&  0.07 & 0.85 & 0.08 & 0.08 \\
	 40   & -2.0 & 0.70	& 0.15	&  0.15 & 0.61 & 0.18	& 0.21  \\
	 100   & -2.0 & 0.61	& 0.18	&  0.21 & 0.46 & 0.22	& 0.32  \\
	1   & -2.3 & 0.96	& 0.03 & 0.02  & 0.95 & 0.04 & 0.02  \\
	 40   & -2.3 & 0.79	& 0.12	&  0.08 & 0.70 & 0.17	& 0.13  \\
	 100   & -2.3 & 0.69	& 0.18	&  0.14 & 0.53 & 0.23	& 0.24  \\ \hline
	& & & & $\Mcl$ Simulations\tablenotemark{a} \\ \hline
	 $20M_\odot$	& -1.8 & 0.56	& 0.15 & 0.29 & 0.53 & 0.16 & 0.31 \\
	 $50M_\odot$	& -1.8 & 0.54	& 0.16 & 0.30 & 0.44  & 0.19 & 0.37 \\
	 $20M_\odot$	& -2.0 & 0.67	&  0.14 & 0.19  & 0.63 & 0.16 & 0.21 \\
	 $50M_\odot$	& -2.0 & 0.64	&  0.16 & 0.21  & 0.52 & 0.20 & 0.28 \\
	 $20M_\odot$	& -2.3 & 0.78  & 0.12 & 0.10 & 0.74 & 0.14 & 0.12 \\
	 $50M_\odot$	& -2.3 & 0.75  & 0.14 & 0.12 & 0.61 & 0.20 & 0.19 \\ \hline
	 & & & & SMC Observed\tablenotemark{b} 	\\ \hline
	 ... & ... & 0.61$\pm$0.08	&  0.19$\pm$0.04   &  0.19$\pm$0.04 & 0.65$\pm$0.04 & 0.15$\pm$0.02 & 0.19$\pm$0.02  \\
    \enddata
    \tablenotetext{a} {Errors for simulated values are $\leq 0.01$}
    \tablenotetext{b} {From Oey et al. (2004).}
    \label{oclusterlaw}
\end{deluxetable*}

Table~\ref{oclusterlaw} reveals three trends: (1)
steepening the power law slope of the simulation results in an
increasing fraction of O or OB star clusters containing a single O or
OB star, (2) the fraction of
OB star clusters with a single OB star is lower than the fraction of O
star clusters with a single O star and (3) the $N_*$ simulations are more sensitive to the lower
limit truncation than the $\Mcl$ simulations.  The final trend is due to our cluster population method.  In the case where $\Mcl < m_{\rm up}$, stars with mass greater than $\Mcl$ will not be allowed to form in such a cluster in the $\Mcl$ simulations.  However, in the $N_*$ simulations, each cluster can form stars up to $m_{\rm up}$.

In the observations by Oey et al. (2004), 61--65\% of clusters having
at least one massive star contain only a single massive star.  We find
best agreement with this fraction in our simulations following a --2
power law slope and a truncation of $N_*$ = 40 or
$20M_\odot$.  Table \ref{oclusterlaw} also shows good agreement with $\beta = -1.8$ for the $N_*$ simulation
having a truncation of $N_*$ = 20.  For $\beta = -2.3$, the
steepness of the slope causes the OB cluster sample and O star cluster
sample to behave differently enough from each other that they cannot both be in
agreement with the observations, regardless of the lower limit.  Similarly, at a truncation of $N_*$ = 100 or $\Mcl =
50M_\odot$, the OB cluster sample and O star cluster sample never come
into agreement simultaneously at the same power law slope. 

\subsubsection{Comparison with Galactic Isolated O Star Fraction}
\label{isoostar}

We also compare our simulations to the de Wit et al. (2005)
result that 4$\pm$2\% of all Galactic O stars are found in isolation.
The simulations by Parker \& Goodwin (2007) are in agreement with
this result.  Since we modeled our $\Mcl$ simulations after their work,
our results should match theirs quite closely, although
the exact parameters of the simulations differ slightly.  Parker \&
Goodwin (2007) set the lower mass limit of the stellar IMF to
$0.1M_\odot$ , while
our simulations use a value of $0.08M_\odot$.  Also,
Parker \& Goodwin use a lower limit of $\Mcl = 50M_\odot$, while we
vary the limit as discussed above.  For comparison with this study, we adopt the de Wit et al. (2005) definition of O stars as having mass $\geq 17.5M_\odot$.  This comparison
provides a good check for our results and also quantifies the effect
of the lower limit, $\Mcllo$ or $\Nlo$, on the isolated O star fraction.  

Parker \& Goodwin's definition of an isolated O star is twofold: first,
the cluster contains no B stars with mass $\geq 10M_\odot$ and second, the
total cluster mass $\Mcl < 100M_\odot$.  These constraints were
set according to the detection limits of the de Wit et al. (2005)
result and are intended to mimic the sensitivity of those
observations.  For the following analysis, we follow Parker \& Goodwin's definition of O star isolation.  The results of this analysis are shown in Table
\ref{singleo}, which has rows divided between the $N_*$ simulations, the $\Mcl$ simulations, and the de Wit et al. (2005) observations.  $\Nlo$ or $\Mcllo$ for the simulations is listed in column 1, the power law slope for the simulations is listed in column 2, 
and the fraction of isolated O stars as a fraction of all O stars is listed in
column 3.  This table reveals trends similar to those seen in
Table~\ref{oclusterlaw}, that steepening the 
power law slope increases the fraction of isolated O stars, and that
the $N_*$ simulations are more sensitive to the
lower limit truncation than the $\Mcl$ simulations.
Table~\ref{singleo} also shows that the fraction
of isolated O stars matches very well between the $N_*$ and $\Mcl$
simulations having lower limits of $\Nlo$ = 40 and $\Mcllo = 20M_\odot$,
respectively.  

For their simulation adopting a --2 power law slope, Parker \& Goodwin
find that 4.6\% of O stars are isolated, in good agreement with the
4$\pm$2\% result from de Wit et al. (2005).  We find that many of our simulations
match the de Wit et al. (2005) result, spanning the full range of
lower limits and power law slopes tested, with the exception of $\Nlo$
= 1.  Table \ref{singleo} shows that the --2 power law simulations that best agree with
the isolated O star fraction have the higher values of $\Mcllo$ =
$50M_\odot$ or $\Nlo$ = 100, whereas the simulations that best agree with the SMC O and OB star
clustering have the lower values of $\Mcllo$ = $20M_\odot$ or $\Nlo$ =
40 (Table \ref{oclusterlaw}).  Simulations following a --2.3 power law
slope are least reconcilable with both sets of observations.   

\begin{deluxetable}{ccccccc}
  \tabletypesize{\small}
  \tablewidth{0pc}
  \tablecaption{Fraction of isolated O stars\tablenotemark{a}}
  \tablehead{\colhead{$\Mcllo$ or $\Nlo$} & \colhead{$\beta$} & \colhead{Iso. Fraction\tablenotemark{b}}}
  \startdata
	& $N_*$ Simulations \\ \hline
	 1  	& -1.8 & 0.099	\\
	 40  	& -1.8 & 0.029	\\
	 100	& -1.8 & 0.008	\\
    	 1   	& -2.0 & 0.330	 \\
	 40   	& -2.0 & 0.083	 \\
	 100 	& -2.0 & 0.022	 \\
	 1   	& -2.3 & 0.728	  \\
	 40  	& -2.3 & 0.212	  \\
	 100 	& -2.3 & 0.057	  \\ \hline
	& $\Mcl$ Simulations \\ \hline
	 $20M_\odot$	& -1.8 & 0.027	 \\
	$50M_\odot$	& -1.8 & 0.019	 \\
	 $20M_\odot$	& -2.0 & 0.072	 \\
	 $50M_\odot$ 	& -2.0 & 0.048 \\
	 $20M_\odot$	& -2.3 & 0.190   \\
          $50M_\odot$	& -2.3 & 0.123   \\ \hline
          & Galactic Observed\tablenotemark{c} \\ \hline
	 ... & ... & 0.04$\pm$0.02  \\ 
    \enddata
    \tablenotetext{a} {Here, the definition of isolated O stars is from Parker \& Goodwin (2007).}
    \tablenotetext{b} {Errors for simulated values are $\leq 0.01$.}
    \tablenotetext{c} {From de Wit et al. (2005).}
    \label{singleo}
\end{deluxetable}

\subsubsection{Default Clustering Models}
   
In \S\S \ref{smcstat} and \ref{isoostar}, we find that a number of our
   simulations in both $N_*$ and $\Mcl$ agree with observed statistics
   of SMC massive star clustering and Galactic
   isolated O stars.  The $N_*$ simulations are more
   sensitive to both the power law slope and $\Nlo$, and 
   many models agree with these
   two sets of observations (Tables~\ref{oclusterlaw} and
   \ref{singleo}).    But for the $\Mcl$ simulations, we 
   find that only the $\beta = -2.0$ models match with both
   sets of observations, albeit for different $\Mcllo$ lower limits.  For the --1.8 and --2.3 power law slopes, none of the simulations can appropriately match both the SMC O and OB clustering simultaneously for a given $\Nlo$ or $\Mcllo$, making a strong case for $\beta = -2.0$ models as the best choice for the power law slope.  
   
   We therefore take the --2.0
   power law as the base model for comparison between the $\Mcl$ and
   $N_*$ simulations.  From Table \ref{oclusterlaw} we find that models
   with $\Nlo$ = 40 and
   $20M_\odot$ best match the SMC massive star clustering observations, while Table \ref{singleo}
   shows that models with lower limits at $\Nlo$ = 100 and $\Mcllo$ =
   $50M_\odot$ best match the isolated Galactic O star observations.  Looking further, in Table
   \ref{oclusterlaw} we see that the models with truncations at $N_*$ =
   100 and $\Mcl$ = $50M_\odot$ also match well with SMC O star
   clustering, but do not agree with SMC OB star clustering, with the
   fraction of single OB star clusters being off by $\sim 10\%$.  
This difference is large compared to the fraction of isolated O stars
   for the models with $\Nlo$ = 40 and $\Mcllo$ =
   $20M_\odot$, which agree with the observations within $\sim 1 - 2 \%$.  Thus, we conclude that the simulations that best match both the SMC massive star clustering statistics and the fraction of isolated Galactic O stars are the $N_*$ simulation with $\Nlo = 40$ and the $\Mcl$ simulation with $\Mcllo = 20M_\odot$.  We cannot rule out either $N_*$ or $\Mcl$
  in favor of the other and do not find evidence that either one is a
  more fundamental parameter. 
   
   In the following sections, we compare these two simulations with our observations of minimal O star groups.  For this comparison we use the $\Mcl^{-2}$ and $N_*^{-2}$ simulations with $\Mcllo$ = $20M_\odot$ and $\Nlo$ = 40, respectively, as our default models.

\section{Stellar mass ratio $\mratio$}

A simple and important parameter we can compare between our observations
and simulations is $\mratio$, the mass ratio
of the second-most massive and most massive stars in the
cluster.  This ratio is a directly observable quantity that we can
measure for the minimal O star groups, and as such, provides a
powerful parameter to use as a comparison between our observations and
simulations. 

\begin{figure*}
	\begin{center}
	\includegraphics[scale=.5,angle=0]{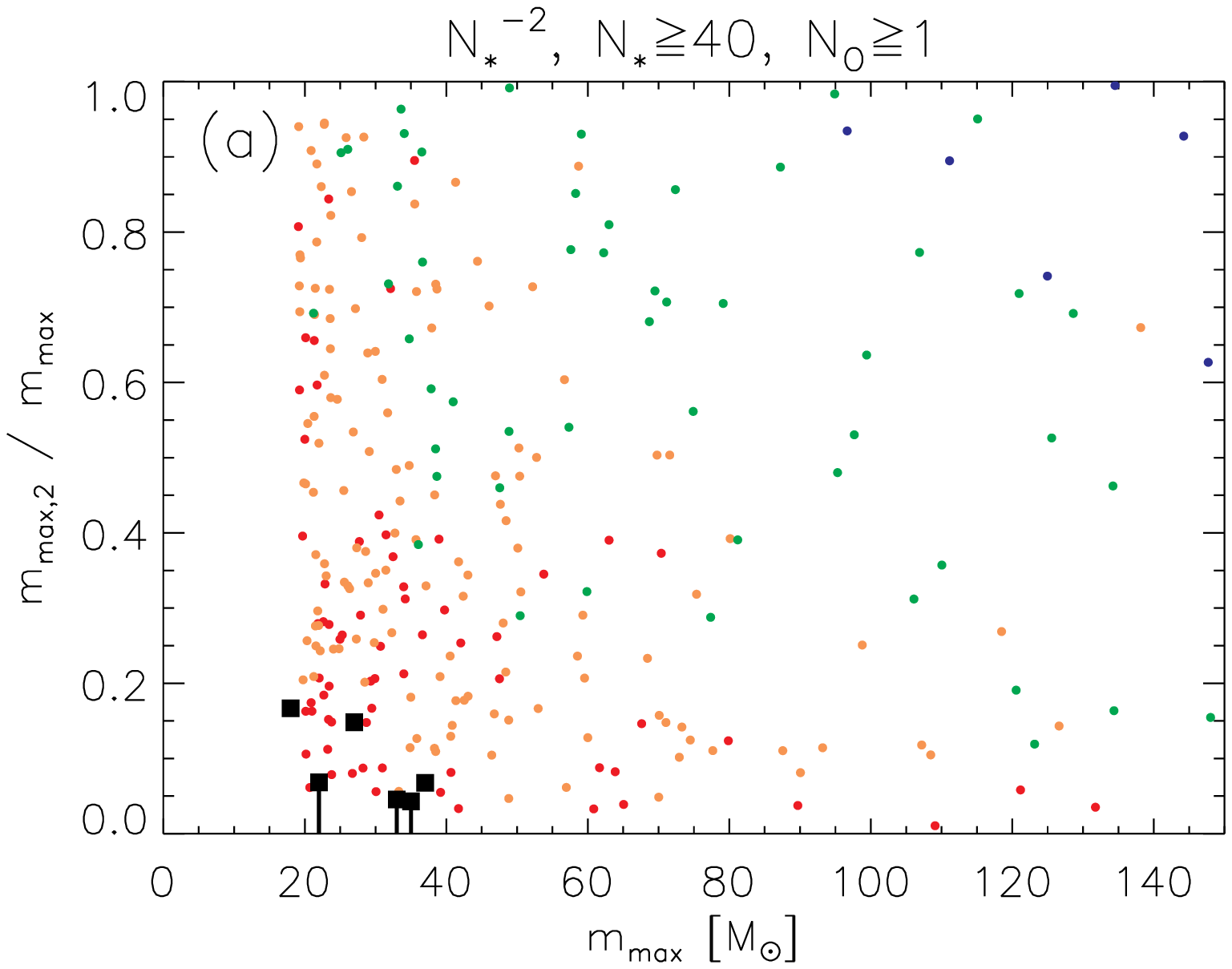}
	\includegraphics[scale=.5,angle=0]{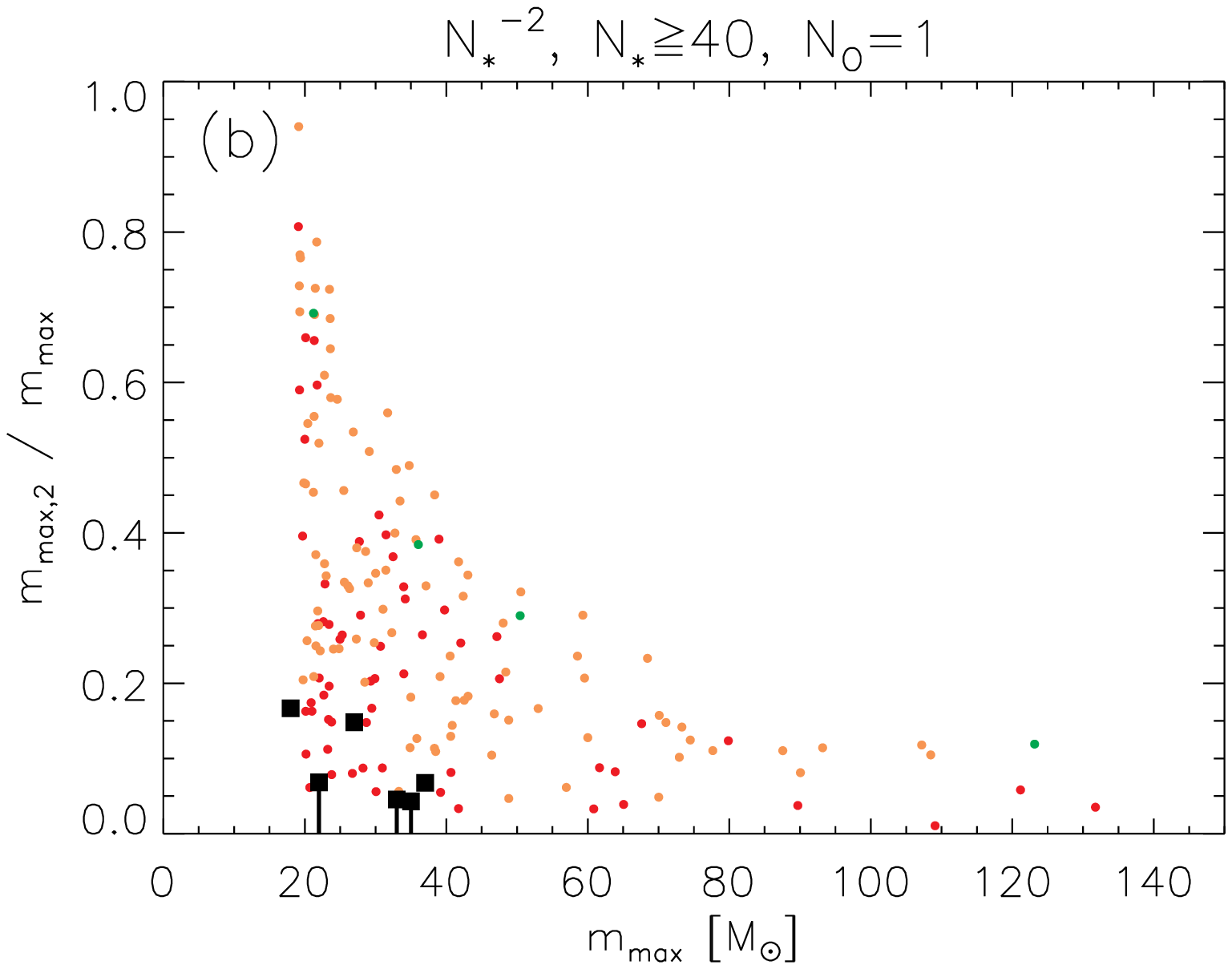}
	\caption{$\mratio$ vs $\mmax$ from the $N_*^{-2}$ simulation with $\Nlo = 40$ for ({\it a}) 
	 all clusters having at least one
	O star, and ({\it b}) all clusters having only a single
	O star.  Data are color coded in
	logarithmic bins of $N_*$, with black = $0 \le \log N_* <
1$, red = $1 \le \log N_* < 2$, orange = $2 \le \log N_* < 3$, green = $3 \le \log N_* < 4$, and blue = $4 \le \log N_* < 5$. Our observations are plotted as black squares.} 
	\label{mratio}
	\end{center}
\end{figure*}

\begin{figure*}
	\begin{center}
	\includegraphics[scale=.5,angle=0]{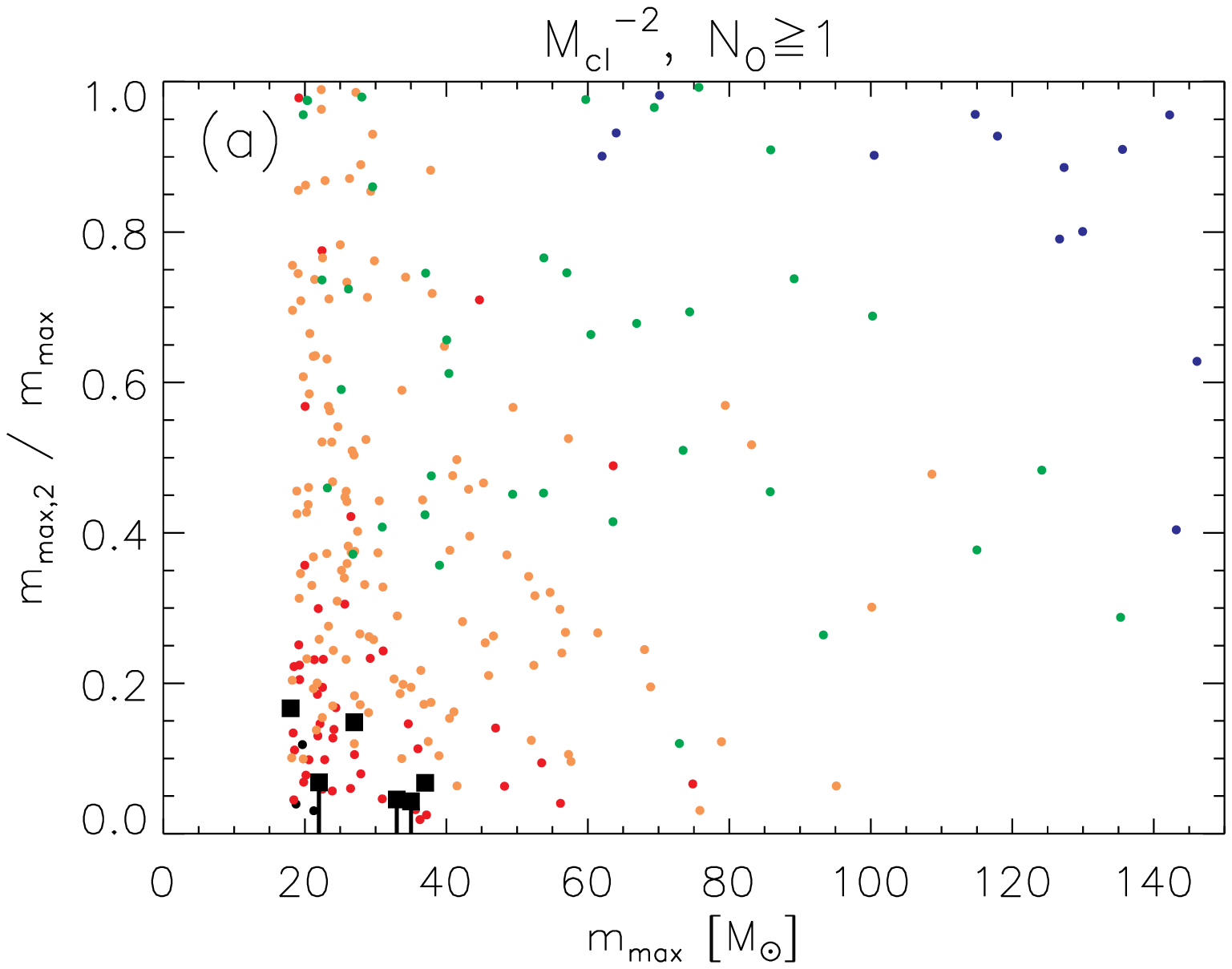}
	\includegraphics[scale=.5,angle=0]{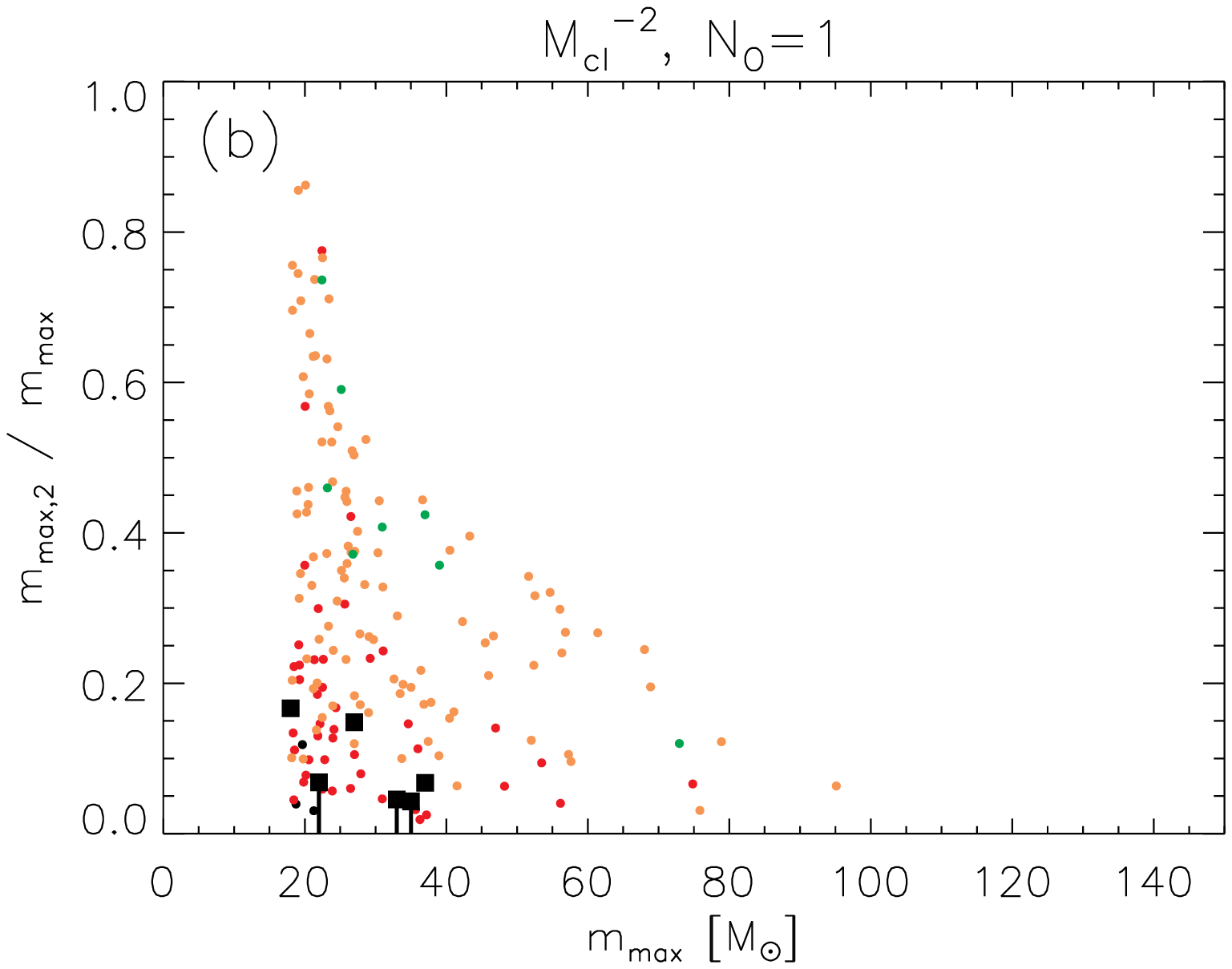}
	\caption{ Same as Figure \ref{mratio}, but for the $\Mcl^{-2}$ simulation with $\Mcllo = 20 M_\odot$.}
	\label{mratioM}
	\end{center}
\end{figure*}

There are
two populations in the simulation that are of interest: (1) clusters
that contain one or more O stars and (2) clusters that contain just
one O star.  We explore the full simulation parameter space of the
mass ratio $\mratio$ as a function of $\mmax$ in Figures
\ref{mratio} and 
\ref{mratioM}.  For Figures \ref{mratio} and 
\ref{mratioM}, panel (a) includes simulated clusters with at least one
O star in the cluster, while panel (b) includes only simulated
clusters with exactly one O star.  For these plots
and all subsequent simulations, we adjust our definition of O stars to
be $m \geq 18M_\odot$ instead of our earlier $20 M_\odot$ definition, allowing
our lowest-mass target stars
to be included in the parameter space when calculating percentile frequencies below.  Since the uncertainty
in our stellar mass estimates is $1M_\odot$ to $3M_\odot$, our adjusted definition is on the order of the uncertainty.  
The simulated clusters plotted in these figures are the O star
clusters taken from a random sample of 573 simulated OB clusters, the 
same number of OB clusters as observed in the SMC, having at least one OB
star ($m\geq 10M_\odot$; Oey et al. 2004).
In these and all subsequent figures, we exclude clusters without O stars from the plots.
The color coding in Figures~\ref{mratio} and \ref{mratioM}
indicates the number of stars per
cluster, with black indicating $0 \le \log N_* <
1$, red indicating $1 \le \log N_* < 2$, orange indicating $2 \le \log
N_* < 3$, green indicating $3 \le \log N_* < 4$, and blue indicating
$4 \le \log N_* < 5$. 
Observations from this paper are 
plotted as black squares.  For our apparently isolated O stars, we
note that undetected companions may exist, having individual masses
up to $1.5M_\odot$, and so we plot their $\mratio$ as upper limits.

Our observations lie in an interesting region of the parameter space
 in both the $\Mcl$ and $N_*$ simulations.  There is a strong drop-off in the population of
simulated clusters having $\mratio < 0.02$.  This drop-off coincides with some of our observed isolated targets.  Thus, the upper limits in $\mratio$ for
the isolated stars approach the extreme lower limit of the parameter space.  All of our observations lie within the parameter space covered by simulated clusters.

To examine this quantitatively, we identify the
subset of simulated clusters that correspond to the
selection criteria of our observed stars and clusters.  For each
target star, we identify all
simulated clusters having $\mmax$ within the uncertainty of the star's
empirically-derived mass (Table~\ref{targets}).  We also eliminate all simulated
clusters containing more than one O star, to obtain our final
sub-sample of simulated clusters.  For $\mmaxtwo$, we use the mass
of the most massive observed, companion main-sequence star.  In the
case of our apparently isolated massive stars, we set $\mmaxtwo$ to an
upper-limit value of $1.5M_\odot$, our F814W 
completeness limit.  In the following analysis, we denote the
observed $\mratio$ as $[\mratio]_{\rm obs}$ and that from
simulated clusters as $[\mratio]_{\rm sim}$.  Table
\ref{simprop} lists the fraction of clusters with [$\mratio]_{\rm sim}
\le [\mratio]_{\rm obs}$ for each star.
Column 1 lists the star ID; column 2 lists the mass of the OB star from Table \ref{targets}; column 3 lists $[\mratio]_{\rm obs}$;
columns 4, 5, and 6 list the fraction of clusters with [$\mratio]_{\rm sim}
\le [\mratio]_{\rm obs}$ for 
the $N_*$ simulations with $\Nlo = 40$, having slopes of --1.8, --2, and --2.3,
respectively; and columns 7, 8, and 9 list these fractions
for the $\Mcl$ simulations with $\Mcllo = 20M_\odot$, as shown.
Columns 5 and 8 correspond to the simulated population in Figures
\ref{mratio}{\it b} and \ref{mratioM}{\it b}, respectively.  Columns 10 and 11 are
the same as Columns 5 and 8, respectively, except that they correspond
to the simulated population in Figures \ref{mratio}{\it a} and
\ref{mratioM}{\it a}, showing all clusters with $\geq 1$ O star. 

\begin{deluxetable*}{ccccccccccc}
  \tabletypesize{\small}
  \tablewidth{0pc}
  \tablecaption{Fraction of clusters with [$\mratio]_{\rm sim} \le [\mratio]_{\rm obs}$}
  \tablehead{\colhead{Star} & \colhead{$\mmax$ ($M_\odot$)} & \colhead{($\frac{m_{\rm max,2}} {m_{\rm max}}$)$_{obs}$} & \colhead{$N_*^{-1.8}$} & \colhead{$N_*^{-2}$} &\colhead{$N_*^{-2.3}$} & \colhead{$\Mcl^{-1.8}$} & \colhead{$\Mcl^{-2}$} & \colhead{$\Mcl^{-2.3}$} & \colhead{$N_*^{-2}$, $N_O \geq 1$} & \colhead{$\Mcl^{-2}$, $N_O \geq 1$}}
  \startdata

    	AzV 58	 & 22 & 0.07 & 0.01 & 0.02 & 0.03 & 0.04 & 0.06 & 0.09 & 0.02 & 0.05 \\
	AzV 67 	 & 37 & 0.07 & 0.07 & 0.10 & 0.11 & 0.08 & 0.10 & 0.11 & 0.07 & 0.06 \\
	AzV 106	 & 18 & 0.17 & 0.11 & 0.14 & 0.18 & 0.16 & 0.21 & 0.27 & 0.14 & 0.20 \\
	AzV 186	 & 33 & 0.05 & 0.01 & 0.02 & 0.02 & 0.03 & 0.04 & 0.05 & 0.01 & 0.03 \\
    	AzV 226	 & 35 & 0.04 & 0.01 & 0.02 & 0.02& 0.03 & 0.04 & 0.05 & 0.01 & 0.03 \\
	AzV 302	 & 27 & 0.15 & 0.20 & 0.24 & 0.29 & 0.19 & 0.24 & 0.30 & 0.19 & 0.19 \\
    \enddata
    \label{simprop}
\end{deluxetable*}

Two trends emerge from these data: (1) steepening the power law slope
increases the fraction of simulated clusters with [$\mratio]_{\rm sim}
\le [\mratio]_{\rm obs}$ and (2) the $N_*$ simulations typically have a slightly
lower fraction of simulated clusters having  [$\mratio]_{\rm sim} \le
[\mratio]_{\rm obs}$ than the $\Mcl$ simulations.  The first effect is
caused by steeper slopes creating a greater fraction of small
clusters, where stochastic effects can result in massive stars forming
with only a few low mass companions.  The second effect is caused by
the fact that some of the clusters fall into the low $\mratio$ regime
in the $\Mcl$ simulation where $N_* < 40$, which is not allowed in the
$N_*$ simulation, due to our lower limit of $\Nlo = 40$.
Looking specifically at the $\beta = -2$ simulations, Table \ref{simprop} shows that the
frequency of single O star simulated clusters having [$\mratio]_{\rm sim} \le
[\mratio]_{\rm obs}$ for all observed groups is below the 25th percentile in both simulations, confirming the impression from Figures \ref{mratio}{\it b} and 
\ref{mratioM}{\it b}.  For all O star clusters from Figures \ref{mratio}{\it a} and \ref{mratioM}{\it a}, our observed groups are all below the 20th percentile.  Some of our isolated star observations are found in the lowest 5th percentile in both simulations, but still within the parameter space covered by the simulations.

That our observations are not well-distributed among the cluster
population in Figures \ref{mratio}{\it b} and \ref{mratioM}{\it b}
is primarily due to our sample selection.  Our selection process included a visual
inspection of our targets using ground-based imaging to ensure they
appeared isolated, thereby ensuring that our objects have
extremely low values of $\mratio$.
These targets were drawn to qualify for both the field O star and field
OB star samples from Oey et al. (2004), 
defined to have no other stars having $m \geq 20M_\odot$ and $m\geq
10M_\odot$ within a clustering length, respectively.  Thus by
definition, $\mmaxtwo < 10M_\odot$ for our sample.  On the other hand,
27 of the 91 stars (30\%) in the field O star sample of Oey et al. are not
members of the field OB sample, implying that for these stars $\mmaxtwo \geq 10M_\odot$.
The remaining 70\% of isolated O stars having companions with masses below $10M_\odot$ can be compared to a
simulated fraction of 57\% in our $\Mcl^{-2}$ simulation and 64\% in our $N_*^{-2}$ simulation.
However, the SMC field O star sample is contaminated by runaway stars which will inflate the observed 
fraction of O stars with $\mmaxtwo \leq 10M_\odot$.

However, we also note that the distribution of $\mratio$ in the
simulations does depends somewhat on the cluster population parameters
and populating algorithm.  For example, if we do not discard and repopulate
clusters when the total cluster mass exceeds 105\% of the
target mass (see \S \ref{numsims}), then our test simulations
show that the increased production of massive stars can affect the
percentiles by up to a factor of two, so that our objects fall
in the lowest 50th percentile, for the default $\Mcl$ simulation.
Thus in this case, our observed clusters are nearer to the median and more
well-distributed overall.

\section{The relation between $\mmax$ and $\Mcl$} 
\subsection{Observations and Simulations}

\begin{figure*}
	\begin{center}
	\includegraphics[scale=.5,angle=0]{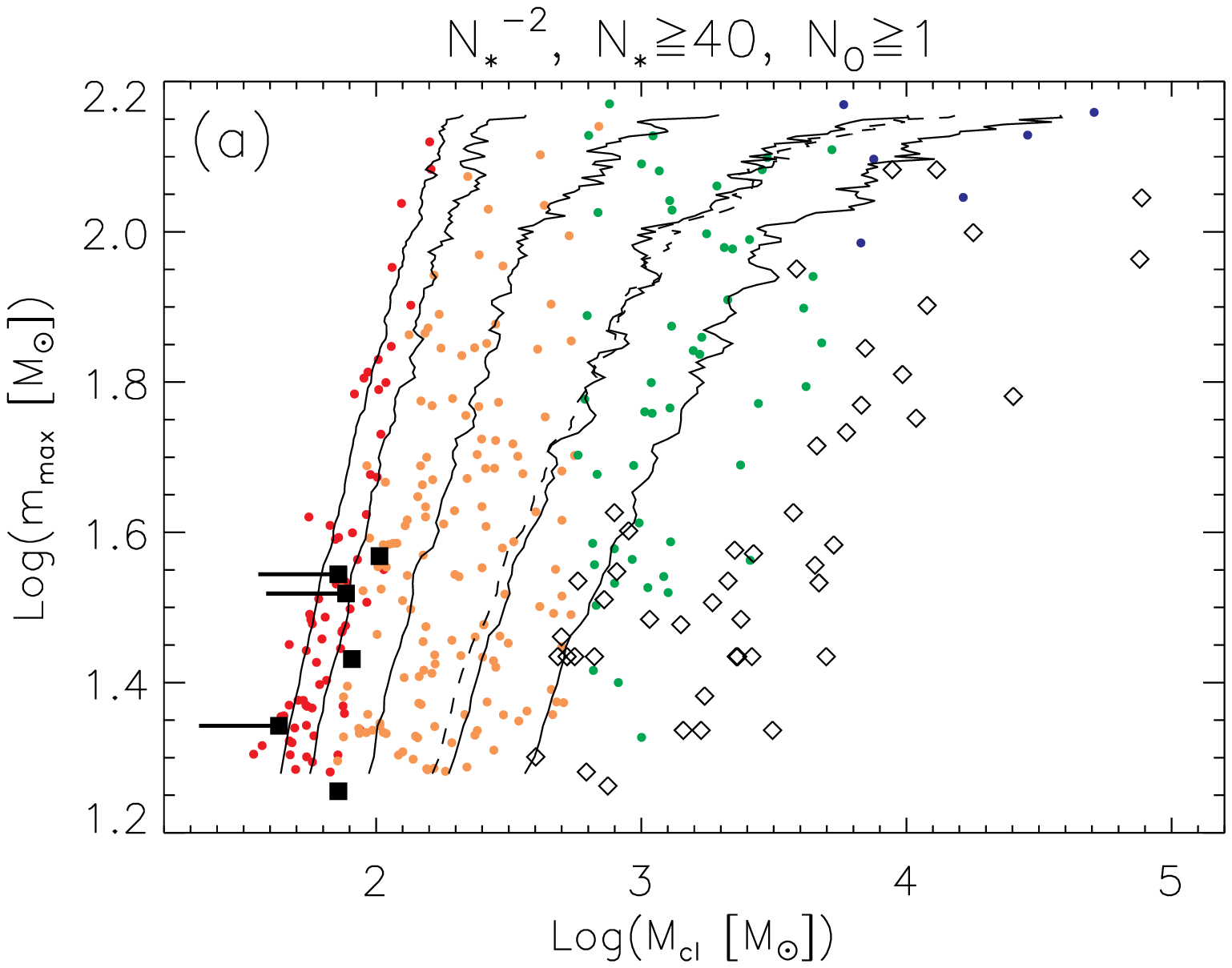}
	\includegraphics[scale=.5,angle=0]{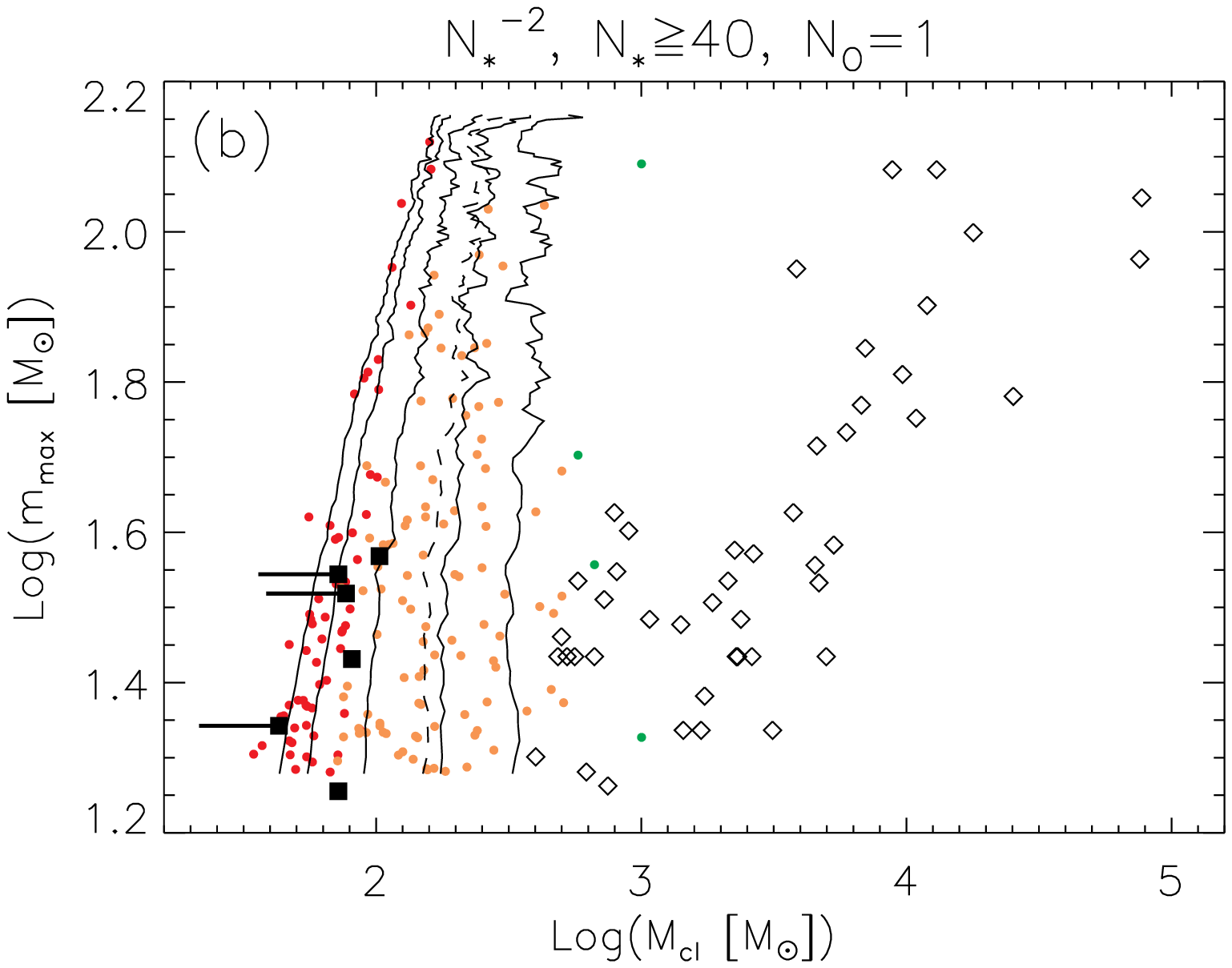}
	\caption{log($\mmax$) vs. log($\Mcl$) from the $N_*$ simulations for ({\it a}) all clusters having at least one O star, and ({\it b}) all clusters having only a single O star.   Colors are as in Figure \ref{mratio}, with diamonds representing Galactic clusters tabulated by Weidner et al. (2010a).  The solid lines represent the 10th, 25th, 50th, 75th, and 90th percentile of $\Mcl$ as a function of $\mmax$ from the simulations.  The dashed line represents the average value from the simulations. }
	\label{msmcm}
	\end{center}
\end{figure*}

\begin{figure*}
	\begin{center}
	\includegraphics[scale=.5,angle=0]{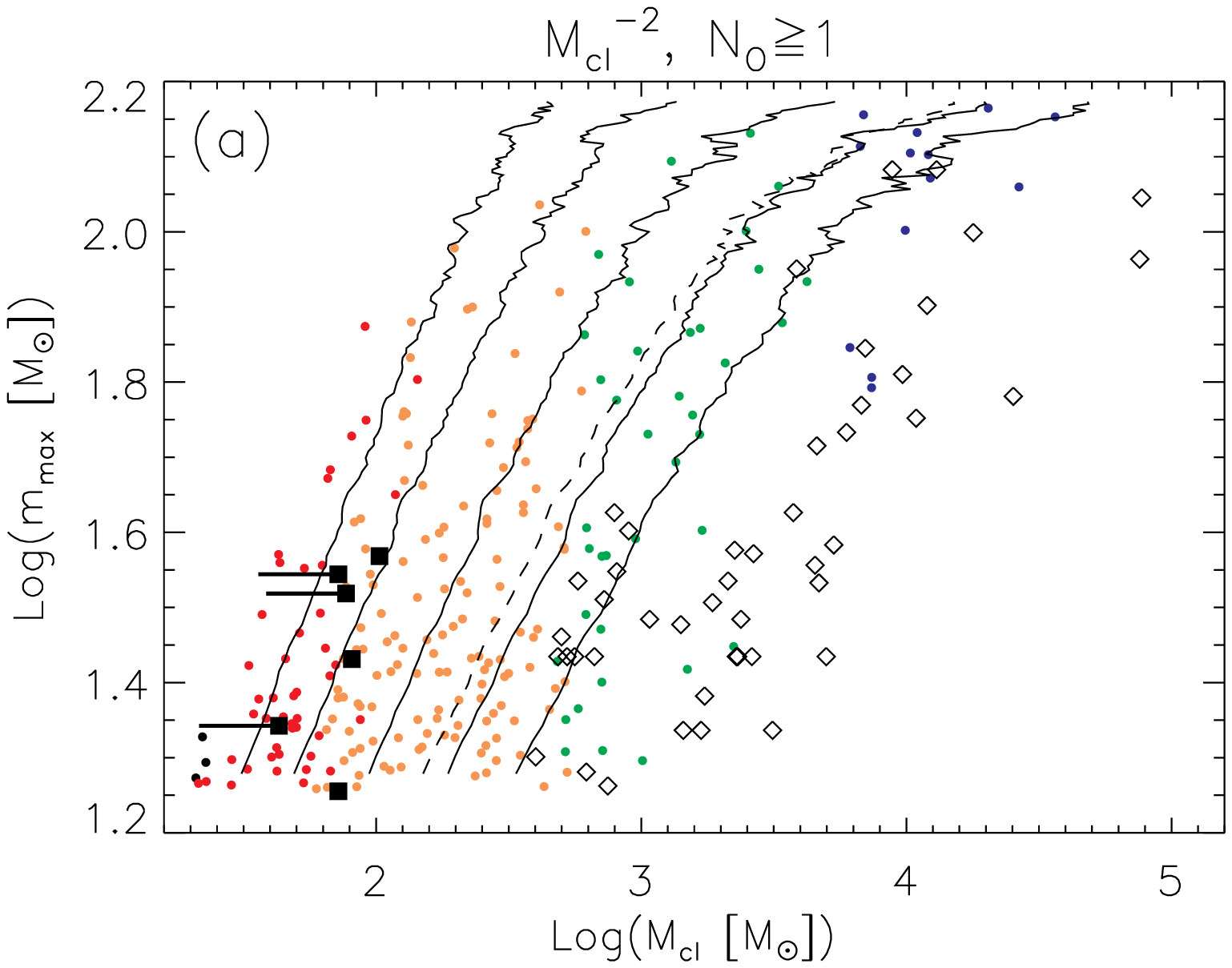}
	\includegraphics[scale=.5,angle=0]{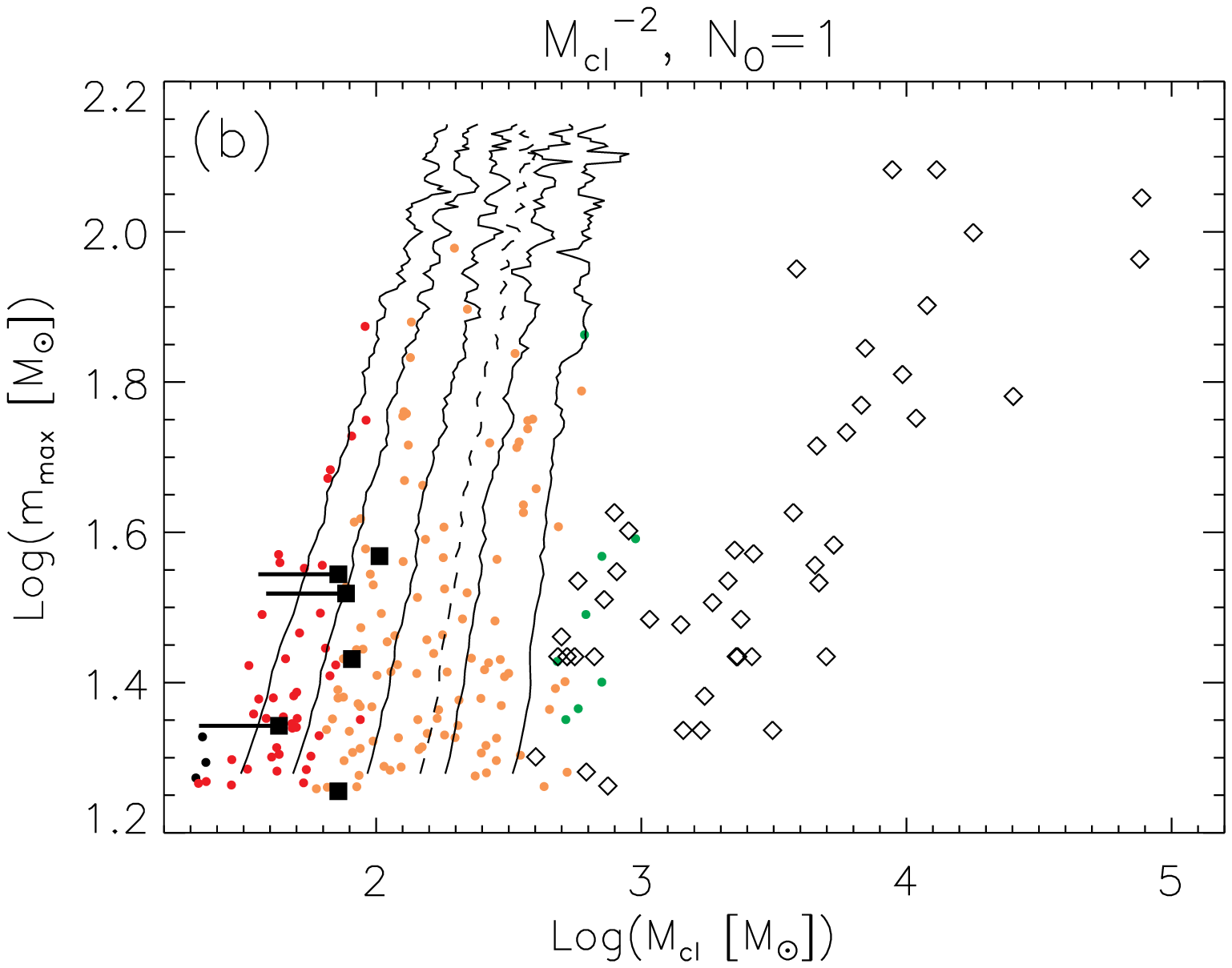}
	\caption{Same as Figure \ref{msmcm} but for the $\Mcl$ simulations.}
	\label{msmcmM}
	\end{center}
\end{figure*}

In Figures \ref{msmcm} and \ref{msmcmM}, we compare the relationship
between the cluster mass $\Mcl$ and maximum stellar mass $m_{\rm
max}$.  As in Figure \ref{mratio}, panel (a) includes simulated clusters with $\ge 1$
O star, while panel (b) includes only simulated
clusters with exactly one O star. The color coding scheme is also the same as in Figure~\ref{mratio}.  The solid lines show contours for the 10th,
25th, 50th, 75th, and 90th percentiles of $\Mcl$ as a function of
$\mmax$ in the simulation, while the dashed line represents the mean.
These percentiles are computed from a much larger set of $10^6$ ($\Mcl$ simulations) or $10^7$ ($N_*$ simulations)
modeled clusters, to reduce stochastic scatter.
For our observations (black squares), we calculate $\Mcl$ as described in \S \ref{aprobsec} (Table \ref{analytic}).  The diamonds show observed Galactic clusters
whose $\Mcl$ and $\mmax$ are tabulated by Weidner et
al. (2010a).

Figures \ref{msmcm}{\it a} and \ref{msmcmM}{\it a} show that the majority of the
Galactic cluster sample lies above the 90th percentile in $\Mcl$ for a
given $\mmax$ in both the $N_*$ and $\Mcl$ simulations.  In contrast,
our observed objects all occur below the 50th
percentile in both simulations and are more representative of the
single O star cluster sample in Figures \ref{msmcm}{\it b} and
\ref{msmcmM}{\it b}.  Indeed, Figures \ref{msmcm}{\it b} and \ref{msmcmM}{\it b} show that the
majority of the Galactic cluster sample lies outside the single O star
cluster parameter space.  This indicates that the Galactic cluster
sample is comprised of clusters with a well-populated IMF, probably
due to selection effects, since more fully populated clusters
preferentially tend to be observed (Maschberger \& Clarke 2008).

\begin{figure*}
	\begin{center}
	\includegraphics[scale=.5,angle=0]{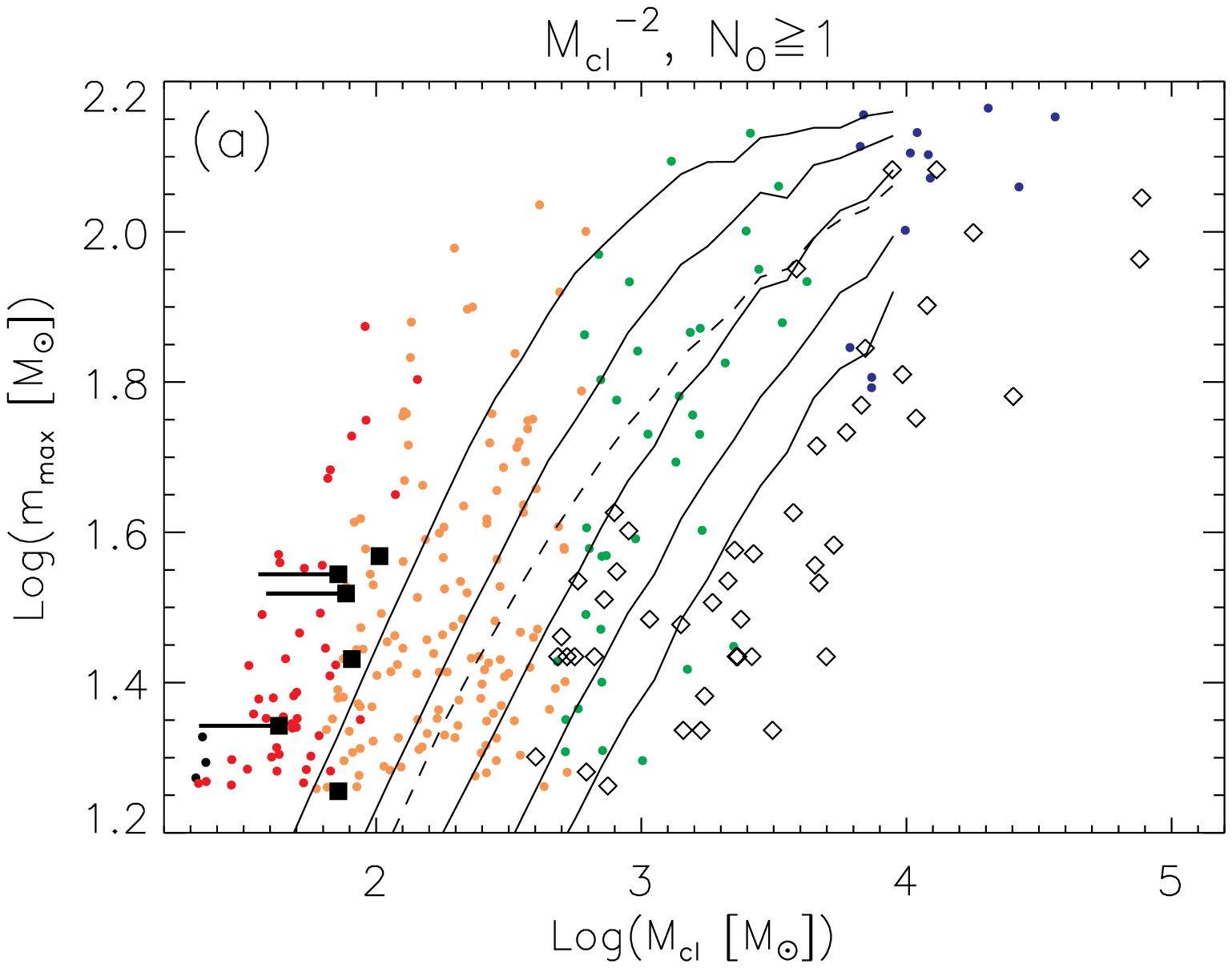}
	\includegraphics[scale=.5,angle=0]{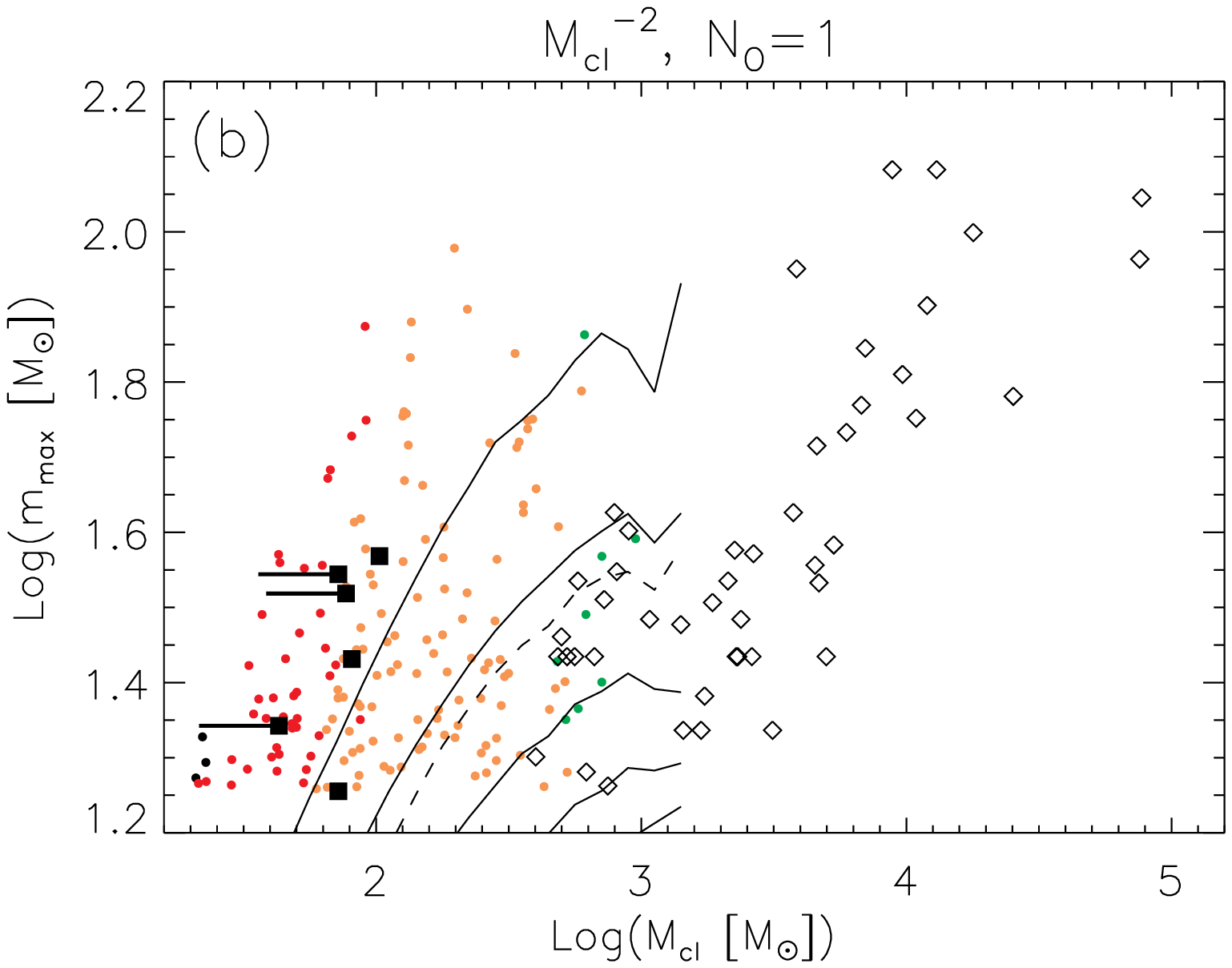}
	\caption{Same as Figure \ref{msmcmM}, except with percentile and mean lines showing $\mmax$ as a function of $\Mcl$.}
	\label{msmcmrelation}
	\end{center}
\end{figure*}

In Figure \ref{msmcmrelation}, we
plot the mean $\mmax$ value as a function of $\Mcl$ (dashed line) for
our $\Mcl^{-2}$ simulation.  The solid lines now show contours for the
10th, 25th, 50th, 75th, and 90th percentiles of $\mmax$ as a function
of $\Mcl$ in the simulation, again calculated from the larger sets of
simulated clusters.  The dashed line represents the mean.  We
note that in Figure \ref{msmcmrelation}{\it a}, the percentiles are
calculated from all clusters having a given $\Mcl$, while the simulated
clusters plotted here are only those which contain at least one O
star.  In Figure \ref{msmcmrelation}{\it b}, the percentiles exclude all
clusters with multiple O stars, and the plotted clusters are those
with a single O star.  

Here, we see that nearly all of our observed
clusters are above the 90th percentile of $\mmax$ as a function of
$\Mcl$.  In Figure \ref{msmcmrelation}{\it a}, all the Galactic clusters are below
the 50th percentile, with the majority below the 10th percentile.  Figure \ref{msmcmrelation}{\it b}
again demonstrates that the Galactic O star cluster sample largely falls outside the parameter
space of the single O star clusters.  Comparing the percentiles plotted in Figure \ref{msmcmrelation}{\it b} 
with those in Figure \ref{msmcmrelation}{\it a} shows little difference for our observed $\Mcl$ values; however,
as $\Mcl$ increases, the transition from single O star clusters to clusters with $>1$ O star is revealed in the turnover of 
the percentiles in Figure \ref{msmcmrelation}{\it b}.  This confirms that the Galactic cluster sample is comprised mainly of clusters with multiple O stars.  

\subsection{Does $\Mcl$ determine $\mmax$?}

We now compare our observations with numerical simulations that
are limited by a relation between the maximum
stellar mass within a cluster and total cluster mass ($\mmax$-$\Mcl$). As mentioned in
\S \ref{intro}, various forms of this relation have been proposed,
based on both theory (Bonnell et al. 2004) 
and observations (WK06), that invoke a physical relation between
$\mmax$ and $\Mcl$.  This is different from the purely statistical
relation between the average $\mmax$ and $\Mcl$ (Oey \& Clarke 2005).
In the latter case it is simply improbable to form a massive star in a
small cluster, whereas in the integrated galaxial initial mass 
function (IGIMF) proposed by Weidner \& Kroupa (2005), the $\mmax$-$\Mcl$
relation is modeled deterministically, such that $\mmax$ never exceeds
the value derived from this $\mmax$-$\Mcl$  relation.  For reference,
see Figure 1 of Weidner \& Kroupa (2005), which plots various
$\mmax$-$\Mcl$  relations from the literature.   

If we adopt the mean in Figure~\ref{msmcmrelation}{\it a}  as a
simple $\mmax$-$\Mcl$ relation, then all simulated clusters above the
dashed line are in violation of such a relation.  We note that the
mean corresponds to somewhat
lower-mass clusters for a given $\mmax$ than the WK06 $\mmax$-$\Mcl$
relation.  Even so, our observed minimal O star groups do not fit
within the framework of a steepened IGIMF as presented by Weidner \&
Kroupa (2005), although we note that statistical variation of the
$\mmax$-$\Mcl$ relation is not included in their work. 

\begin{figure*}
	\begin{center}
	\includegraphics[scale=.5,angle=0]{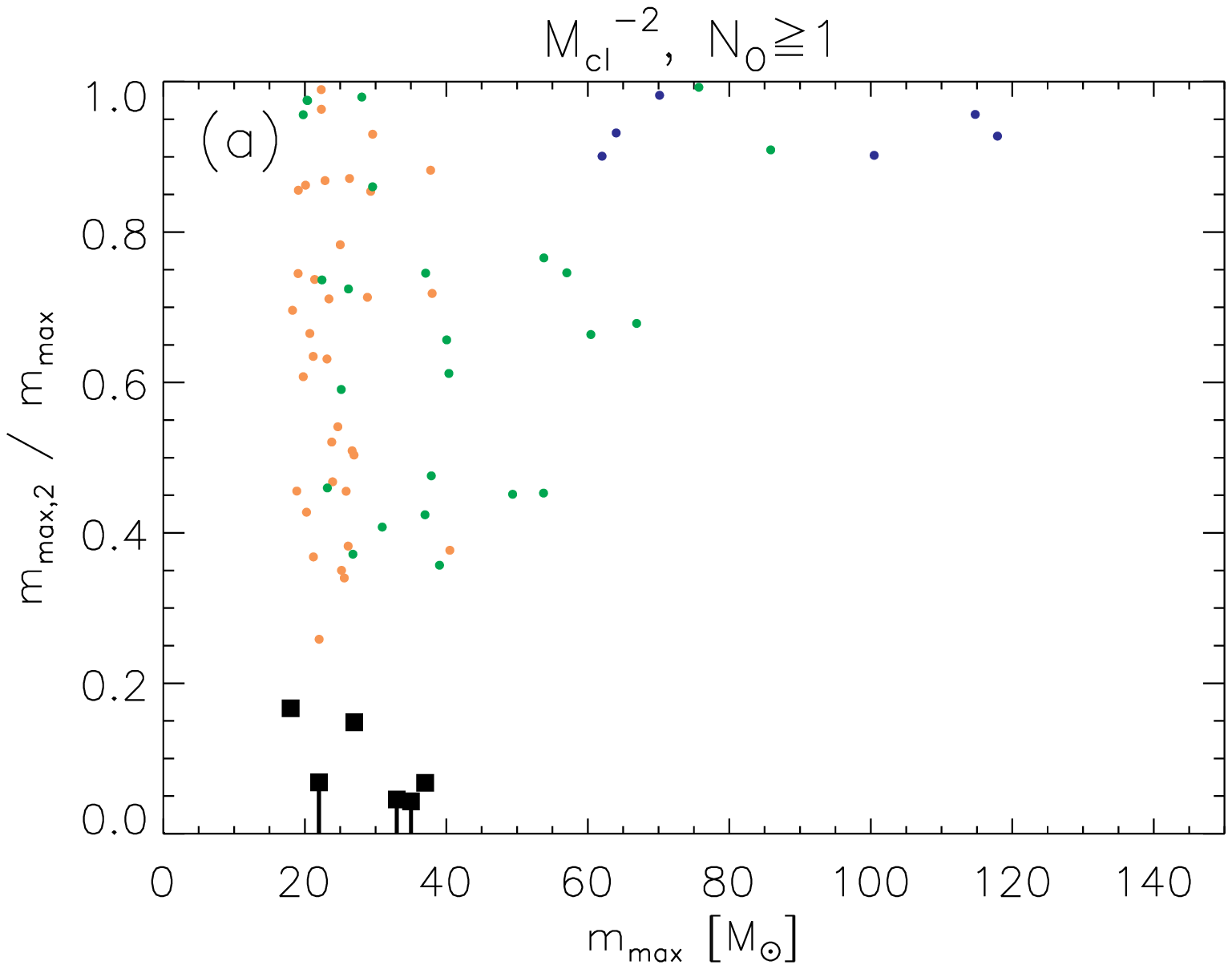}
	\includegraphics[scale=.5,angle=0]{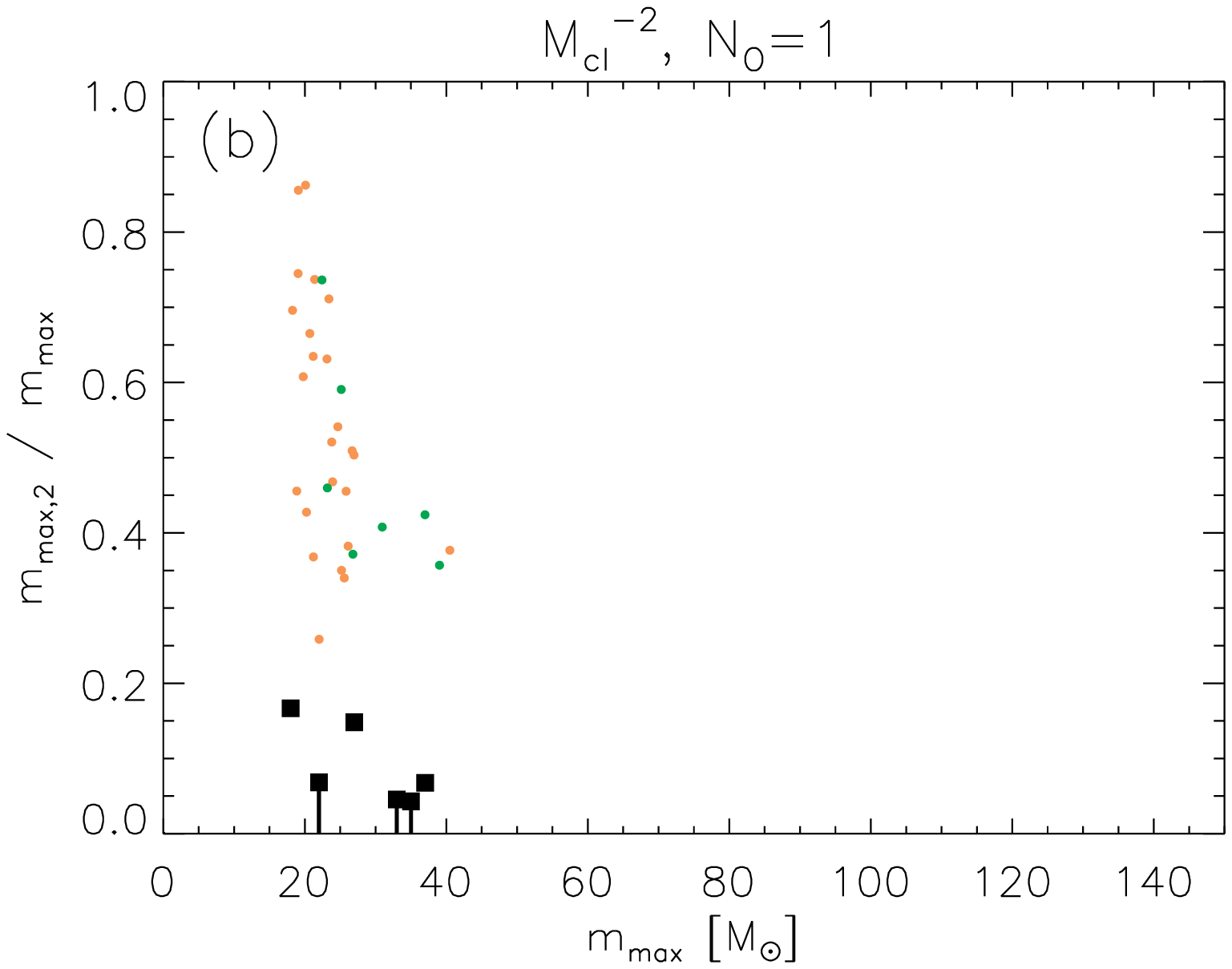}
	\caption{Same as Figure \ref{mratioM}, but applying a fixed $\mmax$ - $\Mcl$ relation.}
	\label{mfK}
	\end{center}
\end{figure*}

Figure \ref{mfK} shows $\mratio$ as a function of $\mmax$ for
clusters that correspond to our imposed  $\mmax$-$\Mcl$ relation,
which are those below the mean $\mmax$ plotted in Figure
\ref{msmcmrelation}{\it a}.  The color coding and panel samples are the same
as in Figure \ref{mratioM}.  Figure \ref{mfK}{\it a} illustrates that all simulated O
star clusters now exhibit $\mratio \geq 0.2$. 
With this imposed $\mmax$-$\Mcl$ 
relation, our observations appear to fall completely outside the parameter space
of the simulations in both Figure~\ref{mfK}{\it a} and \ref{mfK}{\it b}, given the
number of clusters corresponding to the 
SMC cluster population.  This form of the deterministic $\mmax$-$\Mcl$
relation is therefore poorly supported by our observations of minimal
O star groups. 

One possible interpretation of these results is that our observed
objects are remnants of clusters affected by ``infant weight
loss'' (e.g., Bastian \& Goodwin, 2006).  However, even if the objects have
been reduced by a factor of a few in $\Mcl$ or $N_*$, they would still be
discrepant from the Galactic cluster sample of Weidner et al. (2010a).  Recent
simulations have shown that cluster mass segregation can occur on
timescales of $\sim$1 Myr for the most massive cluster stars (Allison
et al. 2009).  Massive stars segregated to the cluster core are unlikely to be
evaporated by ``infant weight loss'', and so clusters are likely to
retain the two most massive stars.  Therefore,
the observed $\mratio$ values are unlikely to be affected by
dynamical evaporation.  Moreover, ``infant weight loss'' is
associated with the rapid removal of gas, which, however, is still
present in the majority of our objects (\S \ref{isosec}).
Furthermore, $N$-body 
simulations of sparse, young clusters by Weidner et al. (2010b) show
that $< 15$\% of the cluster's mass is removed within 5 Myr,
suggesting that significant mass loss is relatively unimportant in
such objects.

Our results are similar to those found by Maschberger \& Clarke
(2008), who examined a sample of clusters from the literature using
studies that focused on low $N_*$ clustering around high-mass stars.
The observations of isolated Herbig Ae/Be stars by Testi et
al. (1997, 1999), as well as our observations of isolated OB 
stars, show that massive stars may form in even the most sparse environments.
As we showed above,
these observations are not consistent with a strictly-defined,
deterministic $\mmax$-$\Mcl$ relation.  At the very least, the minimal
OB groups, along with clusters compiled by Weidner et al. (2010a), 
imply huge deviation from a direct $\mmax$-$\Mcl$ relation (Figure~\ref{msmcmrelation}).
While a steepened IGIMF will occur even without a $\mmax$-$\Mcl$
relation, this result has problematic implications for the magnitude
of proposed steepening for the 
IGIMF in aggregate galactic stellar populations (Weidner \& Kroupa 2005).
In addition, the competitive accretion theory of star formation is also
linked to an $\mmax$-$\Mcl$ relation (Bonnell et al. 2004), although
we note that Maschberger et al. (2010) find that competitive accretion
simulations are nevertheless able to produce massive field stars of at least $9M_\odot$.    
On the other hand, simulations of star formation based on core accretion (e.g.,
Krumholz et al. 2010) show that under specific cloud conditions,
radiative heating can prevent fragmentation, perhaps more directly
forming minimal O star groups similar to those observed in this paper.

\section{Conclusions}

We carried out a SNAP program with {\it HST's} Advanced Camera for
Surveys that yielded high resolution observations of eight field
OB stars in the SMC.  These stars range in spectral type from O7 to
B0.5, and in mass from $18M_\odot$ to $37M_\odot$.  Radial velocities
for two stars (AzV 223 and smc16) show them to be runaways, and we may expect
one or two more to be transverse runaways.  There is 
no evidence of clustering down to a $1.5M_\odot$ detection limit
in three of the six non-runaway cases.  The non-runaway,
isolated stars (AzV 58, AzV 186, and AzV 226) remain candidates for isolated OB star
formation.  Two of
these isolated OB stars reside within HII regions, indicating that
these stars may still be located in the region of their formation,
and strengthening the possibility that these
O stars have formed alone.  For the other three non-runaway OB stars (AzV 67, AzV 106, and AzV 302),
we detect an associated population of stars using a stellar density
analysis and a separate friends-of-friends algorithm.  After
accounting for field contamination, we find eight to ten stars associated
with each OB star, ranging in mass from our lower detection limit of
$1M_\odot$ to $4M_\odot$. 

The three observations that do show evidence of clustering exhibit a
flat IMF with slope of $\Gamma = 0.1 \pm 1.0$ to $-0.2 \pm 0.9$ 
when combining their
populations, although due to the small sample size, the IMF may be
consistent with a Salpeter IMF, which has $\Gamma$ = -1.35.  The flat
IMF is due to a lack of low-mass companions that ordinarily are expected to form
along with these O stars. 

Assuming that each of our non-runaway stars is still in
the location where it formed, we infer cluster membership of $N_* =
19-171$ based upon the companion population $\geq 1.5M_\odot$, integrated over the full stellar mass range for
our Kroupa IMF of $0.08M_\odot - 150M_\odot$.  Given their
inferred $N_*$, we calculate that a small fraction, only 0.01--0.2, of clusters will form a
star with mass $m \geq \mmax$ observed.

We have conducted Monte Carlo simulations to explore where our
observations fall within the cluster parameter space of typical star
formation, assuming a Kroupa IMF and either a clustering law based on
$N_*$ or cluster mass function based on $\Mcl$.   The power law slopes and lower limits of $\Mcl$ and $N_*$ for these simulations were constrained  using observations of massive star clustering
in the SMC by Oey et al. (2004) and the fraction of isolated Galactic O stars by de Wit et al. (2005).
Together, these observational constraints resulted in a --2 power law slope with lower limits of $\Nlo$ = 40 or $\Mcllo =  20M_\odot$
as the best fit models, which we adopted as the default simulations for this study.  These default $N_*$ and $\Mcl$ models match equally well
with the observations, thus neither one is established as a more suitable metric for modeling cluster distribution.

We find that observed mass ratios of the two highest-mass stars are below the 25th percentile of single O star clusters generated by the default simulations.  This result is due to choosing targets which appeared isolated in ground-based imaging.  Our observations also lie below the 50th percentile when comparing total cluster mass (either $\Mcl$ or $N_*$) as a function of $\mmax$, whereas a sample of Galactic clusters from Weidner et al. (2010a) are nearly all above the 90th percentile.  These numbers suggest that our observations are more typical examples of O star clusters than the Galactic cluster sample, which contains clusters with well-populated IMFs.  

We show evidence that our observed minimal O star groups are inconsistent
with a deterministic $\mmax$-$\Mcl$ relation.
By extension, our observations are also inconsistent with the $\mmax$-$\Mcl$
relation proposed for the IGIMF effect (Weidner \& Kroupa 2005) and the relation that
$\Mcl \propto \mmax^{1.5}$ (Bonnell et al. 2004), predicted by the
competitive accretion model of massive star formation.  We argue that
in most cases, an observed $\mmax$-$\Mcl$ relation is simply a product
of the stochastic, probabilistic nature of a universal IMF, rather
than an IMF with a variable upper limit corresponding to cluster mass.   
We conclude that our observations of minimal O star groups are consistent with a universal stellar mass function, including a constant stellar upper-mass limit, without the need to invoke a $\mmax$-$\Mcl$ relation.

\acknowledgments
We thank the referee, Simon Goodwin, and Cathie Clarke for helpful
comments on the manuscript.  We thank Fred Adams, Oleg
Gnedin, Wen-hsin Hsu, and Mario Mateo for useful 
discussions.  We are grateful to Jason Harris and Dennis Zaritsky for access to the MCPS $V$-band
imaging data and thank Andrew Graus for helpful data checking. This work was supported by
program HST-GO-10629.01, provided by NASA through a
grant from the Space Telescope Science Institute, which is operated by the
Association of Universities for Research in Astronomy, Inc., under
NASA contract NAS5-26555. We also recognize support from NSF grant AST-0907758.

\bibliography{gc}

\end{document}